\documentclass[11pt, reqno]{amsart}
\usepackage[pdfstartview=FitB]{hyperref}
\usepackage{url}
\usepackage[dcucite]{harvard}
\usepackage{float}

\usepackage{graphicx,amsmath,amssymb,epsfig,harvard}
\usepackage{colortbl}
\usepackage{tabularx}

\usepackage{mathrsfs}
\long\def\comment#1{}
\long\def\comment#1{}

\newtheorem{theorem}{Theorem}

\newtheorem{proposition}{Proposition}

\theoremstyle{definition}

\newcommand{\citen}{\citeasnoun}

\newcommand{\be}{\begin{eqnarray}}
\newcommand{\ee}{\end{eqnarray}}

\newcommand{\betahat}{\hat \beta}
\newcommand{\ind}{ \overset{d}{\longrightarrow}}

\newcommand{\ba}{\begin{array}}
\newcommand{\ea}{\end{array}}
\newcommand{\bs}{\begin{align}\begin{split}\nonumber}
\newcommand{\bsnumber}{\begin{align}\begin{split}}
\newcommand{\es}{\end{split}\end{align}}

\renewcommand{\(}{\left(}
\renewcommand{\)}{\right)}
\renewcommand{\[}{\left[}
\renewcommand{\]}{\right]}
\renewcommand{\hat}{\widehat}

\newcommand{\Ep}{{\mathrm{E}}}

\newcommand{\EnT}{{\frac{1}{nT}\sum_{i=1}^n \sum_{t=1}^T}}

\newcommand{\ddotEp}{{  \frac{1}{nT}  \sum_{i=1}^n \sum_{t=1}^T} \Ep }

\newcommand{\EnTT}{{\frac{1}{nT}\sum_{i=1}^n \sum_{t=1}^T}\sum_{t'=1}^T}

\newcommand{\PX}{\mathcal{P}_{\hat I}}

\newcommand{\MX}{\mathcal{M}_{\hat I}}

\renewcommand{\Pr}{{\mathrm{P}}}

\def\RR{ {\mathbb{R}}}

\newcommand{\ceil}[1]{\left\lceil #1 \right\rceil}
\newcommand{\semin}[1]{\varphi_{{\rm min}}(#1)}
\newcommand{\semax}[1]{\varphi_{{\rm max}}(#1)}
\renewcommand{\hat}{\widehat}
\renewcommand{\leq}{\leqslant}
\renewcommand{\geq}{\geqslant}

\usepackage{epstopdf}

\begin{document}
\title[Inference in High Dimensional Panel Data]{ Inference in High Dimensional Panel Models with an Application to Gun Control}
\author[Alexandre Belloni]{Alexandre Belloni}
\author[Victor Chernozhukov ]{Victor Chernozhukov}
\author[Christian Hansen ]{Christian Hansen  }
\author[Damian Kozbur ]{Damian Kozbur }

\date{First version:  April 2013.  This version is of  \today. }

\thanks{We thank St\'ephane Bonhomme, Elena Manresa, and seminar participants at UIUC for helpful comments.  We gratefully acknowledge financial support from the ETH Postdoctoral Fellowship.}

\sloppy

\begin{abstract}

We consider estimation and inference in panel data models with additive unobserved individual specific heterogeneity in a high dimensional setting.  The setting allows the number of time varying regressors to be larger than the sample size.  To make informative estimation and inference feasible, we require that the overall contribution of the time varying variables after eliminating the individual specific heterogeneity can be captured by a relatively small number of the available variables whose identities are unknown.  This restriction allows the problem of estimation to proceed as a variable selection problem.  Importantly, we treat the individual specific heterogeneity as fixed effects which allows this  heterogeneity to be related to the observed time varying variables in an unspecified way and allows that this heterogeneity may be non-zero for all individuals.  Within this framework, we provide procedures that give uniformly valid inference over a fixed subset of parameters in the canonical linear fixed effects model and over coefficients on a fixed vector of endogenous variables in panel data instrumental variables models with fixed effects and many instruments.  An input to developing the properties of our proposed procedures is the use of a variant of the Lasso estimator that allows for a grouped data structure where data across groups are independent and dependence within groups is unrestricted.  We provide formal conditions within this structure under which the proposed Lasso variant selects a sparse model with good approximation properties.  We present simulation results in support of the theoretical developments and illustrate the use of the methods in an application aimed at estimating the effect of gun prevalence on crime rates.

\vspace{.25in}


\end{abstract}

\maketitle

\section{Introduction}

The use of panel data is extremely common in empirical economics.  Panel data is appealing because it allows researchers to estimate the effects of variables of interest while accounting for time invariant individual specific heterogeneity in a flexible manner.  For example, the most widespread model employed in empirical analyses using panel data in economics is the linear fixed effects model which treats individual specific heterogeneity as a set of additive fixed effects to be estimated jointly with other model parameters.  This approach is attractive because it allows the researcher to estimate the common slope parameters of the model without imposing any structure over the individual specific heterogeneity.

Many panel data sets also have a large number of time varying variables available for each observation; i.e. they are ``high dimensional'' data.  The large number of available variables may arise because the number of measured characteristics is large.  For example, many panel data analyses in economics make use of county, state, or country level panels where there is a large set of measured characteristics and aggregates such as output, employment, demographic characteristics, etc. available for each observation.  A large number of time varying variables may also be present due to a researcher wishing to allow for flexible dependence of an outcome variable on a small set of observed time varying covariates and thus considering a variety of transformations and interactions of the underlying set of variables.  Identification of effects of interest in panel data contexts is also often achieved through a strategy where identification becomes more plausible as one allows for flexible trends that may differ across treatment states.  Allowing for flexible trends that may differ based on observable characteristics may then be desirable but potentially introduces a large number of control variables.

A difficulty in high dimensional settings is that useful predictive models and informative inference about model parameters is complicated by the presence of the large number of explanatory variables.  The problem with building a model for prediction can easily be seen when one considers the example of forecasting using a linear regression model in which there are exactly as many linearly independent explanatory variables as there are observations.  In this case, the ordinary least squares estimator will fit the data perfectly, returning an $R^2$ of one.  However, the estimated model is likely to provide very poor out-of-sample prediction properties because the model estimated by unrestricted least squares is overfit.  The least squares fit captures not just the signal about how the predictor variables may be used to forecast the outcome but also perfectly captures the noise in the given sample which is not useful for generating out-of-sample predictions.  Constraining the estimated model to avoid perfectly fitting the sample data, or ``regularization,'' is necessary for building a useful predictive model.  Similarly, informative inference about parameters in a linear regression model is clearly impossible if the number of explanatory variables is larger than the sample size if one is unwilling to impose additional model structure.

A useful structure which has been employed in the recent econometrics literature focusing on inference in high dimensional settings is approximate sparsity; see, for example, \citen{BellChernHans:Gauss}, \citen{BellChenChernHans:nonGauss}, and \citen{BelloniChernozhukovHansen2011}.  A leading example is the approximately sparse linear regression model which is characterized by having many covariates of which only a small number are important for predicting the outcome.\footnote{There are many statistical methods designed for doing prediction in exactly sparse models in which the relationship between an outcome and a large dimensional set of covariates is perfectly captured by a model using only a small number of non-zero parameters; see \citen{elements:book} for a review.  Approximately sparse models generalize exactly sparse models by allowing for approximation errors from using a low-dimensional approximation.}  Approximately sparse models nest conventional parametric regression models as well as standard sieve and series based nonparametric regression models.  In addition to nesting standard econometric models, the framework is appealing as it reduces the problem of finding a good predictive model to a variable selection problem.  Sensible estimation methods appropriate for this framework also yield models with a relatively small set of variables which aids interpretability of the results and corresponds to the usual approach taken in empirical economics where models are typically estimated using a small number of control variables.

There are a variety of sensible variable selection estimators that are appropriate for estimating approximately sparse models.  For example, $\ell_1$-penalized methods such as the Lasso estimator of \citen{FF:1993} and \citen{T1996} have been proposed for model selection problems in high dimensional least squares problems in part because they are computationally efficient. Many $\ell_1$-penalized methods and related methods have been shown to have good estimation properties with i.i.d. data even when perfect variable selection is not feasible; see, e.g., \citen{CandesTao2007}, \citen{MY2007}, \citen{BickelRitovTsybakov2009},  \citen{horowitz:lasso}, \citen{BC-PostLASSO} and the references therein. Such methods have also been shown to extend to nonparametric and non-Gaussian cases as in \citen{BickelRitovTsybakov2009} and \citen{BellChenChernHans:nonGauss}, the latter of which also allows for conditional heteroscedasticity.

While the models and methods mentioned above are useful in a variety of contexts, they do not immediately apply to standard panel data models.  There are two key points of departure between conventional approximately sparse high dimensional models and conventional panel data models used in empirical economics.  The first is that the approximately sparse framework seems highly inappropriate for usual beliefs about individual specific heterogeneity in fixed effects models.\footnote{There are other natural alternatives to dimension reduction over individual specific heterogeneity.  For example, one could assume that individual specific heterogeneity is drawn from some common distribution as in conventional random effects approaches.  This distribution could also be allowed to be flexibly specified as in \citen{altonji:matzkin} or \citen{bester:hansen:index}.  Alternatively, one could consider a grouped structure over heterogeneity as in \citen{bester:hansen:group} or \citen{bonhomme:manresa}.} Specifically, the approximately sparse structure would imply that individual specific heterogeneity differs from some constant level for only a small number of individuals and may be completely ignored for the vast majority of individuals.  A seemingly more plausible model for individual specific heterogeneity is one in which it is allowed to be relevant for every individual, related to observed variables in an unrestricted manner, and different for each individual.  Under these beliefs, naively applying a method designed for approximately sparse models may result in a procedure with poor estimation and inference properties.

The second key difference is that the assumption of independent observations is inappropriate for many panel data sets used in economics.  Many economic panels appear to exhibit substantial correlation between observations within the same cross-sectional unit of observation.  It is well-known that failing to account for this correlation when doing inference about model parameters in panel data with a small number of covariates may lead to tests with substantial size distortions.  This concern has led to the routine use of ``clustered standard errors'' which are robust to within-individual correlation and heterogeneity across individuals in empirical research.\footnote{See \citen{arellano:feinf}, \citen{bdm:cluster}, and \citen{hansen:cluster} among others.}  In the context of variable selection in high dimensional models, failing to account for this correlation may result in substantial understatement of sampling variability.  This understatement of sampling variability may then lead to a variable selection device selecting too many variables, many of which have no true association to the outcome of interest.  The presence of these spuriously selected variables may have a substantial negative impact on the resulting estimator as the spuriously selected variables are, by construction, the most strongly correlated to the noise within the sample.

A key contribution of this paper is offering a variant of the Lasso estimator that accommodates a clustered covariance structure (Cluster-Lasso).  We provide formal conditions under which the estimator performs well in the sense of returning a sparse estimate and having good forecasting and rate of convergence properties.  By providing results allowing for a clustered error structure, we are also able to allow for the presence of unrestricted additive individual specific heterogeneity which are treated as fixed effects that are partialed out of the model before variable selection occurs.  Accommodating this structure requires partialing out a number of covariates that is proportional to the sample size under some asymptotic sequences we consider.  In general, partialing out a number of variables proportional to the sample size will induce a non-standard, potentially highly dependent covariance structure in the partialed-out data.  The structure of the fixed effects model is such that partialing out the fixed effects cannot induce correlation across individuals, though it may induce strong correlation within the observations for each individual.  Because this structure is already allowed for in the clustered covariance structure, partialing out the fixed effects poses no additional burden after allowing for clustering.

The second contribution of this paper is taking the derived performance bounds for the proposed Lasso variant and using them to provide methods for doing valid inference following variable selection in two canonical models with high dimensional components:  the linear instrumental variables (IV) regression with high dimensional instruments and additive fixed effects and the partially linear treatment model with high dimensional controls and additive fixed effects.  Inference in these settings is complicated due to the fact that variable selection procedures inevitably make model selection mistakes which may result in invalid inference following model selection; see \citen{potscher} and \citen{leeb:potscher:pms} for examples.  It is thus important to offer procedures that are robust to such model selection mistakes. To address this concern, we follow the approach of \citen{BellChenChernHans:nonGauss} in the IV model and \citen{BelloniChernozhukovHansen2011} in the partially linear model making use of the Cluster-Lasso to accommodate within-individual dependence and partialing out of fixed effects.  We show that standard inference following these procedures results in inference about model parameters of interest that is uniformly valid within a large class of approximately sparse regression models as long as a clustered covariance estimator is used in estimating the parameters' asymptotic variance.  The results of this paper thus allow valid inference about a prespecified, fixed set of model parameters of interest in canonical panel data models with additive fixed effects in the realistic scenario where a researcher is unsure about the exact identities of the relevant set of variables to be included in addition to the variables of interest and the fixed effects.

In addition to theoretical guarantees, we illustrate the performance of the proposed methods through simulation examples.  In the simulations, we consider a fixed effects IV model and a conventional linear fixed effects model.  The IV model has a single endogenous variable whose coefficient we would like to infer and a large number of instruments which satisfy the IV exclusion restriction only after eliminating the fixed effects.  Given the large number of instruments, we use the Cluster-Lasso to select a small set of instruments to use in an IV estimator as in \citen{BellChenChernHans:nonGauss}.  In the linear model, we have a single treatment variable of interest that is related to a set of fixed effects and additionally have a large number of potential confounding variables.  We estimate the effect of the variable of interest using the double selection procedure of \citen{BelloniChernozhukovHansen2011} with the Cluster-Lasso used as the variable selector.  In both cases, we find that the Cluster-Lasso-based procedures perform well in terms of both estimation risk and inference properties as measured by size of tests.  The most interesting feature of the simulation results is that the Cluster-Lasso-based procedures perform markedly better than variable selection procedures that do not allow for clustering.  This difference in performance suggests that additional modifications of Lasso-type procedures to account for other dependence structures may be worthwhile.

We also provide results from an empirical example that looks at estimating the effect of guns on crime using data from a panel of U.S. counties following the analysis of \citen{cook:ludwig:guns}.  In their analysis, \citen{cook:ludwig:guns} use a conventional linear fixed effects model with a small number of time varying county level control variables and find a positive and statistically significant effect of their proxy of gun prevalence on the overall homicide rate and the gun homicide rate and a negative but insignificant effect on the non-gun homicide rate.  We extend this analysis by considering a broad set of county-level demographic characteristics and a set of flexible trends that are interacted with baseline county-level characteristics.  We then use the methods developed in this paper to select a small, data-dependent set of variables that it is important to control for if one wants to hold fixed important sources of confounding variation.  Interestingly, our findings are largely consistent with those of \citen{cook:ludwig:guns} despite allowing for a much richer set of conditioning information.  Specifically, we find a strong positive relationship between gun prevalence and gun homicides, a small and statistically insignificant relationship between gun prevalence and non-gun homicides, and a borderline significant positive effect of gun prevalence on the overall homicide rate.

\section{Dimension Reduction and Regularization via Lasso Estimation in Panels}\label{ClusterLassoSetUp}

There are many regularization or dimension reduction techniques available in the statistics and econometrics literature.  An appealing method for estimating sparse high dimensional linear models is the Lasso.  Lasso estimates regression coefficients by minimizing a least squares objective plus an $\ell_1$ penalty term.  We begin with an informal discussion of Lasso in linear models with fixed effects before proceeding  with more precise specifications and modeling assumptions.  In particular, our goal for this section is to outline a Lasso procedure for estimating the model
$$y_{it} = x_{it}'\beta + \alpha_i + \epsilon_{it}, \quad i =1,..., n, \quad t = 1,..., T,$$
where $y_{it}$ is an outcome of interest, $x_{it}$ are covariates, $\alpha_i$ are individual specific effects, and $\epsilon_{it}$ is an idiosyncratic disturbance term which is mean zero conditional on covariates but may have dependence within an individual.\footnote{We abstract from issues arising from unbalanced panels for notational convenience but note that the arguments go through immediately provided that the missing observations are missing at random.}

\subsection{Cluster-Lasso Estimation in Panel Models}

The first step in our estimation strategy is to eliminate the fixed effect parameters.  
For simplicity, we will always consider removing the fixed effects by within individual demeaning but note that removing the fixed effects using other differencing methods could be accommodated using similar arguments.   To this end, we define$$\ddot x_{it}  = x_{it} - \frac{1}{T}\sum_{t=1}^T x_{it}.$$  We define the quantities $\ddot y_{it}$ and $\ddot \epsilon_{it}$ similarly and note that the double dot notation will signify deviations from within individual means throughout the paper. Eliminating the fixed effects by substracting individual specific means leads to the ``within model'':

$$\ddot y_{it} = \ddot x_{it}' \beta + \ddot \epsilon_{it}.$$

The Cluster-Lasso coefficient estimate  $\hat \beta_L$ is defined by the solution to the following penalized minimization problem on the within model:

\begin{align}\label{Lassoprob}
\betahat_{L} & \in \text{arg} \min_{b} \frac{1}{nT} \sum_{i=1}^n \sum_{t=1}^T (\ddot y_{it} - \ddot x_{it}'b)^2 + \frac{\lambda}{nT} \sum_{j=1}^p \hat \phi_j |b_j|.
\end{align}

\noindent Solving the problem (\ref{Lassoprob}) requires two user-specified tuning parameters:  the main penalty level, $\lambda$, and covariate specific penalty loadings, $\{\hat \phi_j\}_{j=1}^{p}$.  The main penalty parameter dictates the amount of regularization in the Lasso procedure and serves to balance overfitting and bias concerns.  The covariate specific penalty loadings $\{\hat \phi_j\}_{j=1}^{p}$
are introduced to allow us to handle data which may be dependent within individual, heteroscedastic, and non-Gaussian.  
We provide further discussion of the specific choices of penalty parameters in the next subsection.  Correct choice of these penalty parameters is particularly important when Lasso is used and the ultimate goal is inference about parameters of a ``structural'' model.

We will also make use of a post model selection estimator; see for example \citen{BC-PostLASSO}.  The Post-Cluster-Lasso estimator is defined with respect to the variables selected by Cluster-Lasso: $\hat I = \{ j : \betahat_{Lj} \neq 0\}$. The Post-Cluster-Lasso estimator is simply the least squares estimator subject to the constraint that covariates not selected in the initial Cluster-Lasso regression must have zero coefficients:
\begin{align}\label{PostLassoprob}
\betahat_{PL} = \underset{b : \ b_j = 0 \ \forall \ j \notin \hat I }{\text{argmin}}  \ \ \ \frac{1}{nT} \sum_{i=1}^n \sum_{t=1}^T (\ddot y_{it} - \ddot x_{it}'b)^2.
\end{align}
As discussed in Section \ref{ClusterLassoTheory}, the selected model $\hat I$ has good properties under regularity conditions and approximate sparsity of the coefficient $\beta$.  Just as in \citen{BC-PostLASSO}, good properties of the selected set of variables will then translate into good properties for the Post-Cluster-Lasso estimator.

\subsection{Clustered Penalty Loadings}

An important condition used in proving favorable performance of Cluster-Lasso and inference following Cluster-Lasso-based model selection is the use of penalty loadings and penalty parameters that dominate the score vector in the sense that

\begin{align}\label{RegularizationEvent}
\frac{\lambda \hat \phi_j} {nT} \geq 2 c \left |  \frac{1}{{ nT}}\sum_{i=1}^n \sum_{t=1}^T \ddot x_{itj} \ddot \epsilon_{it} \right | \hspace{2mm} \text{for each } 1 \leq j \leq p,
\end{align}

\noindent for some constant slack parameter $c>1$. \citen{BelloniChernozhukovHansen2011} refer to condition (\ref{RegularizationEvent}) as the ``regularization event''.  Note that the term $\frac{1}{nT}\sum_{i=1}^n \sum_{t=1}^T \ddot x_{itj} \ddot \epsilon_{it}$ intuitively captures the sampling variability in learning about coefficient $\beta_j$.  The regularization event thus corresponds to selecting penalty parameters large enough to dominate the noise in estimating model coefficients.   Looking at the structure of the Lasso optimization problem, (\ref{Lassoprob}), we can see that (\ref{RegularizationEvent}) leads to settings all coefficients whose magnitude is not big enough relative to sampling noise exactly to zero in the Lasso solution.  This property makes Lasso-based methods appealing for forecasting and variable selection in sparse models where many of the model parameters can be taken to be zero and it is desirable to exclude any variables from the model that cannot reliably be determined to have strong predictive power.

Given the importance of event (\ref{RegularizationEvent}) in verifying desirable properties of Lasso-type estimators, it is key that penalty loadings and the penalty level are chosen so that (\ref{RegularizationEvent}) occurs with high probability.  The intuition for suitable choices can be seen by considering $\hat \phi_j = \phi_j$
where
$$\phi_j^2 = \frac{1}{nT} \sum_{i=1}^n \left ( \sum_{t=1}^T  \ddot x_{itj}  { \ddot \epsilon}_{it} \right)^2 = \frac{1}{nT} \sum_{i=1}^n \sum_{t=1}^T \sum_{t' = 1}^T \ddot x_{itj} \ddot x_{it'j}  { \ddot \epsilon}_{it} { \ddot \epsilon}_{it'}.$$
Note that the quantity $\phi_j^2$ is a natural measure for the noise in estimating $\beta_j$ that allows for arbitrary within-individual dependence.   
With these loadings, we can apply the moderate deviation theorems for self-normalized sums due to \citen{jing:etal} to conclude that $$\frac{P( \phi_j^{-1} \frac{1}{\sqrt{nT}} \sum_{i=1}^n \sum_{t=1}^T \ddot x_{itj} \ddot \epsilon_{it}>m)}{P(N(0,1)>m)}  = o(1),    \quad \text{uniformly in }  |m| = o(n^{1/6}), \ j \in {1,...,p}. $$  It follows from this result and the union bound that setting $\lambda$ large enough to dominate $p$ standard Gaussian random variables with high-probability, specifically as in (\ref{eq: set lambda}), will implement condition (\ref{RegularizationEvent}) using $\hat \phi_j = \phi_j$.

This form of loadings is an extension of the loadings considered in \citen{BellChenChernHans:nonGauss} which apply in settings with non-Gaussian and heteroscedastic but independent data to the present setting where we need to accommodate strong within-individual dependence.
Formally verifying the validity of this approach requires control of the tail behavior of each of the sums $\phi_j^{-1} \frac{1}{{nT}} \sum_{i=1}^n  \sum_{t=1}^T \ddot x_{itj} \ddot \epsilon_{it}$ uniformly over $j \leq p$. \citen{BellChenChernHans:nonGauss} verify the appropriate uniform tail control under independence across observations using loadings suitable for independent observations and $\lambda =  2 c \sqrt{nT} \Phi^{-1}(1-o(1)/2p)$ by using results from the theory of moderate deviations of self-normalized sums from \citen{jing:etal}. We extend these results to allow for a clustered dependence structure in the observations which is appropriate for panel data and well-suited to fixed effects applications.

In practice, the values $\{\phi_j\}_{j=1}^{p}$ are infeasible since they depend on the unobservable $\ddot \epsilon_{it}$.  To make estimation feasible, we use preliminary estimates of $\ddot \epsilon_{it}$, denoted $\hat {{\epsilon}}_{it}$, in forming feasible loadings:

\begin{align}\label{FeasibleLoadings}
\hat \phi_j ^2 = \frac{1}{nT} \sum_{i=1}^n \left ( \sum_{t=1}^T  \ddot x_{itj}  { \hat {\epsilon}}_{it}  \right)^2 = \frac{1}{nT} \sum_{i=1}^n \sum_{t=1}^T \sum_{t' = 1}^T \ddot x_{itj} \ddot x_{it'j} \hat {  \epsilon}_{it}    \hat { \epsilon}_{it'}.
\end{align}

\noindent The $\hat{ \epsilon}_{it}$ can be calculated through an iterative algorithm given in Appendix \ref{ClusterImplementation} which follows the algorithm given in \citen{BellChenChernHans:nonGauss} and \citen{BelloniChernozhukovHansen2011}.   We define the Feasible Cluster-Lasso and Feasible Post-Cluster-Lasso estimates as the Cluster-Lasso and Post-Cluster-Lasso estimates using the feasible penalty loadings.   A key property of the feasible penalty loadings needed for validity of the approach is that
 \begin{equation}\label{eq: valid loadings}
\begin{array}{c}
\ell \phi_j  \leq \hat \phi_j \leq u \phi_j,  \quad \text{with probability $1- o(1)$},\\
\text{ for some $\ell \rightarrow 1$ and $u \leq C < \infty$, uniformly for $j=1,...,p$. }
\end{array}
\end{equation}
Under this condition and setting
\begin{equation}\label{eq: set lambda}
\lambda = 2 c \sqrt{n T} \Phi^{-1}(1-\gamma/2p)
\end{equation} with $\gamma = o(1)$, 
the regularization event (\ref{RegularizationEvent}) holds with probability tending to one. 
  
It is worth noting that failure to use the clustered penalty loadings defined in (\ref{FeasibleLoadings}) (or their infeasible version) can lead to an inflated probability of failure of the regularization event.  When this event fails to hold, covariates which are only spuriously related with the outcome have a non-negligable chance of entering the selected model.  In the simulation experiments provided in Section \ref{SimulationSection}, we demonstrate how inclusion of such variables can be problematic for post-model-selection inference essentially due to their introducing a type of endogeneity bias.\footnote{A variable which is spuriously selected must have non-negligible correlation to the errors within sample which results in similar behavior as when an endogenous variable is included in a regression.}

\section{Regularity Conditions and Performance Results for Cluster-Lasso Under Grouped Dependence}\label{ClusterLassoTheory}

This section gives conditions under which Cluster-Lasso and Cluster-Post-Lasso attains favorable performance bounds.  These bounds are useful in their own right and are important elements in establishing the properties of inference following Lasso variable selection discussed in Section \ref{PostSelectionInference}.  In establishing our formal results, we consider the additive fixed effects model
\begin{align}\label{eq: FEmodel}
y_{it} = f(w_{it}) + e_i + \epsilon_{it}, \quad \Ep[\epsilon_{it}|w_{i1},...,w_{iT}] = 0, \quad i=1,...,n, \quad t=1,...,T,
\end{align}
where $e_i$ represents time invariant individual specific heterogeneity that is allowed to depend on $w_{i} = \{w_{it}\}_{t=1}^{T}$ in an unrestricted manner.  Throughout, we will assume that $\{y_{it},w_{it}\}_{t=1}^{T}$ are i.i.d. across $i$ but do not restrict the within individual dependence.\footnote{Note that we impose that data are i.i.d. across $i$ for notational convenience.  We could allow for data that are i.n.i.d. across $i$ at the cost of complicating the notation and statement of the regularity conditions.}
Our results will hold under $n \rightarrow \infty$, $T$ fixed asymptotics and $n \rightarrow \infty$, $T \rightarrow \infty$ joint asymptotics.\footnote{Because we want to accommodate both $n \rightarrow \infty$, $T$ fixed asymptotics and $n \rightarrow \infty$, $T \rightarrow \infty$ joint asymptotics in a simple and unified manner, we maintain strict exogeneity of $w_i$ throughout and do not consider time effects.  We note that time effects may be included easily under $n \rightarrow \infty$, $T$ fixed asymptotics but that some modification of the formal results would be needed under $T \rightarrow \infty$ sequences.}

A key distinction between the analysis in this paper and previous work on Lasso is allowing for within-individual dependence.  To aid discussion of this feature, we let
$$
\imath_T :=  T \min_{1 \leq j \leq p }\frac{  \Ep [ \frac{1}{T} \sum_{t=1}^T  \ddot x^2_{itj} \ddot \epsilon^2_{it}]}{ \Ep[ \frac{1}{T} (\sum_{t=1}^T \ddot x_{itj} \ddot \epsilon_{it})^2] } = T \min_{1 \leq j \leq p}  \frac{\Ep [ \frac{1}{T} \sum_{t=1}^T  \ddot x^2_{itj} \ddot \epsilon^2_{it}] }{\Ep [\phi^2_j]}
$$
be the\textit{ index of information} induced by the ``time" or ``within-group" dimension.  This time information index,  $\imath_T$, is inversely related to the strength of within-individual dependence and can vary between two extreme cases:
\begin{itemize}
\item $\imath_T = 1$, no information, corresponding to perfect dependence within the cluster $i$,
\item $\imath_T = T$, maximal information, corresponding to perfect independence within  $i$.
\end{itemize}
There are many interesting cases between these extremes.  A leading case is where $\imath_T \propto T$ which occurs when there is weak dependence within clusters and results in clustering only affecting the constants in the Lasso performance bounds.  The case where $\imath_T \propto T^{a}$ for some $0 \leq a < 1$ corresponds to stronger forms of dependence within clusters which could be generated, for example, by fractionally integrated data.  Our results will allow for the two extreme cases as well as those falling between these two extremes.

We begin the presentation of formal conditions by defining approximately sparse models.  Note that the model $f$ as well as the set of covariates $w_{it}$ may depend on the sample size, but we suppress this dependence for notational convenience.

\textbf{Condition ASM.} (Approximately Sparse Model).   \textit{The function $f(w_{it})$ is well-approximated by a linear combination of a dictionary of transformations, $x_{it} = X_{nT}(w_{it})$, where $x_{it}$ is a $p \times 1$ vector with $p \gg n$ allowed, and $X_{nT}$ is a measurable map.   That is, for each $i$ and $t$}
$$ f(w_{it}) = x_{it}' \beta + r(w_{it}),$$
\textit{where the coefficient $\beta$ and the remainder term $r(w_{it})$ satisfy}
$$\|\beta\|_0  \leq  s = o(n\imath_T) \ \ \ and \ \ \
 \left [\frac{1}{nT}\sum_{i=1}^n \sum_{t=1}^T r(w_{it})^2 \right ]^{1/2} \leq A_{s} = O_\Pr( \sqrt{s/n\imath_T}).$$

\noindent We note that the approximation error $r(w_{it})$ is restricted to be of the same order as or smaller than sampling uncertainty in $\beta$ provided that the true model were known. Because we will mainly be concerned with the within model, we note that it is straightforward to show that $\ddot y_{it} = \ddot f(w_{it}) + \ddot \epsilon_{it}$ satisfies Condition ASM when the original model does.

The next assumption controls the behavior to the empirical Gram matrix.  Let $\ddot M$ be the $p \times p$ matrix of the sample covariances between the variables $\ddot x_{itj}$.   Thus,
$$
\ddot M = \{M_{jk}\}_{j,k=1}^{p}, \quad  M_{jk} = \frac{1}{nT} \sum_{i=1}^n \sum_{t=1}^T \ddot x_{itj} \ddot x_{itk}.
$$
In standard regression analysis where the number of covariates is small relative to the sample size, a conventional assumption used in establishing desirable properties of conventional estimators of $\beta$ is that $\ddot M$ has full rank.  In the high dimensional setting, $\ddot M$ will be singular if $p \geq n$ and may have an ill-behaved inverse even when $p < n$.  However, good performance of the Lasso estimator only requires good behavior of certain moduli of continuity of $\ddot M$.  There are multiple formalizations and moduli of continuity that can be considered in establishing the good performance of Lasso; see  \citen{BickelRitovTsybakov2009}.  We focus our analysis on a simple eigenvalue condition that is suitable for most econometric applications.  It  controls the minimal and maximal $m$-sparse eigenvalues of $\ddot M$ defined as
\begin{equation}\label{Def:EigSparse}
\semin{m}(\ddot M) = \min_{\delta \in \Delta(m)}  \delta'\ddot  M\delta \ \ \mbox{and} \ \
\semax{m}(\ddot  M) = \max_{\delta \in \Delta(m)}  \delta'\ddot  M\delta.
\end{equation}
where
$$
\Delta(m) = \{ \delta \in \mathbb{R}^p: \|\delta\|_0\leq m, \|\delta\|_2 = 1\},
$$
is the $m$-sparse subset of a unit sphere.

In our formal development, we will make use of the following simple sufficient condition:

\textbf{Condition SE.} (Sparse Eigenvalues) \textit{ For any $C>0$, there exist constants
$0< \kappa' <  \kappa'' < \infty$, which do not depend on $n$ but may depend on $C$, such
that with probability approaching one, as $n \to \infty$,
$\kappa' \leq \semin{Cs}(\ddot M) \leq \semax{Cs}(\ddot M) \leq \kappa''$.
}

\noindent Condition SE requires only that certain ``small" $Cs \times Cs$ submatrices of the large $p \times p$ empirical Gram matrix are well-behaved.  This condition seems reasonable and will be sufficient for the results that follow.  Note that we prefer to write the eigenvalue conditions in terms of the demeaned covariates as it is straightforward to show that the conditions continue to hold under data generating processes where the covariates have nonzero within-individual variation.  Condition SE could be shown to hold under more primitive conditions by adapting arguments found in \citen{BC-PostLASSO} which build upon results in \citen{ZhangHuang2006} and \citen{RudelsonVershynin2008}; see also \citen{RudelsonZhou2011}.

The final condition collects various rate and moment restrictions.  Again, the conditions are expressed in terms of demeaned quantities for convenience.  Also consider the following third moment:
$$
\varpi_j = \(\Ep\left [ \left | \frac{1}{\sqrt{T}} \sum_{t=1}^T \ddot x_{itj} \ddot \epsilon_{it}\right |^3 \right]\)^{1/3}.
$$

\textbf{Condition R.} (Regularity Conditions) \textit{Assume that for data $\{y_{it},w_{it}\}$ that are i.i.d. across $i$, the following conditions hold with $x_{it}$ defined as in Condition ASM with probability $1-o(1)$:}

\textit{(i)}  $\(\frac{1}{T} \sum_{t=1}^ T \Ep[\ddot x_{itj}^2 \ddot \epsilon_{it}^2]\) + \(\frac{1}{T} \sum_{t=1}^ T \Ep[\ddot x_{itj}^2 \ddot \epsilon_{it}^2]\)^{-1}  = O(1)$,

\textit{(ii)} $1 \leq \max_{1 \leq j \leq p } \phi_j/\min_{1 \leq j \leq p} \phi_j = O(1)$,  

\textit{(iii)} $1 \leq \max_{1 \leq j \leq p} \varpi_j/\sqrt{\Ep \phi_j^2} = O(1),$ 

\textit{(iv)}  $\log^3 (p) = o(nT) $ and $\ s \log (p\vee nT) = o(n\imath_T)$,

\textit{(v)}  $\max_{1 \leq j \leq p}|\phi_j - \sqrt{\Ep \phi_j^2}|/\sqrt{\Ep \phi_j^2} = o(1)$.

This condition is sufficient under the high-level assumption (\ref{eq: valid loadings}) on the availability of the valid feasible data loadings. In the appendix we provide additional conditions under which we exhibit validity of data-dependent loadings constructed using an iterative algorithm of the type proposed in \citen{BellChenChernHans:nonGauss}.

 With the above conditions in place, we can state the asymptotic performance bounds of Cluster-Lasso.

\begin{theorem}[Model Selection Properties of Cluster-Lasso
and Post-Cluster-Lasso]\label{ClusterLassoRates} Let $\{P_{n,T}\}$ be a sequence of  probability laws, such that $ \{(y_{it}, w_{it}, x_{it}) \}_{t=1}^T \sim P_{n,T}$, i.i.d. across $i$ for which  $n, \ T \rightarrow \infty$ jointly or $n \rightarrow \infty, \ T$ fixed.  Suppose that Conditions  ASM, SE and R hold for probability measure $\Pr =\Pr_{P_{n,T}}$ induced by $P_{n,T}$. Consider a feasible Cluster-Lasso estimator with penalty level (\ref{eq: set lambda}) and loadings obeying (\ref{eq: valid loadings}).  Then the data-dependent model $\widehat I$ selected by a feasible Cluster-Lasso estimator satisfies with probability $1-o(1)$, $\hat s = | \widehat I | \le Ks$ for some constant $K > 0$ that does not depend on $n$.  In addition, the following relations hold for the Cluster-Lasso estimator ($\betahat = \betahat_L$) and Post-Cluster-Lasso estimator ($\betahat = \betahat_{PL}$):
 {\small $$ \frac{1}{nT}\sum_{i=1}^n \sum_{t=1}^T (\ddot x_{it}'\hat \beta   -  \ddot x_{it}'\beta)^2   =O_{\Pr} \( {\frac{s \log (p\vee nT )}{ n \imath_T}} \), $$ $$\ \
\|\hat{\beta} - \beta \|_2  =O_{\Pr} \(\sqrt{\frac{s \log (p\vee nT )}{ n \imath_T}}\)
$$
$$  \|\hat{\beta} - \beta \|_1  =O_{\Pr} \(\sqrt{\frac{s^2 \log (p\vee nT )}{ n \imath_T }} \).
$$}
\end{theorem}

Theorem \ref{ClusterLassoRates} shows that Cluster-Lasso and Post-Cluster-Lasso continue to have good model selection and prediction properties allowing for a clustered dependence structure when penalty loadings that account for this dependence are used.
Establishing these results is important as applied researchers in economics typically assume the data have a clustered dependence structure and because it allows us to accommodate partialing out a large number of fixed effects which in general will induce a clustered error structure in the demeaned data.  The simulation results in Section \ref{SimulationSection} also illustrate the importance of allowing for clustering not just in calculating standard errors but also in forming penalty loadings when selecting variables using Lasso, showing that inference about coefficients of interest following variable selection using Lasso may be poor when this dependence is ignored when forming penalty loadings.

\section{Applications of Cluster-Lasso}\label{PostSelectionInference}

The bounds derived in Section \ref{ClusterLassoTheory} allow us to derive the properties of inference methods following variable selection with the Cluster-Lasso.  In this section, we use these results to provide two different applications of using Cluster-Lasso to select variables for use in causal inference.

\subsection{Selection of Instruments}

Instrumental variables techniques are widely used in applied economic research.  While these methods give an important tool for calculating structural effects, they are often imprecise.  One way to improve the precision of instrumental variables estimators is to use many instruments or to try to approximate the optimal instruments as in \citen{amemiya:optimalIV}, \citen{chamberlain}, and \citen{newey:optimaliv}.

In this section, we follow \citen{BellChenChernHans:nonGauss} who consider using Post-Lasso to estimate optimal instruments.  Using Lasso-based methods to form first-stage predictions in IV estimation provides a practical approach to obtaining the efficiency gains from using optimal instruments while dampening the problems associated with many instruments.  We prove that Cluster-Lasso-based procedures produce first-stage predictions that provide good approximations to the optimal instruments when controlling for individual heterogeneity through fixed effects.  We consider the following model:
\begin{align}
\label{IVmodel}
y_{it} &= \alpha d_{it} + e_i + \epsilon_{it} \\
\label{IVfirststage}
d_{it} &= h(w_{it}) + f_i + u_{it}
\end{align}
where $\Ep[\epsilon_{it} u_{it}] \ne 0$ but $\Ep[\epsilon_{it}|w_{i1},...,w_{iT}] = \Ep[u_{it}|w_{i1},...,w_{iT}] = 0$.\footnote{The extension to $d_{it}$ an $r \times 1$ vector with $r \ll nT$ fixed is straightforward and omitted for convenience.  The results also carry over immediately to the case with a small number of included exogenous variables $y_{it} = \alpha d_{it} + x_{it}'\beta + e_i + \epsilon_{it}$ where $x_{it}$ is a $k \times 1$ vector with $k \ll nT$ fixed that will be partialed out with the fixed effects.}  

We consider estimation of the parameter of interest $\alpha$, the coefficient on the endogenous regressor, using Cluster-Lasso to select instruments.  We assume that the first-stage follows an approximately sparse model with $h(w_{it}) = z_{it}'\pi + r(w_{it})$ where we let $z_{it} = z(w_{it})$ denote a dictionary of transformations of underlying instrument $w_{it}$ and $\pi$ be a sparse coefficient as in Condition ASM. After eliminating the fixed effect terms through demeaning, the model reduces to 
\begin{align}
\label{IVmodelDemeaned}
\ddot y_{it} &= \alpha \ddot d_{it} +\ddot  \epsilon_{it} \\
\label{IVfirststageDemeaned}
\ddot d_{it} &= \ddot h(w_{it}) + \ddot u_{it} = \ddot D_{it} + \ddot u_{it} = \ddot z_{it}'\pi + \ddot r(w_{it}) + \ddot u_{it}
\end{align}
where we set $D_{it}=h(w_{it})$ for notational convenience.
By Theorem 1, the Cluster-Lasso estimate of the coefficients on $\ddot z_{it}$ when we use $\ddot z_{it}$ to predict $\ddot d_{it}$, $\hat \pi$, will be sparse with high probability.  Letting $\hat  I_\pi = \{ j : \hat \pi_j \neq 0 \}$, the Cluster-Lasso-based estimator of $\alpha$ may be calculated by standard two stage least squares using only the instruments selected by Cluster-Lasso: $\ddot z_{it \hat I_\pi} := \(\ddot z_{itj} \)_{ j \in \hat I_\pi}$.  That is, we define the Post-Cluster-Lasso IV estimator for $\alpha$ as
\begin{align}\label{LassoIV}
\hat \alpha = \hat Q^{-1}  \EnT \hat D_{it} \ddot y_{it} \ \textnormal{where} \ \hat Q = \EnT \ddot d_{it} \hat D_{it},
\end{align}

\noindent $\hat {D_{it}}$ is the fitted value from the regression of $\ddot d_{it}$ on $\(\ddot z_{itj} \)_{ j \in \hat I_\pi}$,\footnote{That is, $\hat D_{it}$ is the Post-Cluster-Lasso forecast of $\ddot d_{it}$ using $\ddot z_{it}$ as predictors.} and $\hat \epsilon_{it} = \ddot y_{it} - \hat \alpha \ddot d_{it}$.  
We then define an estimator of the asymptotic variance of $\hat \alpha$, which will be used to perform inference for the parameter $\alpha$ after proper rescaling, as
\begin{align}\label{LassoIVVAR}
\hat V =  \hat Q ^{-1} \(  \EnTT \ddot d_{it} \ddot d_{it'} \hat \epsilon_{it} \hat \epsilon_{it'} \) \hat Q^{-1}.
\end{align}

\noindent Scaled appropriately, the estimate $\hat V$ will be close to the quantity
$$V = \frac{\imath_T^D}{T} Q^{-1} \Omega Q^{-1}, \ \ \text{with  probability $1-o(1)$} $$
where
$$Q=\Ep[ \frac{1}{T} \sum_{t=1}^T \ddot D_{it}^2], \ \ \ \Omega = \frac{1}{nT} \sum_{i=1}^n \sum_{t=1}^T \sum_{t'=1}^T \Ep[ \ddot D_{it} \ddot D_{it'} \ddot \epsilon_{it} \ddot \epsilon_{it'}].$$

Finally, it is convenient to define the following quantities that are useful in discussing formal conditions for our estimation procedure.  We define appropriate moments and information indices analogous to those used to derive properties of Cluster-Lasso and Post-Cluster-Lasso.  For any arbitrary random variables, $A=\{A_{it}\}_{i\leq n,t\leq T}$, define

$$
\phi^2(A) = \frac{1}{n}\sum_{i=1}^n \( \frac{1}{\sqrt T} \sum_{t=1}^T A_{it} \)^2, \ \ 
 \varpi (A)= \Ep\left [ \left | \frac{1}{\sqrt{T}} \sum_{t=1}^T A_{it}\right |^3 \right]^{1/3}, \ \
\imath_T(A) = T \frac{\Ep \[\frac{1}{T}\sum_{t=1}^T A_{it}^2 \] } {\Ep \[ \phi^2(A) \]}.
$$
For use in the instrumental variables estimation, we let
\begin{align*}
&\phi_j^2 = \phi^2(\{\ddot z_{itj} \ddot u_{it}\}), \ \ \
\varpi_j = \varpi(\{\ddot z_{itj} \ddot u_{it}\}), \ \ \ \ \
 \imath_T = \min_{1 \leq j \leq p} \imath_T(\{\ddot z_{itj} \ddot u_{it}\})\\
&\phi_D^2 = \phi^2(\{\ddot D_{it} \ddot \epsilon_{it}\}),  \ \ 
\varpi_D = \varpi(\{\ddot D_{it} \ddot \epsilon_{it}\}), \  \ \ \ \
\imath_T^D = \imath_T(\{\ddot D_{it} \ddot \epsilon_{it}\})\\
&\phi_{z_jd}^2 = \phi^2(\{\ddot z_{itj} \ddot d_{it} \}),  \ \  
\varpi_{z_jd} = \varpi(\{\ddot z_{itj}\ddot d_{it} \}), \ \ 
\imath_T^{z_jd} = \imath_T(\{\ddot z_{itj} \ddot d_{it} \})\\
&\phi_{z_j\epsilon}^2 = \phi^2(\{\ddot z_{itj} \ddot \epsilon_{it} \}),  \ \ 
\varpi_{z_j\epsilon} = \varpi(\{\ddot z_{itj} \ddot \epsilon_{it}\}), \ \ \
\imath_T^{z_j \epsilon} = \imath_T(\{\ddot z_{itj} \ddot \epsilon_{it}\})\\
\end{align*}



To derive asymptotic properties of these estimators, we will make use of the following condition in addition to those assumed in Section \ref{ClusterLassoTheory}.

\textbf{Condition SMIV}  

\noindent \textit{ (i) Sufficient conditions for Post-Cluster-Lasso:  ASM, SE, R hold for model \ref{IVfirststage}.}


\noindent \textit{ (ii) Sufficient conditions for asymptotic normality of $\hat \alpha$ and consistency of $\frac{\imath_T^D}{T}\hat V$: }

\textit{ (a) } $\Ep\[ \frac{1}{T} \sum_{t=1}^T \ddot D_{it}^2\],$  $ \Ep\[\frac{1}{T} \sum_{t=1}^T \ddot \epsilon_{it}^2 \ddot D_{it}^2\]$, $\Ep \[ \( \frac{1}{T} \sum_{t=1}^T \ddot d_{it}^2 \)^2 \] $   \textit{ are bounded uniformly from above and away from zero, uniformly in } $n, T.$ \textit{Additionally, }$\Ep\[ \( \frac{1}{T}\sum_{t=1}^T \ddot \epsilon_{it}^2 \)^q\] = O(1)$ \textit{ for some }$q>4$.

\textit{ (b) } $ \varpi_D / \sqrt{ \Ep \phi_D^2 } = O(1)$, $\max_{1 \leq j \leq p} \varpi_{z_j\epsilon}/\sqrt{\Ep \phi_{z_j \epsilon}^2} = O(1),$

\textit{ (c) } $\max_j  \frac{\imath_T^{z_j \epsilon}}{T}  \phi_{z_j \epsilon}^2 = O_P(1) $,  $\frac{1}{T} \phi_{dD}^2 = O_\Pr(1),$  $\max_j \frac{1}{T}   \phi_{z_j d}^2 = O_P(1)$

\textit{ (d) }$\frac{s^2 \log^2 (p\vee nT) }{n\imath_T} \max \{ 1, \max_{1 \leq j \leq p} \frac{\imath_T^D}{\imath_T^{z_j \epsilon}} \} =o(1)$\textit{ and }$ \frac{\imath_T^D}{\imath_T}n^{2/q}\frac{s \log(p \vee nT)}{n}=o(1)$

The conditions assumed in Condition SMIV are fairly standard.  Outside of moment conditions, the main restriction in Condition SMIV is condition (ii)(a) that guarantees that the parameter $\alpha$ would be strongly identified if $\ddot D_{it}$ could be observed.  Coupled with the approximately sparse model, this condition implies that using a small number of the variables in $z_{it}$ is sufficient to strongly identify $\alpha$ which rules out the case of weak-instruments as in \citen{ss:weakiv} and many-weak-instruments as in \citen{NeweyEtAl-JIVE}.\footnote{See also \citen{RJIVE} who consider many-weak-instruments in a $p > n$ setting.}

With the model and conditions in place, we provide the following results which can be used to perform inference about the structural parameter $\alpha$.

\begin{theorem}
Uniformly over all sequences $\{\Pr_{n,T}\}$ for which $\{(y_{it},x_{it}, z_{it} ) \}_{t=1}^T \sim \Pr_{n,T}$, i.i.d. across $i$, for which the instrumental variable model holds, and for which condition SMIV holds,\footnote{More precisely, the convergence holds uniformly over sequences satisfying Condition SMIV, with the same implied constants with $n, \ T \rightarrow \infty$ jointly or $n \rightarrow \infty, \ T$ fixed.} the IV estimator $\hat \alpha$ satisfies
$$\sqrt{n \imath_T^D}V^{-1/2}(\hat \alpha - \alpha) \ind N(0,1).$$  In addition, 
$$ V - \frac{\imath_T^D}{T} \hat V \overset P \rightarrow 0.  $$
\end{theorem}

This theorem verifies that the IV estimator formed with instruments selected by Cluster-Lasso in a linear IV model with fixed effects is consistent and asymptotically normal.
In addition, one can use the result with $\hat V$ defined in (\ref{LassoIVVAR}), which is simply the usual clustered standard error estimator \cite{arellano:feinf}, to perform valid inference for $\alpha$ following instrument selection.  Note that this inference will be valid uniformly over a large class of data generating processes which includes cases where perfect instrument selection is impossible.

\subsection{Selection of Control Variables}
A second strategy for identifying structural effects in economic research is based on assuming that variables of interest are as good as randomly assigned conditional on time varying observables and time invariant fixed effects.  Since this approach relies on including the right set of time varying observables, a practical problem researchers face is the choice of which control variables to include in the model.  The high dimensional framework provides a convenient setting for exploring data-dependent selection of control variables.  In this section, we consider the problem of selecting a set of variables to include in a linear model from a large set of possible control variables in the presence of unrestricted individual specific heterogeneity.

The structure of the Lasso optimization problem ensures that any estimated coefficient that is not set to zero can be reliably differentiated from zero relative to estimation noise when (\ref{RegularizationEvent}) holds while any coefficient that can not be distinguished reliably from zero will be estimated to be exactly zero.  This property makes Lasso-based methods appealing for variable selection in sparse models.  However, this property also complicates inference after model selection in approximately sparse models which may have a set of variables with small but non-zero coefficients in addition to strong predictors.  In this case, satisfaction of condition (\ref{RegularizationEvent}) will result in excluding variables with small but non-zero coefficients which may lead to non-negligible omitted variables bias and irregular sampling behavior of estimates of parameters of interest.  This intuition is formally developed in \citen{potscher} and \citen{leeb:potscher:pms}.  Offering solutions to this problem with fully independent data is the focus of a number of recent papers; see, for example, \citen{BellChernHans:Gauss}; \citen{BellChenChernHans:nonGauss}; \citen{ZhangZhang:CI};  \citen{BCH2011:InferenceGauss}; \citen{BelloniChernozhukovHansen2011}; \citen{vdGBRD:AsymptoticConfidenceSets}; \citen{JM:ConfidenceIntervals}; and \citen{BCFH:Policy}.\footnote{These citations are ordered by date of first appearance on arXiv.} In this section, we focus on extending the approach of  \citen{BelloniChernozhukovHansen2011} to the panel setting with dependence within individuals.

To be precise, we consider estimation of the parameter $\alpha$ in the partially linear additive fixed effects panel model:
\begin{align}\label{eq: PL1}
  y_{it}  &= d_{it}\alpha + g(z_{it}) + e_i + \zeta_{it},  \qquad  \Ep[\zeta_{it} \mid z_{i1},...,z_{iT},d_{i1},...,d_{iT},e_i]= 0,\\
\label{eq: PL2}
  d_{it}  &= m(z_{it}) + f_i + u_{it},   \qquad   \Ep[u_{it} \mid z_{i1},...,z_{iT},f_i] = 0,
\end{align}
where $y_{it}$ is the outcome variable, $d_{it}$ is the policy/treatment variable whose impact $\alpha$ we would like to infer,\footnote{The analysis extends easily to the case where $d_{it}$ is an $r \times 1$ vector where $r$ is fixed and is omitted for convenience.} $z_{it}$ represents confounding factors on which we need to condition, $e_i$ and $f_i$ are fixed effects which are invariant across time, and $\zeta_{it}$ and $u_{it}$ are disturbances that are independent of each other.  Data are assumed independent across $i$, and dependence over time within individual is largely unrestricted.

The confounding factors $z_{it}$ affect the policy variable via the function $m(z_{it})$ and the outcome variable via the function $g(z_{it})$. Both of these functions are unknown and potentially complicated. We use linear combinations of control terms $x_{it} = P(z_{it})$ to approximate $g(z_{it})$ and $m(z_{it})$ with $ x_{it}'\beta_{g}$ and $x_{it}'\beta_{m}$.  In order to allow for a flexible specification and incorporation of pertinent confounding factors, we allow the dimension, $p$, of the vector of controls, $x_{it} = P(z_{it})$, to be large relative to the sample size.\footnote{High dimensional $x_{it}$ typically occurs in either of two ways.  First, the baseline set of conditioning variables itself may be large so $x_{it} = z_{it}$.  Second, $z_{it}$ may be low-dimensional, but one may wish to entertain many nonlinear transformations of $z_{it}$ in forming $x_{it}$.  In the second case, one might prefer to refer to $z_{it}$ as the controls and $x_{it}$ as something else, such as technical regressors.   For simplicity of exposition and as the formal development in the paper is agnostic about the source of high dimensionality, we call the variables in $x_{it}$ controls or control variables in either case.}
Upon substituting these approximation into (\ref{eq: PL1}) and (\ref{eq: PL2}), we are essentially left with a conventional linear fixed effects model with a high dimensional set of potential confounding variables:
\begin{align*}
  y_{it}  &= d_{it}\alpha + x_{it}'\beta_g + e_i + r_g(z_{it}) + \zeta_{it},\\
  d_{it}  &= x_{it}'\beta_m + f_i + r_m(z_{it}) + u_{it},
\end{align*}
where $r_g(z_{it})$ and $r_m(z_{it})$ are approximation errors.  The fixed effects can again be eliminated by subtracting within group means yielding
\begin{align}\label{LinearFE1}
  \ddot y_{it}  &= \ddot d_{it}\alpha + \ddot x_{it}'\beta_g + \ddot r_g(z_{it}) + \ddot \zeta_{it},\\
\label{LinearFE2}
  \ddot d_{it}  &= \ddot x_{it}'\beta_m + \ddot r_m(z_{it}) + \ddot u_{it}.
\end{align}

Informative inference about $\alpha$ is not possible in this model without imposing further structure since we allow for $p > n$ elements in $x_{it}$.  The additional structure is added by assuming that condition ASM applies to both $g(z_{it})$ and $m(z_{it})$ which implies that exogeneity of $d_{it}$ may be taken as given once one controls linearly for a relatively small number, $s < n$, of the variables in $x_{it}$ whose identities are \textit{a priori} unknown.\footnote{Note that this condition is stronger than necessary but convenient; see, e.g., \citen{farrell:JMP}.  We briefly explore this in the simulation example in Section \ref{SimulationSection} where we consider a design where Condition ASM holds only in one equation and show that our procedure still yields good results in that setting.}  Under this condition, estimation of $\alpha$ may then proceed by using variable selection methods to choose a set of relevant control variables from among the set $\ddot x_{it}$ to use in estimating (\ref{LinearFE1}).


To estimate $\alpha$ in this environment, we adopt the post-double-selection method of \citen{BelloniChernozhukovHansen2011}.
This method proceeds by first substituting (\ref{LinearFE2}) into (\ref{LinearFE1}) to obtain predictive relationships for the outcome $\ddot y_{it}$ and the treatment $\ddot d_{it}$ in terms of only control variables:
\begin{align}\label{PLMReducedForm}
  \ddot y_{it}  &= \ddot x_{it}'\pi + \ddot r_{RF}(z_{it}) + \ddot v_{it},\\
\label{PLMFirstStage}
  \ddot d_{it}  &= \ddot x_{it}'\beta_m + \ddot r_m(z_{it}) + \ddot u_{it}.
\end{align}
We then use two variable selection steps.  Cluster-Lasso is applied to equation (\ref{PLMReducedForm}) to select a set of variables that are useful for predicting $\ddot y_{it}$; we collect the controls $x_{itj}$ for which $\hat \pi_{j} \neq 0$ in the set $\hat I_{RF}$.
Cluster-Lasso is then applied to equation (\ref{PLMFirstStage}) to select a set of variables that are useful for predicting $\ddot d_{it}$; we again collect the controls $x_{itj}$ for which $\hat \beta_{m,j} \neq 0$ in the set $\hat I_{FS}$.  The set of controls that will be used is then defined by the union $\hat I = \hat I_{FS} \cup \hat I_{RF}$.  Estimation and inference for $\alpha$ may then proceed by ordinary least squares estimation of $\ddot y_{it}$ on $\ddot d_{it}$ and the set of controls in $\hat I$ using conventional clustered standard errors \cite{arellano:feinf}.

\citen{BelloniChernozhukovHansen2011} develop and discuss the post-double-selection method in detail.  They note that including the union of the variables selected in each variable selection step helps address the issue that model selection is inherently prone to errors unless stringent assumptions are made.  As noted by \citen{leeb:potscher:pms}, the possibility of model selection mistakes precludes the possibility of valid post-model-selection inference based on a single Lasso regression within a large class of interesting models.  The chief difficulty arises with covariates whose effects in (\ref{LinearFE1}) are small enough that the variables are likely to be missed if only (\ref{LinearFE1}) is considered but have large effects in (\ref{LinearFE2}).  The exclusion of such variables may lead to substantial omitted variables bias if they are excluded which is likely if variables are selected using only (\ref{LinearFE1}).\footnote{The argument is identical if only (\ref{LinearFE2}) is used for variable selection exchanging the roles of (\ref{LinearFE1}) and (\ref{LinearFE2}).  The argument also holds if considering only one of (\ref{PLMReducedForm}) or (\ref{PLMFirstStage}) for variable selection.} Using both model selection steps guards against such model selection mistakes and guarantees that the variables excluded in both model selection steps have a neglible contribution to omitted variables bias under Condition ASM.

We present additional moment and rate conditions before stating a result which can be used for performing inference about $\alpha$.  Results will be valid uniformly over the large class of models that satisfy the following conditions as well as appropriate conditions from Section \ref{ClusterLassoTheory}.  We again define several moments using the same notation introduced before condition SMIV.  

\begin{align*}
\phi_{x_ju}^2 &= \phi_{j,FS}^2 = \phi^2(\{\ddot x_{itj} \ddot u_{it}\}), \
\varpi_{x_ju} = \varpi_{j,FS} = \varpi(\{\ddot x_{itj} \ddot u_{it}\}), \
\imath_T^{x_j u} = \imath_T(\{\ddot x_{itj} \ddot u_{it}\}), \
\imath_T^{FS} = \min_{1 \leq j \leq p} \imath_T^{x_j u} \\
\phi_{x_jv}^2 &= \phi_{j,RF}^2 = \phi^2(\{\ddot x_{itj} \ddot v_{it}\}), \
\varpi_{x_jv} = \varpi_{j,RF} = \varpi(\{\ddot x_{itj} \ddot v_{it}\}), \
\imath_T^{x_j v} = \imath_T(\{\ddot x_{itj} \ddot v_{it}\}), \ 
\imath_T^{RF} = \min_{1 \leq j \leq p} \imath_T^{x_j v} 
\end{align*}
\begin{align*}
\phi_{u \zeta}^2 &= \phi^2(\{\ddot u_{it} \ddot \zeta_{it}\}), \ 
\varpi_{u \zeta} = \varpi(\{\ddot u_{it} \ddot \zeta_{it}\}), \
\imath_T^{u \zeta} = \imath_T(\{\ddot u_{it} \ddot \zeta_{it}\}) \\ 
\phi_{x_j\zeta}^2 &= \phi^2(\{\ddot x_{itj} \ddot \zeta_{it} \}), \
\varpi_{x_j\zeta} = \varpi(\{\ddot x_{itj}\ddot \zeta_{it} \}), \
\imath_T^{x_j \zeta} = \imath_T(\{\ddot x_{itj} \ddot \zeta_{it}\}) \\
\phi_{ud}^2 &= \phi^2(\{\ddot u_{it} \ddot d_{it} \}),  \ 
\varpi_{ud} = \varpi(\{\ddot u_{it} \ddot d_{it}\}), \
\imath_T^{ud} = \imath_T(\{\ddot u_{it} \ddot d_{it}\}) \\
\end{align*}

\textbf{Condition SMPLM}  

\noindent \textit{ (i) Sufficient conditions for Post-Cluster-Lasso:  ASM, SE, R hold for models \ref{PLMReducedForm} and \ref{PLMFirstStage}.}

\noindent \textit{ (ii) Sufficient conditions for asymptotic normality of $\hat \alpha$ and consistency of $\frac{\imath_T^D}{T}\hat V$: }

\textit{ (a) } $Q=\Ep\[ \frac{1}{T} \sum_{t=1}^T \ddot u_{it}^2\]$, $ \Ep\[\frac{1}{T} \sum_{t=1}^T \ddot u_{it}^2 \ddot \zeta_{it}^2\] $, $ \Ep\[\( \frac{1}{T} \sum_{t=1}^T \ddot u_{it}^2 \)^2\] $ \textit{ are bounded uniformly from above and away from zero, uniformly in } $n,T$.  \textit{Additionally, }$\Ep\[ \(\frac{1}{T} \sum_{t=1}^T \ddot \zeta_{it}^2\)^q \] = O(1), \ \Ep\[ \(\frac{1}{T} \sum_{t=1}^T \ddot u_{it}^2\)^q \] = O(1)$ and $\Ep\[ \(\frac{1}{T} \sum_{t=1}^T \ddot d_{it}^2\)^q \] = O(1)$ for some $q>4$.  $|\alpha| \leq B < \infty$.

\textit{ (b) } $ \varpi_{u\zeta} / \sqrt{ \Ep \phi_{u\zeta}^2} = O(1)$, $\max \limits_{1 \leq j \leq p} \varpi_{x_j\zeta}/\sqrt{\Ep \phi_{x_j \zeta}^2} = O(1),$ $\max \limits_{1 \leq j \leq p} \varpi_{x_ju}/\sqrt{\Ep \phi_{x_j u}^2} = O(1)$.

\textit{ (c) } $ \max_j  \frac{\imath_T^{x_j \zeta}}{T} \phi_{x_j \zeta}^2 = O_P(1) $,  $\frac{\imath_T^{u\zeta}}{T} \phi_{u\zeta}^2 = O_\Pr(1),$  $\max_j   \frac{\imath_T^{x_j u}}{T} \phi_{x_j u}^2 = O_P(1)$, $\frac{1}{T} \phi_{ud}^2 = O_\Pr(1)$. 

\textit{ (d) }$\frac{\imath_T^{u\zeta}}{\min \{\imath_T^{RF},\imath_T^{FS},\min_j \{\imath_T^{x_j \zeta}\}\}}\left(s +  n^{2/q}\right)\left(\max_{i,t,j} \ddot x_{itj}^2\right) \frac{s \log^2(p \vee nT)}{n} =o_\Pr(1)$.

Finally, we define the following variance estimators for the post double selection procedure:

  $$\hat V_n = \hat Q^{-1} \hat \Omega \hat Q^{-1}$$ 
 $$\hat Q = \EnT \hat { u}_{it}^2, \ \ \
\hat \Omega = \EnTT \hat {u}_{it} \hat{ u}_{it'} \hat{ \zeta}_{it} \hat{ \zeta}_{it'},$$
where 
\begin{align*}
&\hat u_{it} = \ddot d_{it} - \ddot x_{it}'\hat\beta_m,  \ \ \hat \zeta_{it} = \ddot y_{it} - \hat\alpha \ddot d_{it} - \ddot x_{it}' \hat \beta_g \\
& \hat\beta_m = \underset{b : \ b_j = 0 \ \forall \ j \notin \hat I }{\textnormal{argmin}} \sum_{i=1}^n \sum_{t=1}^T (\ddot d_{it} - \ddot x_{it}'b)^2 \\
& (\hat\alpha,\hat\beta_g')' = \underset{(a,b) : \ b_j = 0 \ \forall \ j \notin \hat I }{\textnormal{argmin}} \sum_{i=1}^n \sum_{t=1}^T (\ddot y_{it} - a \ddot d_{it} - \ddot x_{it}'b)^2.\\
\end{align*}

Given the above conditions, we have the following central limit theorem for the Post-Double Cluster-Lasso estimator $\hat \alpha$.

\begin{theorem}[Estimation and Inference on Treatment Effects]\label{theorem:inference}   Uniformly over all sequences $\{\Pr_n \}$ for which $\{ (y_{it}, d_{it}, x_{it}) \}_{t=1}^T \sim \Pr_n$, i.i.d. across $i$, for which Condition SMPLM holds,\footnote{More precisely, the convergence holds uniformly over sequences satisfying Condition SMPLM, with the same implied constants with $n, \ T \rightarrow \infty$ jointly or $n \rightarrow \infty, \ T$ fixed.} the Post-Double-Cluster-Lasso estimator $\hat \alpha$ satisfies
$$
\sqrt{n\imath_T^{u\zeta}} V^{-1/2}  (\hat \alpha - \alpha) \ind N(0,1),
$$
In addition,
$$V - \frac{\imath_T^{u\zeta}}{T} \hat V \overset{ \Pr }{\rightarrow} 0.$$
\end{theorem}

This theorem verifies that the OLS estimator which regresses $\ddot y_{it}$ on $\ddot d_{it}$ and the union of variables selected by Cluster-Lasso from (\ref{PLMReducedForm}) and (\ref{PLMFirstStage}) is consistent and asymptotically normal with asymptotic variance that can be estimated with the conventional clustered standard error estimator.
Inference based on this result will be valid uniformly over a large class of data generating processes which includes cases where perfect variable selection is impossible.

\section{Simulation Examples}\label{SimulationSection}

The results in the previous sections suggest that Cluster-Lasso based estimates should have good estimation and inference properties in panel models with individual specific heterogeneity provided the sample size $n$ is large.  In this section, we provide simulation evidence about the performance of our asymptotic approximation for inference about structural parameters in IV models with fixed effects and many instruments and linear fixed effects models when Cluster-Lasso is used for variable selection.  We also provide a comparison with several other standard estimators.  The Cluster-Lasso based procedures compare favorably to all other feasible approaches considered.

\subsection{Simulation 1: IV}\label{IVSimulation}

The first simulation illustrates the performance of the Cluster-Lasso based IV estimator in a simple instrumental variables model with fixed effects and many instruments.  In our simulation experiments, we generate data from the linear IV model
\begin{align*}
y_{it} &= \alpha d_{it} + e_i + \epsilon_{it} \\
d_{it} &= z_{it}'\pi + f_i + u_{it}.
\end{align*}
We generate disturbances according to
\begin{align*}
\epsilon_{it} &= \rho_{\epsilon} \epsilon_{it-1} + \nu_{1,it} \\
u_{it} &= \rho_u u_{it-1} + \nu_{2,it} \\
\left(\begin{array}{c} \nu_{1,it} \\ \nu_{2,it} \end{array}\right)
&\sim N\left(\left(\begin{array}{c} 0 \\ 0 \end{array}\right) , \left(\begin{array}{cc} 1 & \rho_{\nu} \\  \rho_{\nu} & 1 \end{array}\right)\right) \quad \textnormal{iid}
\end{align*}
with initial conditions for $\epsilon_{it}$ and $u_{it}$ drawn from their stationary distribution.
We generate the individual heterogeneity $e_i$ for $i = 1,...,n$ as correlated normal random variables with $\Ep[e_i] = 0$, Var$(e_i) = \frac{4}{T}$, and Corr$(e_i,e_j) = .5^{|i-j|}$ for all $i$ and $j$.  We set $f_i = e_i$.  We draw the instruments conditional on the fixed effects from
\begin{align*}
z_{i1j} &= \frac{e_i}{1-\rho_z} + \sqrt{\frac{1}{1-\rho_z^2}} \varphi_{i1j} \\
z_{itj} &= e_i + \rho_z z_{i(t-1)j} + \varphi_{itj}  \quad t > 1
\end{align*}
where $\varphi_{itj}$ are normal random variables with $\Ep[\varphi_{itj}] = 0$, Var$(\varphi_{itj}) = 1$, and Corr$(\varphi_{itj},\varphi_{itk}) = .5^{|j-k|}$ that are independent across $i$ and $t$.  In all of our simulations, we set $\rho_{\epsilon} = \rho_{u} = \rho_z = .8$, and we set $\rho_{\nu} = .5$.  We also set $\alpha = .5$.  We redraw the disturbances $\epsilon$ and $u$ at each simulation replication but condition on one realization of the fixed effects and instruments.  We consider different sample sizes set to $n=50,100,150,200$ all with $T=10$.

Note that the instruments are not valid without conditioning on the fixed effects within this structure.  The fixed effects are also dense in the sense that most of the generated effects will be small but non-zero.  This feature would lead to a failure of variable selection methods that included the set of fixed effects in the variables over which selection will occur.  Such methods would fail to include the majority of the effects which would then result in invalidity of the instruments and substantial bias in the resulting estimates of $\alpha$.  A simple and widely-used way to bypass this problem is removing the entire set of fixed effects as considered in this paper.

The final features of the design are the number of instruments and the structure of the coefficients on the instruments, $\pi$.  We consider three different coefficient vectors $\pi_1, \pi_2,$ and $\pi_3$ defined as
\begin{align*}
\text{Design 1: } \ \ \ \ \pi_{1j} &= (-1)^{j-1} \left(\frac{1}{\sqrt{s}}1_{\{j \leq s\}} + \frac{1}{j^2} 1_{\{j > s\}}\right), \ \ s = \lfloor \frac{1}{2}n^{1/3} \rfloor  \\
\text{Design 2: } \ \ \ \ \pi_{2j} &= (-1)^{j-1} \left(\frac{1}{\sqrt{s}}1_{\{j \leq s\}} + \frac{1}{\sqrt{p-s}} 1_{\{j > s\}} \right), \ \ s = \lfloor \frac{1}{2}n^{1/3} \rfloor \\
\text{Design 3: } \ \ \ \ \pi_{3j} &= (-1)^{j-1} \left(\frac{1}{\sqrt{s}}1_{\{j \leq s\}} \right) , \ \ s = 2\lfloor \frac{1}{2}n^{1/3} \rfloor
\end{align*}
for $1 \leq j \leq p$ where $\lfloor a \rfloor$ returns the integer part of $a$.  We refer to designs using $\pi_{1}$, $\pi_{2}$, and $\pi_3$ as Design 1, Design 2, and Design 3 respectively.  Design 1 is approximately sparse and should be the most favorable setting as the majority of the signal concentrates in the smallest number of variables among the designs considered.  Design 2 does not satisfy Condition ASM, but a substantial amount of the signal is captured by the first few variables.  Cluster-Lasso should be able to reliably identify these variables which should result in reasonable properties of the Cluster-Lasso based IV estimator, though there should be a substantial loss of efficiency relative to the infeasible setting where the exact values of $\pi_2$ are known.  Design 3 is exactly sparse but should be the most difficult design because the signal is diffused equally over twice as many variables as in Design 1.  This spreading of the signal will make it harder to reliably detect any of the instruments.
Finally, we consider two different numbers of instruments, $p=n \times (T-2)$ and $p = n \times (T+2)$, for each sample size and design of first-stage coefficients.

For each setting, we report results from five different estimators.  We consider IV estimates based on variables selected using the clustered penalty loadings developed in this paper (Clustered Loadings).  As a comparison, we also consider IV estimates based on variables selected using the loadings that are valid with heteroscedastic and independent data from \citen{BellChenChernHans:nonGauss} (Heteroscedastic Loadings).  In cases with $p<nT$, we report estimates using 2SLS on the full set of instruments (All).   Finally, we consider two different infeasible oracle estimators.  The first oracle knows the value of the coefficients $\pi$ (Oracle) while the second also knows the exact values of the fixed effects (FE Oracle).  Thus, both oracle estimators use a single instrument that uses the true values of the first stage coefficients, $z_{it}'\pi$.  The difference between the two is that the fixed effects are removed by taking differences of all variables from within-individual means in the Oracle results while the true values of the FE are directly subtracted from $y_{it}$ and $d_{it}$ in the FE Oracle results.

The results are based on 1000 simulations for each setting described above.  For results based on All, Heteroscedastic Loadings, Clustered Loadings, and Oracle, the fixed effects are treated as unknown parameters and eliminated by taking deviations from within-individual means.  For each estimator, we report mean bias, root mean squared error, and rejection rates for a 5\%-level test of $H_0:\alpha=.5$ using both clustered standard errors and heteroscedastic standard errors.\footnote{Since moments of IV estimators may not exist, we calculate truncated bias and truncated RMSE, truncating at $\pm 10,000$.}  In some of the simulation replications, the IV estimator using variables selected by Lasso is undefined as Lasso sets all coefficients to zero.   In such a case, we record a failure to reject the null which is a conservative alternative to applying the Sup-score statistic described in \citen{BellChenChernHans:nonGauss}.  Mean bias and root-mean-square-error for Lasso-based estimates are calculated conditional on Lasso selecting at least one instrument.

The results for estimation of $\alpha$ with first stage coefficients $\pi_1$, $\pi_2$, and $\pi_3$ are reported respectively in Tables \ref{IVDesign11}, \ref{IVDesign21}, and \ref{IVDesign31} when $p = n \times (T-2)$ and Tables \ref{IVDesign12}, \ref{IVDesign22}, and \ref{IVDesign32} when $p = n \times (T+2)$.  The two oracle estimators provide infeasible benchmarks.  Looking at these results, we do see that IV based on the infeasible instruments formed using the true values of the first-stage coefficients perform well in the designs considered.  As expected given the well-known properties of 2SLS, the 2SLS estimates using the full set of instruments when $p < nT$ exhibit large bias relative to standard error, large RMSE, and produce tests that suffer from large size distortions.

The Lasso-based results where we do variable selection using loadings that are appropriate under independence but ignore within-individual dependence are quite interesting.  This approach performs relatively well compared to naive 2SLS using all of the instruments.  However, using instruments selected by Lasso with loadings that ignore the dependence produces an IV estimator of $\alpha$ that has a substantial bias and results in tests that have large size distortions even when clustered standard errors are applied.  The presence of this bias  illustrates the point that care must be taken when selecting instruments for a post model selection analysis.  In general, $\Ep[z_{itj}\epsilon_{it}| \ \ j \ \ \text{selected}] \neq 0$ though the difference from zero is ignorable when (\ref{RegularizationEvent}) occurs.  However, in the absence of the regularization event (\ref{RegularizationEvent}), this conditional expectation can be large which introduces a type of ``endogeneity'' bias as the selected instruments are effectively invalid.  We see this behavior when using the heteroscedastic loadings in the designs we consider as these loadings produce smaller penalty levels than the appropriate clustered loadings which results in the spurious inclusion of instruments.

Finally, we see that IV based on instruments selected by Cluster-Lasso clearly dominates the other feasible procedures in the simulation designs considered.  In Designs 1 and 2, using this procedure produces tests that have approximately correct size that is comparable to size of tests based on both oracle models considered.  We also see that the performance for Bias, RMSE, and size of tests is similar to the infeasible Oracle benchmark in Design 1.  Designs 2 and 3 were both designed to be difficult.  As expected, we see a substantial loss in RMSE relative to the infeasible oracles in Design 2 as the Lasso-based variable selection is unable to consistently identify and exploit the signal available in the variables with small, non-zero coefficients.  It is reassuring that performance is still reasonable in this setting.  We also see that the more diffuse signal in Design 3 poses challenges when $n$ is small as the contribution each variable makes to the overall signal is too weak to be reliably detected. Thus, there are many replications where no instruments are selected and replications where only a subset of the relevant variables are selected, effectively resulting in weak identification.  For the larger sample sizes, this problem is diminished though performance still deviates from the oracle benchmark.  Overall, these results are favorable in that using Cluster-Lasso to select instruments outperforms the other methods explored here and performs reasonably well even in somewhat adversarial conditions.

\subsection{Simulation 2: Linear Model}\label{LMSimulation}
In this simulation, we consider estimation of a coefficient on a variable of interest in a standard linear fixed effects model.  Specifically, we generate data according to the model
\begin{align*}
y_{it} &= \alpha d_{it} + z_{it}'\beta+ e_i + \epsilon_{it}  \\
d_{it} &= z_{it}'\gamma + f_i + u_{it}.
\end{align*}
We generate disturbances according to
\begin{align*}
\epsilon_{it} &= \rho_{\epsilon} \epsilon_{it-1} + \nu_{1,it} \\
u_{it} &= \rho_u u_{it-1} + \nu_{2,it} \\
\left(\begin{array}{c} \nu_{1,it} \\ \nu_{2,it} \end{array}\right)
&\sim N\left(\left(\begin{array}{c} 0 \\ 0 \end{array}\right) , \left(\begin{array}{cc} 1 & 0 \\  0 & 1 \end{array}\right)\right) \quad \textnormal{iid}
\end{align*}
with initial conditions for $\epsilon_{it}$ and $u_{it}$ drawn from their stationary distribution.
We generate $e_i$, $f_i$, and $z_{it}$ exactly as in Section \ref{IVSimulation}, so we omit the details for brevity.
We again set $\rho_{\epsilon} = \rho_{u} = .8$ and set $\alpha = .5$.  We redraw the disturbances $\epsilon$ and $u$ at each simulation replication but condition on one realization of the fixed effects and controls.  We use sample sizes set to $n=50,100,150,200$ with $T=10$.
As in Section \ref{IVSimulation}, failure to condition on the full set of fixed effects may result in substantial biases in estimation of $\alpha$.  The structure of the fixed effects will also invalidate methods that include the set of fixed effects in the variables over which selection will occur as illustrated in the results below.

As in Section \ref{IVSimulation}, we consider three different specifications for the coefficient vectors $\beta$ and $\gamma$:  \begin{align*}
\textnormal{Design 1:}& \quad \gamma_{1j} = \beta_{1j} = (-1)^{j-1} \left(\frac{1}{\sqrt{s}}1_{\{j \leq s\}} + \frac{1}{j^2} 1_{\{j > 2\}}\right),  \ \ s = \lfloor \frac{1}{2}n^{1/3} \rfloor  \\
\textnormal{Design 2:}& \quad \gamma_{2j} = (-1)^{j-1} \left(\frac{1}{\sqrt{s}}1_{\{j \leq s\}} + \frac{1}{\sqrt{p-s}} 1_{\{j > s\}}\right),   \\
& \quad \beta_{2j} = (-1)^{j-1} \left(\frac{1}{\sqrt{s}}1_{\{j \leq s\}} + \frac{1}{j^2} 1_{\{j > s\}}\right) ,  \ \ s = \lfloor \frac{1}{2}n^{1/3} \rfloor \\
\textnormal{Design 3:}& \quad \gamma_{3j} = \beta_{3j} = (-1)^{j-1} \left(\frac{1}{\sqrt{s}}1_{\{j \leq s\}}\right),  \ \ s = 2\lfloor \frac{1}{2}n^{1/3} \rfloor
\end{align*}
for $1 \leq j \leq p$ where $\lfloor a \rfloor$ returns the integer part of $a$.  Again, in each design, we consider $p=n \times (T-2)$ and $p = n \times (T+2)$.
Designs 1 and 3 clearly fall within the set of models covered by our theoretical development, and we expect the estimation and inference about $\alpha$ following selection of controls using Cluster-Lasso to perform well in either case, though Design 3 is more difficult than Design 1.  The relationship between $d_{it}$ and $z_{it}$ in Design 2 does not satisfy Condtion ASM, but the relationship between $y_{it}$ and $z_{it}$ satisfies this condition.

For this simulation, we consider six estimators of $\alpha$.  When $p \leq nT$, we use the conventional fixed effects estimator including all the variables in $z_{it}$ (All).  We use the post-double-selection method with penalty loadings appropriate for independent, heteroscedastic data in each Lasso stage (Heteroscedastic Loadings) and with our clustered loadings in each Lasso stage (Clustered Loadings).  We also consider a post-double-selection estimator which includes the fixed effects in the set of variables over which selection occurs (Select over FE) using the approach of \citen{kock:hdpanel}.  We also consider two oracle estimators.  The first oracle knows the values of the coefficients $\beta$ and $\gamma$ (Oracle) while the second also knows the exact values of the fixed effects (FE Oracle).  The Oracle estimate of $\alpha$ is thus obtained by regressing $\ddot y_{it} - \ddot z_{it}'\beta$ onto $\ddot d_{it} - \ddot z_{it}'\gamma$ while the FE Oracle estimate of $\alpha$ is obtained by regressing $y_{it} - z_{it}'\beta - e_i$ onto $d_{it} - z_{it}'\gamma - f_i$.  As before, the results are based on 1000 simulation replications for each setting; and we report mean bias, root mean squared error, and rejection rates for a 5\%-level test of $H_0:\alpha=.5$ using both clustered standard errors and heteroscedastic standard errors for each estimator.

The results for the three partially linear model designs are reported in Tables \ref{LMDesign11}, \ref{LMDesign21}, and \ref{LMDesign31} for $p=n \times (T-2)$ and Tables \ref{LMDesign12}, \ref{LMDesign22}, and \ref{LMDesign32} for $p = n \times (T+2)$.
The two oracle estimators provide infeasible benchmarks and unsurprisingly produce estimators with small bias and RMSE and tests with reasonable size as long as clustered standard errors are used as is conventional in the literature, e.g. \citen{bdm:cluster}.
In all simulations, estimates using the full set of controls when feasible have small bias but large variability leading to large RMSE relative to oracle estimates.  Tests based on estimators using the full set of controls are also badly size distorted regardless of whether heteroscedastic or clustered standard errors are used.  This distortion results from the difficulty in robustly estimating standard errors when many variables are included which is an unresolved topic of current research; see, e.g., \citen{CJN:PLMStandardError}.  This feature suggests that one may not wish to simply including many controls without regularization even when possible.

Estimates based on the double selection method using Lasso with penalty loading appropriate under heteroscedasticity and independence or using Lasso to also select over the fixed effects perform better than simply including all controls in the $p < nT$ case but tend to perform poorly in terms of bias and coverage probabilities.  The bias and poor coverage properties of the estimator that attempts to select over the fixed effects is due to the difficulties in performing selection over the dense part of the model and shows the importance of eliminating fixed effect parameters via demeaning or differencing.  These difficulties arise because sparsity provides a poor approximation to the true fixed effects structure.  Note that a dense model over unobserved heterogeneity where heterogeneity matters differentially for each individual seems quite reasonable in many economic applications and suggests that attempting to select over fixed effects may result in undesirable features at least when inference about model parameters is the goal of the empirical analysis.

We find it more surprising that using heteroscedastic penalty loadings also leads to noticeable bias and a distortion in statistical size.  The heteroscedastic loadings lead to less penalization in our designs which result in inclusion of a few spurious variables.  Usual intuition for linear models suggests that including a few extra variables has little impact on say ordinary least squares estimates of parameters of interest.  The difficulty arises because the spuriously included variables are not included at random but are exactly those variables with little to no impact that are most highly correlated to the noise and are not properly screened out because the penalty is too low for (\ref{RegularizationEvent}) to be a reliable guide.  Choosing the variables most highly correlated to the noise then yields that $\Ep[ x_{itj} \epsilon_{it} | \ \ j \ \ \text{selected} ]$ is not negligible due to the use of incorrect penalty loadings leading to biased estimation just as in the instrumental variables case.

Finally, we again see that basing estimation and inference for $\alpha$ on the post-double-selection method using clustered penalty loadings clearly dominates the other feasible procedures in the simulation designs considered.  This procedure yields an estimator with RMSE comparable to the oracles across all designs considered.  We also see that feasible inference based on this procedure does a relatively good job controlling size across all designs considered.  Overall, these results are favorable to Lasso-based variable selection using clustered penalty loadings after partialing out fixed effects and suggests that these methods may offer useful tools to empirical researchers faced with high dimensional panel data.

\section{Empirical Example:  The Social Cost of Gun Ownership, Cook and Ludwig (2006)}

In the earlier sections, we provided results on the performance of Lasso as a model selection device for panel data models with fixed effects and discussed how to apply Lasso to problems of economic interest in such settings.  In this section, we demonstrate the use of Cluster-Lasso by reexamining the \citen{cook:ludwig:guns} study of the impact of gun ownership on crime.   We briefly review \citen{cook:ludwig:guns} before presenting the results using the methods described in this paper.

\citen{cook:ludwig:guns} give several arguments suggesting that gun ownership levels may impose externalities on a community.  On the one hand, widespread prevelance of guns can act a deterrent to criminal activity.  On the other hand, higher gun prevelance in the general population can lead to higher gun ownership among dangerous people, perhaps through theft or illegal sales, which may lead to an increase in crime.  Thus, it is unclear whether the net effect of guns is positive or negative.  To investigate the impact of guns, \citen{cook:ludwig:guns} estimate the effect of gun prevelance on several measures of crime rates.  In this example, we revisit their estimation of the effect of gun prevelance on homicide rates.

A major contribution of \citen{cook:ludwig:guns} is to provide an improved measure of gun ownership in order to get more accurate estimates of the social costs of gun prevalence.\footnote{Previously, several authors had obtained conflicting estimates for the effect of interest; see, e.g., \citen{lottbook} and \citen{duggan2001}.}  Because exact gun-ownership numbers in the U.S. are difficult to obtain, \citen{cook:ludwig:guns} instead use the fraction of suicides committed with a firearm (abbreviated FSS) within a county as a proxy for county-level gun ownership rates.  \citen{cook:ludwig:guns} argue that if guns are prevalent within a county, then they should be more accessible for the purpose of suicide.  They show that their proxy for gun prevelance, FSS, matches up with survey data directly measuring gun ownership from the General Social Survey better than previously used measures.  
In our analysis, we take it as given that FSS provides a useful measure of gun ownership and that learning the causal effect of FSS is an interesting goal.  We thus abstract from any further measurement issues in order to give a clear illustration of our methods.

The main strategy employed by \citen{cook:ludwig:guns} to estimate causal effects of gun prevalence is to exploit differences in gun ownership across counties and over time.  \citen{cook:ludwig:guns} construct a panel of 195 large United States counties between the years 1980 through 1999 and use this data to estimate linear fixed effects models of the form
\begin{align}\label{CLModel}
\log Y_{it} = \beta_0 + \beta_1 \text{log FSS}_{it-1} + X_{it}'\beta_X + \alpha_i + \delta_t + \epsilon_{it}
\end{align}
where $\alpha_i$ and $\delta_t$ are respectively unobserved county and year effects that will be treated as parameters to be estimated, $X_{it}$ are additional covariates meant to control for any factors related to both gun ownership rates and crime rates that vary across counties and over time, and $Y_{it}$ is one of three dependent variables: the overall homicide rate within county $i$ in year $t$, the firearm homicide rate within county $i$ in year $t$, or the non-firearm homicide rate within county $i$ in year $t$.
\citen{cook:ludwig:guns} consider controls, $X_{it}$, for percent African American, percent of households with female head, nonviolent crime rates, and percent of the population that lived in the same house five years earlier.\footnote{Details about the controls can be found in \citen{cook:ludwig:guns}.}

Interpreting the estimated effect of gun prevelance as measured by FSS as causal relies on the belief that there are no variables associated both to crime rates and FSS that are not included in (\ref{CLModel}).  The inclusion of county and time fixed effects accounts for any aggregate macroeconomic conditions that affect all counties uniformly and any county-level characteristics that do not vary over time.  The additional variables used by \citen{cook:ludwig:guns} in $X_{it}$ are then meant to capture all other sources of variation that are correlated to both FSS and the log of of the homicide rate.  Of course, one might worry that the set of controls included in $X_{it}$ does not adequately capture remaining confounds after controlling for time and county effects.

We extend the analysis performed in \citen{cook:ludwig:guns} by allowing for a much larger set of potential control variables which may strengthen the plausibility of the claim that all sources of confounding variation have been captured.  Specifically, we consider an essentially identical model
$$\log Y_{it} = \beta_0 + \beta_1 \text{log FSS}_{it-1} + W_{it}'\beta_W + \alpha_i + \delta_t + \epsilon_{it}$$
which differs from (\ref{CLModel}) by our consideration of a large set of variables in $W_{it}$.  We form $W_{it}$ by taking variables compiled by the US Census Bureau.  Basic variables include county-level measures of demographics, the age distribution, the income distribution, crime rates, federal spending, home ownership rates, house prices, educational attainment, voting paterns, employment statistics, and migration rates.\footnote{The exact identities of the variables are available upon request.  The entire dataset is taken from the U.S. Census Bureau USA Counties Database and can be downloaded at http://www.census.gov/support/USACdataDownloads.html.}  We note that $W_{it}$ includes measures meant to capture all the variables controlled for in \citen{cook:ludwig:guns} in their $X_{it}$, though with our data and construction we do not reproduce their results exactly.  However, we show below that we obtain similar results with both sets of variables.
A key concern with the fixed effects model is that there is some feature of the counties that is correlated not just to the level of crime rates and gun ownership but also to the evolution of these variables.  To flexibly allow for this possibility, we also include interactions of the initial (1980) values of all control variables with a linear, quadratic, and cubic term in time.  With the main effects and interactions of initial conditions with a cubic trend, we end up with 978 total control variables.

While controlling for a large set of variables may make the assumption that all relevant confounds have been included in the model more plausible, including too many covariates may lower estimation precision and also complicates estimation of the variance of estimators as illustrated in Section \ref{LMSimulation}.  Thus, a researcher faces a tradeoff between making sure that relevant confounds are included in the model and being able to draw meaningful conclusions from the data.  Using variable selection as outlined in this paper offers one potential resolution to this tension by allowing consideration of a large set of controls while maintaining parsimony and producing valid inferential statements under the assumption that the set of confounds that needs to be included after accounting for the full set of fixed effects is small relative to the sample size.

We present estimation results in Table \ref{CLTableResults} with results for each dependent variable presented across the columns and each row corresponding to a different specification.  As a baseline, we report numbers taken directly from the first row of Table 3 in \citen{cook:ludwig:guns} in the first row of Table 1 (``Cook and Ludwig (2006) Baseline'').  \citen{cook:ludwig:guns} obtained these results by regressing log homicide rates on lagged log FSS, county and time fixed effects, and the baseline set of controls mentioned above and use these numbers as their baseline results.

We report results obtained from our data in Rows 2-4 (labeled ``FSS $+$ Census Baseline'', `` Full Set of Controls'', and ``Cluster Post-Double Selection'').\footnote{All results are based on weighted regression where we weight by the within-county average population over 1980-1999.}  In Row 2 of Table \ref{CLTableResults} (``FSS $+$ Census Baseline''), we attempt to replicate the result from Row 1 using control variables gathered from the census that correspond to the variables indicated as being used in \citen{cook:ludwig:guns} Table 3, Row 1.  Despite using slightly different data, we produce results that are fairly similar to those reported in \citen{cook:ludwig:guns}. Specifically, \citen{cook:ludwig:guns} give point estimates (standard errors) of the coefficient on lagged log FSS of .086 (.038) for overall homicide rates and .173 (.049) for gun homicide rates; and we obtain estimated effects (standard errors) of .070 (.035) for overall homicide rates and of .178 (.046) for gun homicide rates. The discrepancy between the results is somewhat larger for non-gun homicide rates, though the results are still broadly consistent with each other.  \citen{cook:ludwig:guns} report an estimated effect (standard error) of -.033 (.040) while we estimate the effect to be -.071 with a standard error of .038.

We provide the results based on the large set of controls in Rows 3 and 4 of Table \ref{CLTableResults}.  In Row 3, we present the results based on using all 978 potential controls in addition to the full set of county and time effects.  Using all of the controls, the estimated effect of lagged suicide rates is small for each dependent variable.  The estimated coefficients (standard errors) are only -.010 (.033) for overall homicide rates, .00004 (.044) for gun homicide rates, and -.033 (.042) for non-gun homicide rates.  These results are relatively imprecise, and one could not rule out moderate sized positive or negative effects for any of the dependent variables.  In addition, the simulation results illustrate that the estimated standard errors with a large number of controls may be inaccurate, suggesting that one should be hesitant in trusting these results as accurate standard errors may be even larger.  Of course, it is not obvious that one would believe that all 978 controls are necessary though one may not be sure of the exact identities of the variables that should be included.  If this is the case, the methods for variable selection developed in this paper offer one avenue for finding a relevant set of controls that should be included.

In Row 4, we present estimates of the effect of gun prevalence on homicide rates based on the post-double-selection method using Cluster-Lasso to select controls after partialing out the fixed effects.  We also provide the identities of the selected controls in Table \ref{CLSelFS}.  For both overall homicide rates and gun homicide rates, the estimates based on Cluster-Lasso selected controls are very similar to those obtained with the baseline set of controls in our data though standard errors are slightly larger.  For overall homicide results, the Cluster-Lasso estimate (standard error) is .079 (.043) compared to .070 (.035) with the baseline controls; and the Cluster-Lasso estimate (standard error) is .171 (.047) compared to .178 (.046) with the baseline controls when gun homicide is the dependent variable.  This similarity is interesting given that the set of variables selected by Lasso differs substantively from the set of baseline controls.  For the overall homicide rate, we would fail to reject the null hypothesis that gun prevalence as measured by suicide rates is not associated to homicide rates after controlling for a broad set of variables at the 5\% level in the Cluster-Lasso results, though we would reject the hypothesis of no effect of gun prevalence on overall homicide rates at the 10\% level.  The result is stronger when the gun homicide rate is the dependent variable.  In this case, one would draw the conclusion that more guns, as measured by the firearm suicide rate, is strongly positively associated with more homicides committed with firearms.  Under the assumption that the set of controls considered is sufficient to account for relevant confounds, one could also take these estimated effects as causal.  This assumption seems more plausible in the Cluster-Lasso results which allow for consideration of a richer set of controls than the baseline results.

Finally, we turn to the results with non-gun homicide as the dependent variable.  In this case, there is a larger discrepancy between the baseline results and the results using controls selected by Cluster-Lasso, though one would draw the same qualitative conclusion in either case.  With the baseline intuitively selected set of controls, the estimated effect of gun prevalence is -.071 with an estimated standard error of .038; and the estimated effect is smaller in magntitude, at -.019, with an estimated standard error of .040 using the Cluster-Lasso selected controls.  In both cases, we would fail to reject the null hypothesis that gun
prevalence as measured by suicide rates is not associated to non-gun homicide rates after controlling for
a broad set of variables at conventional levels, and one could not rule out
moderate positive or negative effects of gun prevalence on non-gun homicide at conventional
levels using the Cluster-Lasso based results.

Overall, our Cluster-Lasso based results are broadly consistent with the claims of \citen{cook:ludwig:guns}.  We find a strong positive effect of gun prevalence on the firearm homicide rate after allowing for a large set of confounds and including a full set of county and time effects.  We also find some evidence of a positive effect of gun prevalence on overall homicide rates, but produce an imprecise estimate of the effect on non-gun homicides which could be consistent with moderate positive or negative effects.  The similarity to the \citen{cook:ludwig:guns} results adds further credibility to their claims as we allow for a richer set of confounding variables.\footnote{Of course, the estimates are still not valid causal estimates if one does not believe the claim that the lagged log of the firearm suicide rate provides an exogenous measure of gun prevalence after controlling for our large set of county level controls, county, and time fixed effects.  Also, note that the similarity between results following selection and results based on an intuitively selected initital set of controls is not mechanical as evidenced, for example, in the empirical example in \citen{BelloniChernozhukovHansen2011}.}

\section{Conclusion}

In this paper, we have considered variable selection using Lasso in panel data allowing for a clustered error structure.  This structure allows for strong dependence among observations within the same individual and is commonly assumed in panel data applications in economics.  Handling this structure is also important for allowing us to verify good properties of variable selection after partialing out individual specific fixed effects by taking deviations from within individual means which, in general, will induce a clustering structure.  We show that Lasso continues to have good selection and estimation properties allowing for this structure when clustered penalty loadings are used, and we use this good performance to establish results for doing inference following variable selection in IV models with additive fixed effects and partially linear models with additive fixed effects.  We show that these methods perform well in a simulation study and illustrate their use in estimating the effect of gun prevalence on crime as in \citen{cook:ludwig:guns}.  In the empirical example, we find that our results are broadly consistent with results from \citen{cook:ludwig:guns} despite allowing for a broader set of controls.

\appendix

\section{Cluster-Lasso Penalty Loadings Implementation}\label{ClusterImplementation}

We organize implementation details for Cluster-Lasso and establish the asymptotic validity of the proposed algorithm in this appendix.
Feasible options for setting the penalty level and the loadings for $j=1,\ldots,p$ are

\begin{equation}\label{choice of loadings2}\begin{array}{llll}
&\text{Initial:}  &   \hat \phi_j  =   \sqrt{\EnTT \ddot x_{itj} \ddot x_{it'j} \ddot y_{itj}\ddot y_{it'j}},   \\ &  & \lambda =  2c\sqrt{nT} \Phi^{-1}(1- \gamma/(2p) ), \\  \\ 
& \text{Refined: } &   \hat \phi_j  = \sqrt{\EnTT \ddot x_{itj} \ddot x_{it'j} \widehat \epsilon_{i{t}}  \widehat \epsilon_{i{t'}}  },  \\ & & \lambda = 2c\sqrt{nT} \Phi^{-1}(1- \gamma/(2p) ),
\end{array}\end{equation}

\noindent where  $c>1$ is a constant, $\gamma \in (0,1)$, 
and $\hat \epsilon_{it}$ is an estimate of $\ddot \epsilon_{it}$.  Let $K \geq 1$ denote a bounded
number of iterations. We use $c=1.1$, $\gamma = 0.1/\log(p\vee nT)$, and $K = 15$ in our empirical and simulation examples. In what follows, Lasso/Post-Lasso estimator indicates that the practitioner can apply either the Lasso or Post-Lasso estimator. Our preferred approach uses Post-Lasso at each step. 

\noindent \textbf{Algorithm of Cluster-Lasso penalty loadings}

\noindent  (1) Specify penalty loadings according to the initial option in (\ref{choice of loadings2}).  Use these penalty loadings in computing the Lasso/Post-Lasso estimator $\hat \beta$ via equations (\ref{Lassoprob}) or (\ref{PostLassoprob}). Then compute residuals $\hat \epsilon_{it}= \ddot y_{it}- \ddot x_{it}'\hat \beta$ for $i=1,...,n$ and $t = 1,...,T$.

\noindent  (2) If $K > 1$, update the penalty loadings according to the refined option in (\ref{choice of loadings2}) and update the Lasso/Post-Lasso estimator $\hat \beta$.  Then compute a new set of residuals using the updated Lasso/Post-Lasso coefficients $\hat \epsilon_{it}= \ddot y_{it}- \ddot x_{it}'\hat \beta$ for $i=1,...,n$ and $t = 1,...,T$.

\noindent  (3) If $K > 2$, repeat step (2) $K-2$ times.  \qed

The algorithm above yields asymptotically valid penalty loadings in the sense that $\ell \phi_j \leq \hat \phi_j \leq u\phi_j$ for every $j$ with probability $1 - o(1)$, $\ell \overset \Pr \rightarrow 1,$ and $u \leq C < \infty$.  This fact is summarized in the following proposition.  The proposition proceeds under an extended regularity condition:

\textbf{Condition R$'$.} (Extended Regularity) 

\textit{(i)}  $\max_{1 \leq j \leq p}  \left |\frac{1}{n} \sum_{i=1}^n \frac{1}{T} \( \sum_{t=1}^T \ddot x_{itj} \ddot y_{it} \)^2 - \Ep\[ \frac{1}{T}  \(\sum_{t=1}^T \ddot x_{itj} \ddot y_{it} \)^2 \] \right |/\Ep \phi_j^2  = o_\Pr(1).$


\textit{(ii)} $\max_{1 \leq j \leq p}  \left |\frac{1}{n} \sum_{i=1}^n  \frac{1}{T} \( \sum_{t=1}^T \ddot x_{itj} \ddot \epsilon_{it} \)^2 \right |/\Ep \phi_j^2  = O_\Pr(1)$.

\textit{(iii)} $\left(\max_{i,j,t} \ddot x_{itj}^2 / \Ep\[ \phi_j^2\]\right) \frac{s \log(p \vee nT)}{n \imath_T} = o_\Pr(1)$

\begin{proposition} Under the conditions of Theorem 1 and Condition R', the penalty loadings $\hat \phi_j$ constructed by the above algorithm are asymptotically valid.  If $K \geq 2$, then $u \overset \Pr \rightarrow 1$.
\end{proposition}
 
\textbf{Proof:}  The proof of the above proposition is similar to the proof of Lemma 11 in \citen{BelloniChernozhukovHansen2011} but accounts for clustering.  We first consider the basic option in which we let $\hat \epsilon_{it}^{prelim} = \ddot y_{it} - \frac{1}{nT} \sum_{i=1}^n \sum_{t=1}^T \ddot y_{it} \equiv \ddot y_{it}$ yielding
$$\hat \phi_j^2 = \frac{1}{nT} \sum_{i=1}^n \left(\sum_{t=1}^T \ddot x_{itj} \ddot y_{it} \right)^2.$$

\noindent For validity of this initial option, it is sufficient that  
$\max_j \left | \hat \phi_j^2 - \Ep \hat \phi_j^2 \right |/ \Ep \phi_j^2=o_{\Pr}(1)$ 
which is assumed in Condition R'(\textit{i}).

Now we consider the refined option
$$\hat \phi_j^2 = \frac{1}{nT} \sum_{i=1}^n \left(\sum_{t=1}^T \ddot x_{itj} \hat  \epsilon_{it} \right)^2$$

\noindent where $\hat \epsilon_{it} = \ddot y_{it} - \hat \Ep[\ddot y_{it} | \ddot w_{it}] = \ddot y_{it} - \ddot x_{it}' \hat \beta$ with $\hat \beta$ either the Cluster-Lasso or Post-Cluster-Lasso based on the initial option for penalty loadings.  It is sufficient to prove that 
$$\max_j \left |  \frac{1}{nT} \sum_{i=1}^n \left(\sum_{t=1}^T \ddot x_{itj} \hat \epsilon_{it} \right)^2 - \frac{1}{nT} \sum_{i=1}^n \left(\sum_{t=1}^T \ddot x_{itj} \ddot \epsilon_{it} \right)^2 \right | / \Ep \phi_j^2 = o_\Pr(1). $$

\noindent Let $\delta_{it}=\ddot r(w_{it})+ \ddot x_{it}' (\beta - \hat \beta)$.  The expression on the left can then be bounded by 
\begin{align*}
& \max_j \left |  \frac{1}{nT} \sum_{i=1}^n \left(\sum_{t=1}^T \ddot x_{itj} \hat \epsilon_{it} \right)^2 - \frac{1}{nT} \sum_{i=1}^n \left(\sum_{t=1}^T \ddot x_{itj} \ddot \epsilon_{it} \right)^2 \right | / \Ep \phi_j^2\\
& \leq 2 \max_j \left |  \frac{1}{nT} \sum_{i=1}^n \left(\sum_{t=1}^T \ddot x_{itj}\ddot \epsilon_{it} \right)  \left(\sum_{t= 1}^T \ddot  x_{itj}\delta_{it} \right) \right | /\Ep \phi_j^2 + \max_j \left |  \frac{1}{nT} \sum_{i=1}^n \left(\sum_{t=1}^T \ddot x_{itj} \delta_{it} \right)^2 \right | /\Ep \phi_j^2 \\
&\leq 2 \max_j \left | \frac{1}{nT} \left( \sum_{i=1}^n \left(\sum_{t=1}^T \ddot x_{itj}\ddot \epsilon_{it} \right)^2 \right)^{1/2} \left( \sum_{i=1}^n \left(\sum_{t=1}^T \ddot x_{itj} \delta_{it} \right)^2 \right)^{1/2} \right | /\Ep \phi_j^2 \\ & \qquad + \max_j \left |  \frac{1}{nT} \sum_{i=1}^n \left(\sum_{t=1}^T \ddot x_{itj} \delta_{it} \right)^2 \right | /\Ep \phi_j^2 \\
&\leq 2 \left[\max_j \left | \left( \frac{1}{nT} \sum_{i=1}^n \left(\sum_{t=1}^T \ddot x_{itj}\ddot \epsilon_{it} \right)^2 /\Ep \phi_j^2 \right)^{1/2} \right |\right] \left[ \max_j \left| \left(\frac{1}{nT}  \sum_{i=1}^n \left(\sum_{t=1}^T \ddot x_{itj} \delta_{it} \right)^2 /\Ep \phi_j^2 \right)^{1/2} \right |\right]  \\ & \qquad + \max_j \left |  \frac{1}{nT} \sum_{i=1}^n \left(\sum_{t=1}^T \ddot x_{itj} \delta_{it} \right)^2 \right | /\Ep \phi_j^2 \\
&= O_{\Pr}(1)\left[ \max_j \left| \left(\frac{1}{nT}  \sum_{i=1}^n \left(\sum_{t=1}^T \ddot x_{itj} \delta_{it} \right)^2 /\Ep \phi_j^2 \right)^{1/2} \right |\right] + \max_j \left |  \frac{1}{nT} \sum_{i=1}^n \left(\sum_{t=1}^T \ddot x_{itj} \delta_{it} \right)^2 \right | /\Ep \phi_j^2
\end{align*}
using Condition R'(ii).  

It thus suffices to bound $\max_j \left |  \frac{1}{nT} \sum_{i=1}^n \left(\sum_{t=1}^T \ddot x_{itj} \delta_{it} \right)^2 \right | /\Ep \phi_j^2$.  We have that 
\begin{align*}
\max_j \left |  \frac{1}{nT} \sum_{i=1}^n \left(\sum_{t=1}^T \ddot x_{itj} \delta_{it} \right)^2 \right | /\Ep \phi_j^2 
&\leq  \max_j \left | \sum_{i=1}^n \( \frac{1}{T}\sum_{t=1}^T \ddot x_{itj}^2 \) \( \frac{1}{T} \sum_{t=1}^T \delta_{it}^2 \) / \Ep[\phi_j^2] \right | \\
&\leq \left( \max_{j} \max_{i,t} \ddot x_{itj}^2/\Ep[\phi_j^2] \right) \frac{1}{nT} \sum_{i=1}^n \sum_{t=1}^T \delta_{it}^2.
\end{align*}
Note that $\delta_{it}^2 = (\ddot r(w_{it})+ \ddot x_{it}' (\beta - \hat \beta))^2 \le 2\ddot r(w_{it})^2+ 2(\ddot x_{it}' (\hat\beta - \beta))^2$.  Under Condition ASM, we have that $\frac{1}{nT} \sum_{i=1}^n \sum_{t=1}^T \ddot r(w_{it})^2 = O_{\Pr}(\frac{s}{n\imath_T})$, and we have that $\frac{1}{nT} \sum_{i=1}^n \sum_{t=1}^T (\ddot x_{it}' (\hat\beta - \beta))^2 = O_{\Pr}(\frac{s \log(p \vee nT)}{n \imath_T})$ from Theorem 1 because the initial penalty loadings obey (\ref{eq: valid loadings}).  Thus, 
\begin{align*}
\max_j \left |  \frac{1}{nT} \sum_{i=1}^n \left(\sum_{t=1}^T \ddot x_{itj} \delta_{it} \right)^2 \right | /\Ep \phi_j^2 
&\leq \left( \max_j \max_{i,t} \ddot x_{itj}^2/\Ep[\phi_j^2] \right) O_{\Pr}\left(\frac{s \log(p \vee nT)}{n \imath_T}\right) = o_{\Pr}(1)
\end{align*}
under Condition R'(iii).  The result then follows.

\subsection{Comments on Condition R'}

Condition R' imposes a set of high level conditions.  These conditions could be verified under lower level primitive conditions.  Of these conditions, Condition R'(i) will generally require the most stringent conditions as it may require convergence of a $\frac{1}{nT}$ normalized sum over $nT^2$ random elements which do not have mean zero.  In the following two examples, we provide simple sample sets of sufficient primitive conditions under which Condition R' can be established.  The first example covers a $T$ fixed case as well as a case when $T \rightarrow \infty$ and data are strongly dependent in the sense that $\imath_T \propto 1$.  The second case covers a scenario where $T \rightarrow \infty$ and $\imath_T \propto T$ which would be appropriate with weakly dependent data.

\noindent{\textbf{Example 1.}}  \textit{Suppose that $T$ is fixed or that $\imath_T \propto 1$ and that $0 < m \leq \frac{\imath_T}{T} \Ep[\phi_j^2] \leq M < \infty$ for $1 \leq j \leq p$.  Further, assume that the sequences of random variables $\{\ddot y_{it}, \ddot x_{it}\}_{t=1}^{T}$ are iid across i, that regressors are uniformly bounded with $\sup_{i,t,j} |\ddot x_{itj}| \leq B < \infty$, and that $\sup_t \Ep[\ddot y_{it}^4] \leq M < \infty$.  Then Condition R'(i) is satisfied if $\frac{\log(p \vee n)^3}{n} \rightarrow 0$.}

The $T$ fixed and $T \rightarrow \infty$ with $\imath_T \propto 1$ are similar in that essentially no information is accumulating in the time series dimension.  In this case, rates of convergence are completely governed by the cross-sectional dimension and we see that Condition R'(i) may be satisfied when $n$ grows quickly enough relative to $\log(p)$ under moment and boundedness conditions similar to those employed elsewhere in the literature, e.g. Example 3 in \citen{BelloniChernozhukovHansen2011}.  

\noindent{\textbf{Example 2.}}  \textit{Suppose that $T \rightarrow \infty$ with $\imath_T \propto T$ and that $0 < m \leq \frac{\imath_T}{T} \Ep[\phi_j^2] \leq M < \infty$ for $1 \leq j \leq p$.  Suppose that $\{y_{it},x_{it}\}$ is a strictly stationary strongly mixing ($\alpha$-mixing) process with mixing coefficients satisfying $\theta(j) \leq \exp\{-2cj\}$.\footnote{We use $\theta(\cdot)$ to denote strong mixing coefficients rather than the more conventional $\alpha(\cdot)$ to prevent confusion with notation for unobserved heterogeneity used in Section 2 and parameters of interest used in Section 4.}  Further, suppose that the sequences of random variables $\{\{y_{it},x_{it}\}_{t = 1}^{T},\alpha_i\}$ where $\alpha_i$ denotes unobserved individual specific heterogeneity are iid across $i$.  Assume that observed random variables are uniformly bounded with $\sup_{i,t,j} |x_{itj}| \leq B$ and $\sup_{i,t} |y_{it}| \leq B$.  Then Condition R'(i) is satisfied if $\frac{T \log(\max\{n,T,p\})^3}{n} \rightarrow 0$.}  

Example 2 differs interestingly from Example 1 in requiring that $\frac{T}{n} \rightarrow 0$.  The need for having $n$ large relative to $T$ in satisfying Condition R'(i) comes from the use of clustering even though the data is weakly dependent and the fact that $\Ep[x_{itj}y_{it}]$ will not generally be zero.  The clustering estimator in the numerator then behaves like the variance of a strongly dependent process while the term in the denominator $\Ep[\phi^2_j]$ depends only on the $x_{itj}\epsilon_{it}$ process which is weakly dependent.  Keeping the numerator from exploding relative to the denominator then requires a stronger condition on the rate of growth of $T$ relative to $n$.  Without this condition, one could not guarantee that $\ell$ and $u$ (\ref{eq: valid loadings}) would remain bounded; specifically, one could produce $u \rightarrow \infty$ which could result in inflated initial penalty loadings and the failure to select any variables even when there are strong predictors among the set of variables considered.  This feature suggests that there may be a price to pay in ability to select variables in this context in using the clustered variance estimator which is agnostic about dependence structures relative to a covariance estimator more tailored to a weakly dependent setting.

\section{Proofs of main results}\label{SectionProofs}

\subsection{Proof of Theorem 1.}

In this section we give a proof of Theorem 1.  The bounds provided in this section give more information on the performance of the Cluster-Lasso and Post-Cluster-Lasso estimates than does Theorem 1.  In particular, we give more explicit information concerning the actual constants that imply the asymptotic bounds.  The proof proceeds in eight steps.  The first four steps derive bounds on the performance of the Cluster-Lasso estimator.  The next  three steps focus on the analysis of the Post-Cluster-Lasso estimator.  The eighth step brings together the Cluster-Lasso bounds and the Post-Cluster-Lasso bounds to prove the results listed in the statement of the theorem.

\textbf{Step 1.}   Let
$$\displaystyle a: = \min_{1\leq j\leq p} |\phi_j| \leq \max_{1\leq j\leq p} |\phi_j | =: b.$$
We  note here that our goal over the course of the first four steps is to show:
\begin{equation}\label{claim1}\EnT ( \ddot x_{it} \hat \beta - \ddot x_{it}\beta )^2 = O_{\Pr} \(\frac{1}{\kappa_{\bar C} (\ddot M)} \sqrt{\frac{s \log ( p / \gamma )}{n\imath_T }} \)
\end{equation}
\begin{equation}\label{claim2}\| \hat \beta - \beta \|_1 = O_{\Pr}\(\frac{1}{(\kappa_{2\bar C}(\ddot M))^2} \sqrt{ s^2 \log(p/\gamma) / n\imath_T} \),
\end{equation}
where 
$$\bar C = \frac{uc+1}{\ell c -1} \frac{b}{a}, $$
and $$ \kappa_C^2 (M): = \min_{\delta : \|\delta_{I^c}\|_1 \leq C \|\delta_I\|_1, \delta \neq 0 , | I | \leq s }s \frac{ \delta' M \delta}{\| \delta_I \|_1^2}$$
where, for a given vector $\delta \in \mathbb{R}^{p}$ and set of indices $I \subset \{1,...,p\}$ with complement $I^c = \{1,...,p\} \backslash I$, we denote by $\delta_I$ the vector in which $\delta_{I_j} = \delta_j$ if $j \in I$ and $\delta_{I_j} = 0$ if $j \notin I$ with $\delta_{I^c}$ defined similarly.
The latter quantity is the restricted eigenvalue and can be controlled by the sparse eigenvalues discussed in the text.  We will at times abbreviate our notation with $\kappa_C(\ddot M) = \kappa_C$.  We next define weighted restricted eigenvalues, which for some estimates are more convenient to work with,
$$ \bar \kappa^2_{C} := \min_{\delta \in \mathbb{R}^p : \  \|\phi {\delta}_{I^c}\|_{1} \leq C\|\phi {\delta}_{I}\|_1, \|\delta\|_2 \neq 0, |I| \leq s}  s \frac{ \delta' \ddot M \delta}{\| \delta_I \|_1^2}\ \ $$
where
$$\| \phi \delta_I \|_1 : = \sum_{j \in I} |\phi_j \delta_j|.$$
This quantity controls the modulus of continuity between the prediction norm $\sqrt{ \delta' \ddot M \delta } = \sqrt{ \EnT (\ddot x_{it}'\delta)^2 }$ and the $\ell_1$-norm $\|\delta\|_1$ within a restricted region. Note that for every $C>0$, we have
$$\bar \kappa_C \geq (1/b)\kappa_{(bC/a)}(\ddot M).$$
The above inequality holds true because of the inclusion $\{\delta \in \RR^p: \|\phi {\delta}_{I^c}\|_{1} \leq C\| \phi {\delta}_{I}\|_1 \} \subseteq \{\delta \in \RR^p: a\| {\delta}_{I^c}\|_{1} \leq bC\| {\delta}_{I}\|_1 \}$  because $\|\phi{\delta}_{I}\|_1\leq b \| {\delta}_{I}\|_1 $ and $\| \phi \delta_{I^c}\|_1 \geq a \|\delta_{I^c}\|_1.$


%

The main results of Step 1 are the following bounds:

Define $$S_j :=  2\phi_j^{-1} \frac{1}{{ nT}}   \sum_{i=1}^n \sum_{t=1}^T \ddot x_{itj} \ddot \epsilon_{it}.$$
Under Condition ASM, if $\lambda/nT \geq c \max_{1 \leq j \leq p} |S_j|$, and $\widehat \phi_j$ satisfy the condition $\ell\phi_j \leq \hat \phi_j \leq u \phi_j$ with $u \geq 1 \geq \ell > 1/c$, then
$$
\EnT ( \ddot x_{it} \hat \beta - \ddot x_{it}\beta )^2  \leq  \( \(u + \frac{1}{c}\) \frac{\lambda \sqrt{s}}{n T \bar \kappa_{c_0}}+2\sqrt{\EnT \ddot{r}^2(w_{it})} \)^2,
$$
$$ \begin{array}{cll}\sum_{j=1}^p |\phi_j(\hat \beta_j - \beta_j )| & \leq &  3c_0 \frac{\sqrt{s}}{\bar \kappa_{2c_0}}\left( ( u + [1/c])\frac{\lambda \sqrt{s}}{nT \bar \kappa_{c_0}}+2\sqrt{\EnT \ddot{r}^2(w_{it})} \right) \\
& + &  \frac{3c_0nT}{\lambda} \EnT \ddot{r}^2(w_{it}), 
\end{array}$$
where $c_0 = (uc+1)/(\ell c-1)$.
The bounds are finite sample bounds which are established in \citen{BellChenChernHans:nonGauss} Lemma 6.  In particular, the bound is valid regardless of dependence structures in the underlying random variables, and so it applies in our case.

\textbf{Step 2.}
   In this step we control quantiles of the maximum of the scores
$ S_{j} =  2 \EnT \phi_j^{-1} \ddot x_{itj} \ddot \epsilon_{it},$
which implies that the penalty level prescribed in the text implies that the Regularization Event (\ref{RegularizationEvent}) occurs with probability $1-o(1)$.

In \citen{BelloniChernozhukovHansen2011} it was shown, using the results of \citen{jing:etal},  that 
$$\Pr \left( \max_{1 \leq j\leq p }\left |\frac{\sum_{i=1}^n U_{ij}}{\sqrt{ \sum_{i=1}^n U^2_{ij}}}\right |  >  \Phi^{-1}(1- \gamma/2p)  \right) \leq \gamma \(1 + \frac{A}{\ell^3_n}\)$$
for $U_{ij}$ independent random variables across $i$ with mean zero,
where $A$ is an absolute constant (independent of $n,p,U_{ij}$), provided that for $\ell_n>0$
$$0 \leq \Phi^{-1}(1- \gamma/(2p))  \leq \frac{n^{1/6}}{\ell_n} \min_{1\leq j \leq p} \frac{\left( \frac{1}{n} \sum_{i=1}^n E U_{ij}^2\right)^{1/2}}{\left(\frac{1}{n} \sum_{i=1}^n E|U_{ij}^3| \right)^{1/3}}-1.$$
We will apply this result using $U_{ij} = \frac{1}{\sqrt{T}} \sum_{t=1}^T \ddot x_{itj} \ddot \epsilon_{it}$ which gives us the identity $S_j = \frac{2}{\sqrt{nT}} \frac{\sum_{i=1}^n U_{ij}}{\sqrt{ \sum_{i=1}^n U^2_{ij}}}$. To apply the result, we verify the immediately preceding condition for our particular choice of $U_{ij}$. 

$$\frac{n^{1/6}}{\ell_n} \min_{1\leq j \leq p} \frac{\left( \frac{1}{n} \sum_{i=1}^n \Ep \[ \(\frac{1}{\sqrt{T}} \sum_{t=1}^T \ddot x_{itj} \ddot \epsilon_{it}\)^2 \]\right)^{1/2}}{\left(\frac{1}{n} \sum_{i=1}^n \Ep \[ \left |\frac{1}{\sqrt{T}} \sum_{t=1}^T \ddot x_{itj} \ddot \epsilon_{it} \right |^3  \] \)^{1/3}}-1$$
$$= \frac{n^{1/6}}{\ell_n} \min_{1\leq j \leq p} \sqrt{E[\phi_j^2] } / \varpi_j  - 1$$
We note that by Condition R(iii), $n^{1/6}/\ell_n \min_{1\leq j \leq p} \sqrt{E[\phi_j^2] } / \varpi_j  - 1 \geq \frac{1}{C} n^{1/6}/\ell_n -1$ for some constant $C>0$.  By R(iv), it suffices to choose $\ell_n =\frac{1}{C} \log n$ and obtain $n^{1/6}/ \log n - 1 \geq \Phi^{-1}({1-\gamma/2p})$.  
Therefore, 
\begin{align*}\Pr \(\frac{\lambda}{nT} \geq c\max_j |S_j| \) 
& = 1 - \Pr \left( \max_{1 \leq j\leq p }\left |\frac{\sum_{i=1}^n U_{ij}}{\sqrt{ \sum_{i=1}^n U^2_{ij}}}\right |  >  \Phi^{-1}(1- \gamma/2p)  \right) \\
& \geq 1 - \gamma \(1 + \frac{A}{\ell^3_n}\) = 1-o(1).
\end{align*}
As desired, it follows that the regularization event occurs with high probability.

\textbf{Step 3.} Define the expected ``ideal" penalty loadings
$$
 \phi^0_j :=  \sqrt{\Ep\[ \frac{1}{T} \( \sum_{t=1}^T \ddot x_{itj} \ddot \epsilon_{it} \)^2 \]},
$$
where the entries of  $\phi^0_{j}$ are bounded away from zero and from above by
$$
\sqrt{\Ep\[ \frac{1}{T} \sum_{t=1}^T \ddot x^2_{itj} \ddot \epsilon^2_{it} \] } \sqrt{T/\imath_T} = O(\sqrt{T/\imath_T}),
$$
uniformly in $j \leq p$ and in $n$,  by Condition R.
Then the empirical ``ideal" loadings converge to the expected ``ideal" loadings:  $$\max_{1 \leq j \leq p} \frac{|\phi_j - \phi_j^0|}{\phi_j^0} \to_{\Pr} 0.$$ This is assumed
in Condition R.

From this we conclude that
$$
b = \max_{1 \leq j \leq p}  \phi_j = O_P(\sqrt{T/\imath_T}),  \quad  a^{-1} = \( \max_{1 \leq j \leq p}  \phi_j^{-1}\) = O_P(\sqrt{\imath_T/T}).
$$
Using the inequality noted in Step 1 with $C=c_0=(uc+1)/(\ell c-1)$ we have $\bar \kappa_{c_0} \geq (1/b) \kappa_{\bar C}(\ddot M)$, so that
$$\frac{1}{\bar \kappa_{c_0}} \leq \frac{b}{\kappa_{\bar C}(\ddot M)} \leq  O_P\(\sqrt{\frac{T}{\imath_T}}\) \frac{1}{\kappa_{\bar C}(\ddot M)}.$$
Similarly,
$$
\frac{1}{\bar \kappa_{2c_0}} \leq \frac{b}{\kappa_{2\bar C}(\ddot M)} \leq  O_P\(\sqrt{\frac{T}{\imath_T}}\) \frac{1}{\kappa_{2\bar C}(\ddot M)}.
$$
\textbf{Step 4.}
  Combining the results of all the steps above, given that $\lambda=2c\sqrt{nT}\Phi^{-1}(1-\gamma/(2p)) = O \( \sqrt{nT\log(p/\gamma)} \)$ and that penalty loadings satisfy $\ell \phi_j \leq \widehat \phi_j \leq u\phi_j$, $\ell \rightarrow_{\Pr} 1$, $u \rightarrow_{\Pr} C \geq 1$,
and using the bound $\sqrt{\EnT \ddot{r}^2(w_{it})} = O_{\Pr}(\sqrt{s/n\imath_T})$ from Condition ASM, we obtain
$$
\EnT ( \ddot x_{it} \hat \beta - \ddot x_{it}\beta )^2  =O_{\Pr} \(  \left( \sqrt{\frac{T}{\imath_T}} \frac{1}{\kappa_{\bar C}(\ddot M)} \right)^2 {\frac{s \log (p/\gamma)}{ nT }} +  {\frac{s }{ n\imath_T }} \)
$$
which implies that
$$
\EnT ( \ddot x_{it} \hat \beta - \ddot x_{it}\beta )^2  =O_{\Pr} \(  \frac{1}{\kappa^2_{\bar C}(\ddot M)} {\frac{s \log (p/\gamma)}{ n \imath_T}} +  {\frac{s }{ n\imath_T }} \)
$$

In turn, by the triangle inequality and by $\EnT ( \ddot f(w_{it}) - \ddot x_{it}\beta )^2 = O_{\Pr}(s/n\imath_T)$ holding by Condition ASM, we have
$$
\EnT ( \ddot f(w_{it}) - \ddot x_{it}\hat \beta )^2 = O_{\Pr}\(\frac{1}{\kappa^2_{\bar C}(\ddot M)}  {\frac{s \log (p/\gamma)}{ n\imath_T }} \).
$$
The first claim (\ref{claim1}) now follows.

To derive the $\ell_1$-rate we apply the last inequality of Step 1 as follows:
$$\begin{array}{l}  \displaystyle  \|\hat{\beta} - {\beta} \|_1 \displaystyle  \leq \(\max_{j \leq p} | \phi^{-1}_j | \) \sum_{j=1}^n |\phi_j (\hat{\beta_{j}} - {\beta_{j}} )| \\
  \displaystyle = O_{\Pr}\( \(\max_{j \leq p} | \phi^{-1}_j | \) \left (
 \frac{\sqrt{s}}{\bar \kappa_{2c_0}} \left(\frac{1}{\bar \kappa_{c_0}}\sqrt{\frac{s \log (p/\gamma)}{ n \imath_T}} +  \sqrt{\frac{s }{ n\imath_{T} }}\right) +  \frac{1}{\sqrt{\log (p/\gamma)}} \frac{s}{\sqrt{n\imath_T}} \right) \) \\
 \displaystyle
 =O_{\Pr} \(\sqrt{\frac{\imath_T}{T}} \( \sqrt{\frac{T}{\imath_T}} \frac{1}{\kappa_{2\bar C}(\ddot M)}\)^2 \sqrt{\frac{s^2 \log (p/\gamma)}{ n\imath_T }} \) = O_{\Pr} \( \frac{1}{\kappa^2_{2\bar C}(\ddot M)}\sqrt{\frac{s^2 \log (p/\gamma)}{ n \imath_T }} \).\end{array} $$
The second claim, (\ref{claim2}), then follows immediately.

\textbf{Step 5.}  In this step we switch focus away from Cluster-Lasso to deriving performance bounds for Post-Cluster-Lasso.


It is convenient to define a type of projection operator using the following notation.  
Consider $nT$-dimensional vector $v= \mathrm{vec} \{v_{it}\}_{i \in [n], t \in [T]}$, for $[n]=1,...,n$ and $[T]= 1,..., T$, which is created by stacking together vectors $\mathrm{vec}\{v_{it}\}_{i \in [n]}$ with $t \in [T]$.   For a set of indices $S \subset \{1,\ldots,p\}$ and given data, we define
$\mathcal{P}_{S} v$ as the least squares projection of $nT$-vector $\{v_{it} \}_{i \in [n], t \in [T]} $ onto the linear span of $nT$-vectors $ \mathrm{vec}\{\ddot x_{itj}\}_{i \in [n], t \in [T] }$ with $j \in S$.  We also let $\ddot f=  \mathrm{vec}\{\ddot f(w_{it})\}_{i \in [n], t \in [T]}$ and $\ddot \epsilon = \mathrm{vec}\{\ddot \epsilon_{it}\}_{i \in [n], t \in [T]}$.

The goal of this step is to derive the following bounds:

\textit{
Under Conditions ASM and R, let $\widehat I$ denote the support selected by $\widehat \beta = \widehat \beta_{L}$, $I$ denote the support of $\beta$, $\widehat m=
|\widehat I \setminus I|$, and $\widehat \beta_{PL}$ be
the Post-Lasso estimator based on $\widehat I$. Then we have
}

\begin{equation}\label{step5: claim1}
\begin{array}{lll}
 \EnT  (\ddot f(w_{it} )- \ddot x_{it}'\hat \beta_{PL})^2  & =  & O_{\Pr} \Big (  {\frac{s}{n\imath_T}} {\frac{\log(s)}{\semin{s}(\ddot M)}} + \frac{{\widehat m \log(p)}}{{n\imath_T \ \semin{\widehat m}(\ddot M)}}  \Big)  \\
 & +  & O_{\Pr} \Big ( 
 \EnT( \ddot f(w_{it}) - (\mathcal{P}_{\hat I} \ddot f)_{it} )^2 \Big) ,
 \end{array}
 \end{equation}
and also
\begin{eqnarray}\label{step5: claim2}
\begin{array}{lll}
\sqrt{ \EnT (\ddot x_{it}'( \widehat\beta_{PL} -\beta ))^2} & \leq &  \sqrt{\EnT  (\ddot f(w_{it} )- \ddot x_{it}'\widehat \beta_{PL})^2} \\
& + &  \sqrt{\EnT  \ddot{r}^2(w_{it})}
\end{array} \\
 \begin{array}{lll} \sum_{j=1}^p | \phi_j (\widehat\beta_{jPL} -\beta_{j} ) | & \leq  &  \frac{\( \max_{j} |\phi_j |  \)\sqrt{\widehat m + s}}{\sqrt{\semin{\widehat m +s}(\ddot M) }} \\
 & \times & \sqrt{ \EnT (\ddot x_{it}'( \widehat\beta_{PL} -\beta ))^2}\label{step5: claim3}.
 \end{array}
\end{eqnarray}

\textit{
If in addition $\lambda/nT \geq
c\max_{1\leq j \leq p} | S_j|$, and $\hat \phi_j$ satisfy $\ell \phi_j \leq \hat \phi_j \leq u \phi_j$ with $u \geq 1 \geq \ell > 1/c$ in the first stage for Lasso, then we have
 \begin{equation} 
 \label{step5: claim4} 
 {\EnT (\ddot f(w_{it}) - \mathcal P_{\hat T} \ddot f(w_{it}))^2 } \leq  \( \(u + \frac{1}{c}\) \frac{\lambda \sqrt{s}}{nT \bar \kappa_{c_0}}+3\sqrt{\EnT \ddot{r}^2(w_{it})} \)^2.
 \end{equation}
}

To proceed with proving these bounds, note that we have $\ddot f(w_{it}) - \ddot x_{it}'\hat\beta_{PL} = \left( ( 1 - \mathcal{P}_{\hat I}) \ddot f  - \mathcal{P}_{\hat I} \ddot \epsilon\right)_{it}$.
By the triangle inequality and by properties of projections, we have
\begin{align*}
 \sqrt{ \EnT (\ddot f(w_{it}) - \ddot x_{itj}'\hat\beta_{PL} )^2 } &\leq  \sqrt{ \EnT ( ( 1 - \mathcal{P}_{\hat I}) \ddot f)_{it} )^2 }   \\
& +  \sqrt{ \EnT ( (\mathcal{P}_{ I} \ddot \epsilon)_{it})^2 }  \\
& +  \sqrt{ \EnT ( ( \mathcal{P}_{\hat I \setminus I } \ddot \epsilon)_{it} )^2. }  \\
\end{align*}

We next bound the terms in the above expression.  Beginning with the third term, because $\mathcal P_{\hat I \setminus I}$ is a projection onto $\hat m = | \hat I \setminus I|$ vectors, we can use the sparse eigenvalue bound to obtain


$$ \sqrt{ \sum_{i=1}^{n}\sum_{t=1}^{T} \left ( \left( \mathcal{P}_{\hat I \setminus I } \ddot \epsilon\right)_{it} \right )^2 }  \leq  \sqrt{1/\semin{\widehat m}(\ddot M)} \sqrt{ \sum_{j \in \hat I \setminus I} \( \frac{1}{\sqrt{nT}} \sum_{i=1}^{n}\sum_{t=1}^{T} \ddot x_{itj} \ddot \epsilon_{it} \)^2 } $$
Which reduces further to
$$ \sqrt{ \sum_{i=1}^{n}\sum_{t=1}^{T}  (\left( \mathcal{P}_{\hat I \setminus I } \ddot \epsilon\right)_{it} )^2 }  \leq  \sqrt{\hat m /\semin{\widehat m}(\ddot M)} { \max_{j \leq p} \left |\frac{1}{\sqrt{nT}} \sum_{i=1}^{n}\sum_{t=1}^{T}  \ddot x_{itj} \ddot \epsilon_{it} \right | } $$

Under Condition R,  we have
$$\max_{j \leq p}  {\sqrt{nT} \left |\EnT  \ddot x_{itj} \ddot \epsilon_{it} \right | }  \leq  \max_{j \leq p}{ \sqrt{nT} \frac{1}{2}|S_j| \max_{j \leq p} \phi_j } $$ $$= O_{\Pr} \(  \sqrt{\log(p)} \max_{ j\leq p} \phi_j \)= 
O_{\Pr} (\sqrt{ \log (p)} \sqrt{T/\imath_T}).$$


Next, we bound the second term. Proceeding as above, we deduce that
$$  \sqrt{ \sum_{i=1}^{n}\sum_{t=1}^{T} ( \mathcal{P}_{ I} \ddot \epsilon_{it})^2 }= O_{\Pr} \( \log ( s) s/\semin{s}(\ddot M)\max_{j\leq p}  \phi_j \) = O_{\Pr} \( \log ( s) s/\semin{s}(\ddot M)   \sqrt{T/\imath_T}\) $$


These relations yield the first claim (\ref{step5: claim1}) of Step 5. The claim (\ref{step5: claim2}) follows from the triangle inequality. The claims (\ref{step5: claim3})  and (\ref{step5: claim4}) are finite sample results, which are not influenced by the nature of the present problem; their proof is identical to the one given in \citen{BellChenChernHans:nonGauss} (see their Lemma 7).

\textbf{Step 6.} In this step we provide a sparsity bound for Lasso, which is important for establishing various rate results and fundamental to the analysis of Post-Lasso.  Specifically, we gather the following three results proven in \citen{BellChenChernHans:nonGauss} Lemmas 8, 9, and 10 respectively.

Empirical pre-sparsity for Lasso:
Let $\widehat I$ denote the support selected by the Lasso
estimator, $\hat m = |\widehat I\setminus I|$, and assume that
$\lambda/nT \geq c \max_{1 \leq j \leq p } |S_j| $ and $u\geq 1 \geq \ell > 1/c$.
Then, for $c_0 = (uc+1)/(\ell c - 1)$ we have
$$\sqrt{\hat m} \leq \sqrt{\semax{\hat m}(\ddot M)} \( \max_{j \leq p } \phi_j^{-1} \) c_0\[\frac{2\sqrt{s}}{\bar \kappa_{c_0}} + \frac{6nT\sqrt{\EnT \ddot{r}^2(w_{it})}}{\lambda}\].$$

Sub-linearity of maximal sparse eigenvalues:
\label{Lemma:SparseEigenvalueIMP}Let $M$ be any positive semi-definite matrix. For any integer $k \geq 0$ and constant $\ell \geq 1$ we have $ \semax{\ceil{\ell k}}(M) \leq  \lceil \ell \rceil \semax{k}(M).$

 Sparsity bound for Lasso under data-driven penalty:
Let $\widehat m = |\widehat I\setminus I|$ for $\hat I, I$ as above.  Suppose that $\frac{\lambda}{nT} \geq c \max_j |S_j|$. Consider the set $$\mathcal{M}=\left\{ m \in \mathbb{N}:
m > s \ 2\semax{m}  (\ddot M) \( \max_{j \leq p} \phi_j^{-1} \)^2  \[\frac{2c_0}{\bar \kappa_{c_0}} + \frac{6c_0nT\sqrt{\EnT \ddot{r}^2(w_{it})}}{\lambda\sqrt{s}}\]^2 \right\}.$$ Then,
$$ \hat m\leq s \  \left( \min_{m \in \mathcal{M}}\semax{m\wedge nT} (\ddot M) \right) \   \( \max_{j \leq p} \phi_j^{-1}\)^2  \(\frac{2c_0}{\bar \kappa_{c_0}} + \frac{6c_0nT\sqrt{\EnT \ddot{r}^2(w_{it})}}{\lambda\sqrt{s}}\)^2.$$

\textbf{Step 7.} Next we combine the previous steps to establish Theorem 1.
Recall that
$$\frac{1}{\bar \kappa_{c_0}} \leq \frac{b}{\kappa_{\bar C}} \leq  O_P\( \sqrt{\frac{T}{\imath_T}} \right) \frac{1}{\kappa_{\bar C}}.$$
Let $$\mu^2: = \min_{k} \{ \varphi_{\max}(\ddot M)/\varphi_{\min}(\ddot M) : k > 18 \bar C^2 s \varphi_{\max}(k)(\ddot M) / \kappa_{\bar C}^2 \}$$ with $\bar C$ defined as in Step 1.  Let $\bar k$ be the integer that achieves the minimum in the definition of $\mu^2$. Since $\sqrt{\EnT \ddot{r}^2(w_{it})} = O_{\Pr}(\sqrt{s/n \imath_T})$ by Condition ASM and $\lambda \geq c \sqrt{nT} \log (p/o(1)) $ by our choice of $\lambda$ in (\ref{eq: set lambda}),
$$
\max_{j\leq p} \phi_j^{-1} \frac{6c_0nT\sqrt{\EnT \ddot{r}^2(w_{it})}}{\lambda\sqrt{s}}  = O_{\Pr}\( \sqrt{\frac{\imath_T}{T}}\)   O_{\Pr}\( \sqrt{\frac{T}{\imath_T \log (p/o(1))}}\) \to_P 0;
$$
and we have that $\bar k \in \mathcal{M}$ with high probability $1- o(1)$ as $n\to \infty$.

Moreover,  as long as $\lambda/nT \geq c\max_{1\leq j\leq p}|S_j |$, $\ell \to_{\mathrm{P}} 1$ and $c>1$, by Step 6, we have that
\begin{equation}\label{SparsityBound} \widehat m =O_{\Pr} \( s \mu^2 \semin{\bar k+s}/\kappa_{\bar C}^2 \) = O_{\Pr} \(s \mu^2 \semin{\hat m+s}/\kappa_{\bar C}^2  \)  \end{equation}
since  $\bar k \in \mathcal{M}$ implies $\bar k \geq \hat m$.

By the choice of $\lambda=2c\sqrt{nT} \Phi^{-1}(1-\gamma/(2p))$, and since $\gamma \to 0$,
the event $\lambda/nT \geq c\max_{1\leq j \leq p}|S_j|$ holds with probability approaching $1$.
Therefore, by claim (\ref{step5: claim1}) and (\ref{step5: claim4}) of Step 5, we have
$$ \EnT (\ddot f(w_{it}) - \ddot x_{it} \hat \beta_{PL} )^2 = O_{\Pr} \( \(\frac{\mu}{\kappa_{\bar C}}\sqrt{\frac{s\log p}{n \imath_T}} + \sqrt{\EnT \ddot{r}^2(w_{it})} + \frac{\lambda\sqrt{s}}{nT\bar \kappa_{c_0}} \)^2 \).$$
Because $1/\bar \kappa_{c_0}  = O_P\( \sqrt{\frac{T}{\imath_T}} \)/\kappa_{\bar C}$, as noted above, and $\sqrt{\EnT \ddot{r}^2(w_{it})} = O_{\Pr}(\sqrt {s/n\imath_T} )$ by Condition ASM, we have
\begin{equation}\label{BoundPredicNormPL}  \EnT (\ddot f(w_{it}) - \ddot x_{it} \hat \beta_{PL} )^2 = O_{\Pr} \( \frac{\mu}{\kappa_{\bar C}}\sqrt{\frac{s\log (p/\gamma)}{n\imath_T}} \).\end{equation}

Next, to establish a bound for $\| \hat \beta_{PL} - \beta \|_1$, we note that since
$\|\widehat \beta_{PL}-\beta\|_0 \leq \widehat m+s$, we have
$$\|\widehat \beta_{PL}-\beta\|_1\leq
\sqrt{\|\widehat \beta_{PL}-\beta\|_0}\|\widehat \beta_{PL}-\beta\|_2 \leq \sqrt{\widehat m+s}
\frac{ \sqrt{ \EnT (\ddot x_{it} \hat \beta_{PL} - \ddot x_{it}\beta)^2 } }{\sqrt{\semin{\widehat m + s}(\ddot M)}}.$$
The sparsity bound (\ref{SparsityBound}), the prediction norm bound
(\ref{BoundPredicNormPL}), and the relation (\ref{step5: claim2}) yields

$$ \|\hat \beta_{PL} - \beta \|_1 = O_{\Pr} \( \frac{\mu^2}{\kappa_{\bar C}^2 }\sqrt{\frac{s^2 \log (p/\gamma)}{n\imath_T}} \).$$.

\textbf{Step 8.}  This step provides bounds on the quantities $b/a, \bar C, \mu, \kappa_{\bar C}, \phi_{\min}(\bar k + s)$ used in steps 1-7.  These bounds imply the results of the theorem.  Under the stated condition $R$, we have
that with probability approaching 1,
$$
 b/a  =O(1), \quad \bar C  =O( 1).
$$
Our  assumption on the sparse eigenavalues as well as by the standard bounds on restricted eigenvalues in terms of sparse eigenvalues, see e.g.  \citen{BellChenChernHans:nonGauss}, imply that with probability $1-o(1)$:
$$
\mu =O(1),   \quad 1/\kappa_{\bar C} = O(1), \quad  \mu \varphi_{min}(\bar k +s)/\kappa_{\bar C} =O(1).
$$
Thus, the claims of the theorem follow from the bounds established in Step 1 for Cluster-Lasso and Step 7 for  Post-Cluster-Lasso. 
\qed

\subsection{Proof of Theorem 2.}\label{Thm2}
We ignore approximation errors in stating the proof of the theorem for brevity and simplicity.  Incorporating approximation errors satisfying Condition ASM does not substantively affect the argument.

\textbf{Step 0.} Using clustered penalty loadings satisfying the construction given in the text, we have by Theorem 1 that the Post-Lasso estimator obeys:
\begin{eqnarray}\label{eq: Step 0}
&& \frac{1}{nT} \sum_{i=1}^n \sum_{t=1}^T (\widehat D_{i{t}} -\ddot D_{i{t}})^2 = O_{\Pr} \left ( \frac{s \log (p\vee nT) }{n\imath_T} \right ) = o_{\Pr}(1) \\
&& \sqrt{\log p}  \| \widehat \pi- \pi\|_{1} =O_{\Pr} \left ( \sqrt{\frac{s^2 \log^2 (p\vee nT) }{n\imath_T}} \right ) = o_{\Pr}(1).\label{eq: Step 01}
\end{eqnarray}

\textbf{Step 1.}    We have the decomposition
\begin{eqnarray*}
\sqrt{n\imath_T^D}V^{-1/2} (  \hat \alpha - \alpha)
&= & \( \EnT \widehat D_{it} \ddot d_{it} \)^{-1} \( \sqrt{n\imath^D_T} \frac{V^{-1/2} }{nT} \sum_{i=1}^n \sum_{t=1}^T \widehat D_{it} \ddot \epsilon_{it}\)  \\
&=&  \left (\EnT \widehat D_{it} \ddot d_{it}\right )^{-1} \left (\sqrt{n\imath^D_T} \frac{V^{-1/2} }{nT} \sum_{i=1}^n \sum_{t=1}^T \ddot D_{it} \ddot \epsilon_{it}+o_{{P}}(1) \right )  \\
& = & \left ( \ddotEp \ddot D_{it} \ddot d_{it} + o_{{P}}(1)\right )^{-1} \left( \sqrt{n\imath^D_T} \frac{V^{-1/2} }{nT} \sum_{i=1}^n \sum_{t=1}^T  \ddot D_{it} \ddot \epsilon_{it} + o_{{P}}(1) \right),
 \end{eqnarray*}
where by Steps 2 and 3 below:
\begin{eqnarray}
 \EnT \widehat  D_{it} \ddot d_{it} = \ddotEp \ddot D_{it}\ddot d_{it}+ o_{{P}}(1) \label{eq: to show 1} \\
 \sqrt{n\imath_T^D} \frac{V^{-1/2} }{nT} \sum_{i=1}^n \sum_{t=1}^T \widehat D_{it} \ddot \epsilon_{it} =  \sqrt{n\imath^D_T} \frac{V^{-1/2} }{nT} \sum_{i=1}^n \sum_{t=1}^T  \ddot D_{it} \ddot \epsilon_{it} + o_{{P}}(1).\label{eq: to show 2}
 \end{eqnarray}
Note that  $\Ep [\frac{1}{T} \sum_{t=1}^T \ddot D_{it} \ddot d_{it}] = \Ep[ \frac{1}{T} \sum_{t=1}^T \ddot D_{it} \ddot D_{it} ]= Q$ is bounded away from zero
and bounded from above, uniformly in $n,T$ by SMIV(ii)(a) . Moreover,
$$\frac{\imath_T^D}{T} \Omega  = \frac{ \imath_T^D}{T} \Ep \left [ \frac{1}{T} \( \sum_{t=1}^T \ddot \epsilon_{it} \ddot D_{it} \) ^2 \right ]$$
$$ = \Ep \[ \frac{1}{T} \sum_{t=1}^T \ddot \epsilon_{it}^2 \ddot D_{it}^2 \]$$ which implies  that $\frac{ \imath_T^D}{T} \Omega$  is bounded away from zero and from above, uniformly in $n,T$ because of SMIV(ii)(b).  Hence,  $\frac{ \imath_T^D}{T}  V = \frac{ \imath_T^D}{T}  Q^{-1} \Omega Q^{-1}$ is bounded above and away from zero uniformly in $n,T$.
Therefore we can conclude that
$$
\sqrt{n\imath_T^D} V^{-1/2} (\hat \alpha - \alpha) = \sqrt{n\imath_T^D}  V^{-1/2} Q^{-1}  \EnT \ddot D_{it} \ddot \epsilon_{it} + o_{{P}}(1).
$$
We wish to apply a central limit theorem to the above and so we let  $
Z_n := V^{-1/2} \sqrt{n \imath_T^D}(\hat \alpha - \alpha)  =: \frac{1}{\sqrt n}\sum_{i=1}^n z_{i,n} + o_{{P}}(1),$
where 
$$z_{i,n} := V^{-1/2} Q^{-1} \frac{\sqrt{\imath_T^D}}{T} \sum_{t=1}^T \ddot  D_{it} \ddot \epsilon_{it}$$ 
are i.i.d. with mean zero.  In addition, the variance of $z_{i,n}$ is given by

\begin{align*} 
\textrm{Var}(V^{-1/2}Q^{-1}\sqrt{ \imath_T^D}\frac{1}{T} \sum_{t=1}^T \ddot  D_{it} \ddot \epsilon_{it}) 
&=Q^{-2}V^{-1} \imath_T^D \frac{1}{T} \Ep\[ \frac{1}{T} \(\sum_{t=1}^T \ddot D_{it} \ddot \epsilon_{it}\)^2\] \\
&=Q^{-2}V^{-1} T \Ep \[\frac{1}{T} \sum_{t=1}^T \ddot \epsilon_{it}^2 \ddot D_{it}^2 \] \\
&=(\imath_T^D )^{-1} \Omega^{-1} T\Ep \[\frac{1}{T} \sum_{t=1}^T \ddot \epsilon_{it}^2 \ddot D_{it}^2 \] \\
&=(\imath_T^D )^{-1}  \imath_T^D = 1.
\end{align*}
We also have that, for $\delta>0$ and provided the expectation exists,
\begin{align*}
 \Ep \[ |z_{i,n}|^{2+\delta}  \] &= \Ep \[ \left |V^{-1/2} Q^{-1} \frac{\sqrt{\imath_T^D}}{T} \sum_{t=1}^T \ddot  D_{it}\ddot \epsilon_{it} \right |^{2+\delta}   \] \\
& = O(1) \( \frac{\sqrt{\imath_T^D}}{\sqrt{T}}\)^{2+\delta} \Ep \[ \left | \frac{1}{\sqrt T} \sum_{t=1}^T \ddot  D_{it}\ddot \epsilon_{it} \right|^{2+\delta} \] \\
\end{align*}
The choice $\delta = 1$ gives 
$$O(1) \( \frac{\sqrt{\imath_T^D}}{\sqrt{T}}\)^{2+1} \Ep \[ \left | \frac{1}{\sqrt T} \sum_{t=1}^T \ddot  D_{it}\ddot \epsilon_{it} \right|^{2+1} \] 
 = O(1) \(\frac{\varpi_D}{\sqrt{\phi_D^2}}\)^3 = O(1)
$$
 Where $\frac{\varpi_D}{\sqrt{\phi_D^2}} = O(1)$ by SMIV(ii)(b). Then this verifies the conditions in \citen{hansen:cluster}, and application of Lemma 2 therein implies that $Z_n \to_d N(0,1)$ as $n,T\rightarrow \infty$ jointly.   A similar argument verifies the result in the $n \rightarrow \infty, T$ fixed case.  Therefore, we can conclude the first statement of Theorem 2.

\textbf{Step 2.}   To show (\ref{eq: to show 1}), note that
  \begin{align*}
\left | \EnT[(\widehat D_{it} -\ddot D_{it})\ddot d_{it} ]\right | &\leq   \sqrt{\frac{1}{nT} \sum_{i=1}^n \sum_{t=1}^T (\widehat D_{i{t}} -\ddot D_{i{t}})^2 } \sqrt{\EnT\ddot d_{it}^2} \\
& =   o_{\Pr}(1)O_{\Pr}(1) \\
 \end{align*}
\noindent where $ \sqrt{\EnT\ddot d_{it}^2} =O_{\Pr}(1)$ holds by $\Ep [\frac{1}{T}\sum_{t=1}^T \ddot d_{it}^2] =O(1)$ from SMIV(ii)(a) and by Chebyshev inequality, and $ \sqrt{\frac{1}{nT} \sum_{i=1}^n \sum_{t=1}^T (\widehat D_{i{t}} -\ddot D_{i{t}})^2 } =o_{\Pr}(1)$ holds by the Step 0.

\textbf{Step 3.}   To show (\ref{eq: to show 2}), note that $\Ep[\ddot z_{itj} \ddot \epsilon_{it}]=0$, $\Ep[\ddot \epsilon_{it}|D_{i1},...,D_{it}]=0$  and
 \begin{eqnarray*}
 &  &  \frac{\sqrt {n \imath_T^D}}{nT} |\sum_{i=1}^n \sum_{t=1}^T (\widehat D_{it} - \ddot D_{it}) \ddot \epsilon_{it}|  \\
 & & = \frac{\sqrt {n \imath_T^D}}{nT}  | \sum_{i=1}^n \sum_{t=1}^T \ddot \epsilon_{it}  \ddot z_{it}' (\widehat \pi - \pi)  |   \\
& &\leq \frac{\sqrt {n \imath_T^D}}{nT} \|\hat \pi - \pi \|_1 \max_{1 \leq j \leq p} | \sum_{i=1}^n \sum_{t=1}^T \ddot \epsilon_{it}  \ddot z_{itj} | \\
 & & \leq \frac{\sqrt {n \imath_T^D}}{nT}  \|\widehat \pi - \pi\|_{1}   \max_{1 \leq j \leq p}
 \left |\frac{\sum_{i=1}^n \sum_{t=1}^T \ddot z_{itj}\ddot \epsilon_{it}}
 {\sqrt{\sum_{i=1}^n \left ( \sum_{t=1}^T \ddot z_{itj} \ddot  \epsilon_{it} \right )^2 }}  \right |   \max_{1 \leq j \leq p} \sqrt{\sum_{i=1}^n \left ( \sum_{t=1}^T \ddot z_{itj} \ddot  \epsilon_{it} \right )^2 }
 \\
 \end{eqnarray*}

Next we can bound $\max_{1 \leq j \leq p}  \left |\sum_{i=1}^n \sum_{t=1}^T \ddot z_{itj} \ddot \epsilon_{it}/ \sqrt{\sum_{i=1}^n \( \sum_{t=1}^T \ddot z_{itj} \ddot  \epsilon_{it} \)^2} \right |  =O_{\Pr}( \sqrt{ \log p})$ by the bound on moderate deviations of a maximum of a self-normalized sum which can be employed due to SMIV(ii)(b).

Finally,
\begin{align*}
 \frac{\sqrt {n \imath_T^D}}{nT} &  \max_{1 \leq j \leq p} \sqrt{\frac{1}{nT} \sum_{i=1}^n \( \sum_{t=1}^T  \ddot z_{itj} \ddot \epsilon_{it} \)^2  }\\
&  = \max_{1 \leq j \leq p}  \sqrt{ \frac{1}{n} \sum_{i=1}^n \frac{\imath_T^D}{T} \frac{1}{T} \( \sum_{t=1}^T  \ddot z_{itj}
\ddot \epsilon_{it} \)^2 } \\
&  = \sqrt{ \frac{\imath_T^D}{T}}\max_{1 \leq j \leq p}  \sqrt{ \phi_{z_j \epsilon}^2}\\
&= \sqrt{ \max_{1 \leq j \leq p}   \frac{\imath_T^D}{\imath^{z_j \epsilon}_T}}O_\Pr(1), \\
\end{align*}

by Condition SMIV(ii)(c). Thus, combining bounds above with bounds in (\ref{eq: Step 0})-(\ref{eq: Step 01})
$$
 \left |\sqrt{n\imath_T^D}\EnT[(\widehat D_{it} - \ddot D_{it}) \epsilon_i] \right |  =O_{\Pr} \(  \sqrt{ \frac{s^2 \log^2 (p\vee n) }{n \imath_T } \max_{1 \leq j \leq p } \frac{\imath_T^D}{\imath_T^{z_j \epsilon}}  } \) = o_{\Pr}(1)
$$
where the last bound is given by SMIV(ii)(d).

\textbf{Step 4.}   This step establishes
consistency of the variance estimator
of Theorem 1.

 Recall that
$$\hat \Omega :=\EnTT \hat \epsilon_{it} \hat \epsilon_{it'} \hat D_{it} \hat D_{it'}\ \ \ \text{ and} $$
$$\Omega :=\EnTT \Ep[\ddot \epsilon_{it} \ddot \epsilon_{it'} \ddot D_{it} \ddot D_{it'}] .$$   For this step, it is sufficient to show $$\frac{\imath_T^D}{T}\hat \Omega - \frac{\imath_T^D}{T} \Omega \to_{\mathrm{P}} 0.$$


Through appropriate use of the Cauchy-Schwarz and triangle inequalities, we note that
 it is sufficient to show the three following statements:
\begin{align*}
& \left | \frac{\imath_T^D}{T}\EnTT \ddot \epsilon_{it} \ddot \epsilon_{it'} \hat D_{it} \hat D_{it'}  - \frac{\imath_T^D}{T}\EnTT \hat \epsilon_{it} \hat  \epsilon_{it'} \hat  D_{it} \hat D_{it'} \right | =o_\Pr(1), \\
&\left | \frac{\imath_T^D}{T}\EnTT \ddot \epsilon_{it} \ddot \epsilon_{it'} \ddot D_{it} \ddot D_{it'} - \frac{\imath_T^D}{T}\EnTT \ddot \epsilon_{it} \ddot \epsilon_{it'} \hat D_{it} \hat D_{it'} \right | = o_\Pr(1), \\
&\left | \frac{\imath_T^D}{T}\EnTT \ddot \epsilon_{it} \ddot \epsilon_{it'} \ddot D_{it} \ddot D_{it'} - \frac{\imath_T^D}{T}\EnTT \Ep \[ \ddot \epsilon_{it} \ddot \epsilon_{it'} \ddot D_{it} \ddot D_{it'} \] \right | = o_\Pr(1). \\
\end{align*}

Given the above short outline, we bound each term separately starting with the first.  We begin by noting that by making the substitution $\hat \epsilon_{it} = \ddot \epsilon_{it} + \ddot d_{it}(\alpha - \hat \alpha)$, we have
\begin{eqnarray*}
&  & \hspace{-4mm} \left | \frac{\imath_T^D}{T} \EnTT  (\ddot \epsilon_{it} \ddot \epsilon_{it'} - \hat \epsilon_{it} \hat \epsilon_{it'}) \hat D_{it} \hat D_{it'} \right |   \\
& & \leq  \left  |\frac{\imath_T^D}{T} \EnTT \ddot d_{it} \ddot d_{it'}(\hat \alpha - \alpha)^2 \hat D_{it} \hat D_{it'} \right | \\
& & \ \ \ \ +  \left | \frac{\imath_T^D}{T}\EnTT \ddot \epsilon_{it'} \ddot d_{it}(\hat \alpha - \alpha) \hat D_{it} \hat D_{it'} \right | \\
& & \ \ \ \ + \left | \frac{\imath_T^D}{T}\EnTT \ddot \epsilon_{it} \ddot d_{it'}(\hat \alpha - \alpha) \hat D_{it} \hat D_{it'} \right | \   \\
& & = \frac{\imath_T^D}{T} \frac{1}{nT} \left  | \sum_{i=1}^n \sum_{t=1}^T \sum_{t'=1}^T \ddot d_{it} \ddot d_{it'}(\hat \alpha - \alpha)^2 \hat D_{it} \hat D_{it'} \right |\\
& & \ \ \ \ +2  \frac{\imath_T^D}{T}  \frac{1}{nT} \left  | \sum_{i=1}^n \sum_{t=1}^T \sum_{t'=1}^T\ddot \epsilon_{it'} \ddot d_{it}(\hat \alpha - \alpha) \hat D_{it} \hat D_{it'} \right |\\
& & :=A_1 + 2A_2
\end{eqnarray*}
We bound the terms $A_1$ and $A_2$ separately.  Beginning with $A_1$, note that
\begin{eqnarray*}
&A_1&= \frac{\imath_T^D}{T} \frac{1}{nT} \left  | \sum_{i=1}^n \sum_{t=1}^T \sum_{t'=1}^T \ddot d_{it} \ddot d_{it'}(\hat \alpha - \alpha_0)^2 \hat D_{it} \hat D_{it'} \right | \\
& &=  (\hat \alpha - \alpha)^2 \frac{\imath_T^D}{T}\frac{1}{nT} \left  | \sum_{i=1}^n \sum_{t=1}^T \sum_{t'=1}^T \ddot d_{it} \ddot d_{it'} \hat D_{it} \hat D_{it'} \right | \\
& &=: (\hat \alpha - \alpha)^2 A_{1.1} = O_\Pr\(\frac{1}{n \imath_T^D} \) O_\Pr(\imath_T^D) = o_\Pr(1)\\
\end{eqnarray*}
where we have defined $A_{1.1} = \frac{ \imath_T^D}{T} \frac{1}{nT} \sum_{i} ( \sum_{t} \ddot d_{it} \hat D_{it})^2$ and show that $A_{1.1} = O_\Pr(\imath_T^D)$ in the final section of this step (\textbf{Step 4. Bounds}). Next turn to $A_2$:
\begin{eqnarray*}
&A_2&= \frac{\imath_T^D}{T}  \frac{1}{nT} \left  | \sum_{i=1}^n \sum_{t=1}^T \sum_{t'=1}^T\ddot \epsilon_{it'} \ddot d_{it}(\hat \alpha - \alpha_0) \hat D_{it} \hat D_{it'} \right | \\
& & =  (\hat \alpha - \alpha_0) \frac{\imath_T^D}{T}  \frac{1}{nT} \left  | \sum_{i=1}^n \( \sum_{t=1}^T \ddot \epsilon_{it} \hat D_{it} \) \( \sum_{t=1}^T \ddot d_{it} \hat D_{it} \) \right |\\
& &\leq (\hat \alpha - \alpha_0)  \sqrt{\frac{\imath_T^D}{T}  \frac{1}{nT}   \sum_{i=1}^n \( \sum_{t=1}^T \ddot \epsilon_{it} \hat D_{it} \)^2} \sqrt{ \frac{\imath_T^D}{T}  \frac{1}{nT}  \sum_{i=1}^n \( \sum_{t=1}^T \ddot d_{it} \hat D_{it} \)^2} \\
& &:= (\hat \alpha - \alpha_0)\sqrt{A_{2.1}}\sqrt{A_{2.2}} 
\end{eqnarray*}
In \textbf{Step 4. Bounds} below, we collect bounds showing that $A_{2.1} = O_\Pr(1)$ and $A_{2.2}= O_\Pr(\imath_T^D)$.  Therefore, $A_2 = O_\Pr\(\frac{1}{\sqrt{n \imath_T^D}}\) O_\Pr(1) O_\Pr(\sqrt{\imath_T^D}) = o_\Pr(1).$


Next, we note that
\begin{eqnarray*}
&  &\left |\frac{\imath_T^D}{T} \EnTT  \ddot \epsilon_{it} \ddot \epsilon_{it'}\widehat D_{it}\widehat D_{it'}- \frac{\imath_T^D}{T}  \EnTT   \ddot \epsilon_{it} \ddot \epsilon_{it} \ddot D_{it} \ddot D_{it'} \right | \\
& &\leq \left | \frac{\imath_T^D}{T}  \EnTT
\ddot \epsilon_{it} \ddot \epsilon_{it'} \ddot D_{it} (\ddot D_{it'}-\widehat D_{it'}) +   \ddot \epsilon_{it} \ddot \epsilon_{it'} \ddot D_{it'}(\ddot D_{it}-\hat D_{it}) \right | \\
& &  \ \ \ \ +  \left | \frac{\imath_T^D}{T}  \EnTT
\ddot \epsilon_{it} \ddot \epsilon_{it'} (\widehat D_{it}-\ddot D_{it}) (\hat D_{it'}-\ddot D_{it'}) \right |\\
& & = 2A_3 + A_4
\end{eqnarray*}
Now, we bound $A_3$ as
\begin{eqnarray*}
& A_3&= \frac{\imath_T^D}{T}   \left |\EnTT
\ddot \epsilon_{it} \ddot \epsilon_{it'} \ddot D_{it} (\ddot D_{it'}-\widehat D_{it'}) \right | \\
& &= \frac{\imath_T^D}{T}   \left |\frac{1}{nT} \sum_{i=1}^n \( \sum_{t=1}^T  
\ddot \epsilon_{it}   (\ddot D_{it}-\widehat D_{it}) \)   \( \sum_{t=1}^T  
\ddot \epsilon_{it} \ddot  D_{it} \) \right | \\
&  &\leq \sqrt{ \frac{\imath_T^D}{T}  \frac{1}{nT} \sum_{i=1}^n \( \sum_{t=1}^T  
\ddot \epsilon_{it}   (\ddot D_{it}-\widehat D_{it}) \)^2 }\sqrt{ \frac{\imath_T^D}{T} \frac{1}{nT} \sum_{i=1}^n \( \sum_{t=1}^T  
\ddot \epsilon_{it} \ddot  D_{it} \)^2} \\
& &:= \sqrt{A_{3.1}} \sqrt{A_{3.2}}.\\
\end{eqnarray*}
In \textbf{Step 4. Bounds} below, we collect bounds showing that $A_{3.1} = o_\Pr(1)$ and $A_{3.2} = O_\Pr(1)$.  

Next, consider $A_4$.  Define $\hat D_{it} - \ddot D_{it}=:\delta_{it}$ and apply the bound
\begin{eqnarray*}
& A_4 &= \frac{\imath_T^D}{T} \EnTT
\ddot \epsilon_{it} \ddot \epsilon_{it'} (\widehat D_{it}-\ddot D_{it}) (\hat D_{it'}-\ddot D_{it'}) \\
&  &\leq \frac{\imath_T^D}{T} \frac{1}{n} \sum_{i=1}^n \frac{1}{T} \( \( \sum_{t=1}^T \ddot \epsilon_{it}^2 \)^{1/2} \( \sum_{t=1}^T (\hat D_{it} - \ddot D_{it}^2 \)^{1/2} \)^2 \\
&  &= \imath_T^D \frac{1}{n} \sum_{i=1}^n \( \frac{1}{T}\sum_{t=1}^T \ddot \epsilon_{it}^2 \) \( \frac{1}{T} \sum_{t=1}^T  (\hat D_{it} - \ddot D_{it})^2 \) \\
& &\leq \imath_T^D \(  \max_{i} \frac{1}{T}\sum_{t=1}^T \ddot \epsilon_{it}^2 \) \frac{1}{nT} \sum_{i=1}^n \sum_{t=1}^T  (\hat D_{it} - \ddot D_{it})^2 
\\& &= \imath_T^D \(  \max_{i} \frac{1}{T}\sum_{t=1}^T \ddot \epsilon_{it}^2 \) O_\Pr\(\frac{s \log(p \vee nT)}{n\imath_T}\)\\
& &= \imath_T^D O_\Pr\( n^{2/q}  \) O_\Pr\(\frac{s \log(p \vee nT)}{n\imath_T}\)=o_\Pr(1)
\end{eqnarray*}


\noindent where $\max_{i \leq n} \frac{1}{T}\sum_{t=1}^T \ddot \epsilon_{it}^2 =O_{\Pr}(n^{2/q}  )$ follows from $\Ep\[ \(\frac{1}{T}\sum_{t=1}^T \ddot \epsilon_{it}^2 \)^q\] =O(1)$ which holds by SMIV and the last  $o_\Pr(1)$ bound follows from SMIV.

Finally, $\frac{\imath_T^D}{T} \EnTT \ddot \epsilon_{it} \ddot \epsilon_{it'} \ddot  D_{it} \ddot D_{it'} - \frac{\imath_T^D}{T} \EnTT \Ep  \[\ddot \epsilon_{it} \ddot \epsilon_{it'} \ddot  D_{it} \ddot D_{it'}\] \to_{\mathrm{P}} 0$ is a consequence of SMIV(ii)(b).
We conclude that $\frac{\imath_T^D}{T} \hat \Omega - \frac{\imath_T^D}{T} \Omega \rightarrow_{\Pr} \Omega$. \qed

\noindent \textbf{Step 4. Bounds} 

Here, we consider all terms from Step 4 awaiting bound.  First we have that $A_{3.1}=A_4=o_\Pr(1)$,
where $A_4 = o_\Pr(1)$ was shown above.  Next, $A_{3.2} = \frac{\imath_T^D}{T}\phi_D^2 = O_\Pr(1)$ by SMIV(ii)(c). Bounds for $A_{2.1}$: 
\begin{eqnarray*}
& A_{2.1}& = \frac{\imath_T^D}{T} \frac{1}{nT} \sum_{i=1}^n \( \sum_{t=1}^T \ddot \epsilon_{it} \hat D_{it} \)^2 \\ 
&  & = \frac{\imath_T^D}{T} \frac{1}{nT} \sum_{i=1}^n \( \sum_{t=1}^T \ddot \epsilon_{it} \ddot D_{it} + \ddot \epsilon_{it}(\hat D_{it} - \ddot D_{it}) \)^2 \\
&  & \leq  \frac{\imath_T^D}{T} \frac{1}{nT} \sum_{i=1}^n 2\[ \( \sum_{t=1}^T \ddot \epsilon_{it} \ddot D_{it} \)^2 + \(\sum_{t=1}^T \ddot \epsilon_{it}(\hat D_{it} - \ddot D_{it}) \)^2 \] \\
& &=2A_{3.2} + 2A_{3.1} = O_\Pr(1) + o_\Pr(1).
\end{eqnarray*}

\noindent Bounds for $A_{2.2}$:
\begin{eqnarray*}
& A_{2.2}& = \frac{\imath_T^D}{T} \frac{1}{nT} \sum_{i=1}^n \( \sum_{t=1}^T \ddot d_{it} \hat D_{it} \)^2 \\ 
&  & = \frac{\imath_T^D}{T} \frac{1}{nT} \sum_{i=1}^n \( \sum_{t=1}^T \ddot d_{it} \ddot D_{it} + \ddot d_{it}(\hat D_{it} - \ddot D_{it}) \)^2 \\
&  & \leq  \frac{\imath_T^D}{T} \frac{1}{nT} \sum_{i=1}^n 2\[ \( \sum_{t=1}^T \ddot d_{it} \ddot D_{it} \)^2 + \(\sum_{t=1}^T \ddot d_{it}(\hat D_{it} - \ddot D_{it}) \)^2 \] \\
&  & = O_\Pr(\imath_T^D) + 2  \frac{\imath_T^D}{T} \frac{1}{nT} \sum_{i=1}^n  \(\sum_{t=1}^T \ddot d_{it}(\hat D_{it} - \ddot D_{it}) \)^2  \ \ \ \ \text{by SMIV(ii)(c)} \\
& & \leq O_\Pr(\imath_T^D) +  \imath_T^D \(  \max_{i} \frac{1}{T}\sum_{t=1}^T \ddot d_{it}^2 \) \frac{1}{nT} \sum_{i=1}^n \sum_{t=1}^T  (\hat D_{it} - \ddot D_{it})^2 \text{   (as in term $A_{4}$})\\
& &= O_\Pr(\imath_T^D) + \imath_T^D O_\Pr\( n^{2/q}  \) O_\Pr\(\frac{s \log(p \vee nT)}{n\imath_T}\)=O_\Pr(\imath_T^D)+o_\Pr(1) = O_\Pr(\imath_T^D)\\
\end{eqnarray*}

\noindent Lastly,  $A_{1.1} = A_{2.2}$.  Therefore, we have $A_{1.1} = O_\Pr(\imath_T^D)$.

\subsection{Proof of Theorem 3}
We ignore approximation errors in stating the proof of the theorem for brevity and simplicity.  Incorporating approximation errors satisfying Condition ASM does not substantively affect the argument.

First, begin by defining additional notation. Let
$  \ddot Y =[\ddot y_{11},...,\ddot y_{n1},\ddot y_{12},...,\ddot y_{n2},...,\ddot y_{1T},...,\ddot y_{nT}]'$, 
$ \ddot X= [\ddot x_{11},...,\ddot x_{n1}, \ddot x_{12},...,\ddot x_{n2},...,\ddot x_{1T},...,\ddot x_{nT}]'$.  Similarly, define $\ddot D, \ddot U, \ddot \zeta, \ddot m, $ etc.    For $A \subset \{1,...,p\}$,
let the vector $ \ddot X[A] = \{ \ddot X_j, j \in A\}$,
where $\{\ddot X_j, j=1,...,p\}$ are the columns of $ X$.  Let
$\mathcal{P}_A  =  \ddot X[A](\ddot  X[A]' \ddot X[A])^{-}\ddot  X[A]'$ be the projection operator
sending vectors in $\mathbb{R}^n$ onto ${\rm span}[ \ddot X[A]]$, and let $\mathcal{M}_A = {\rm I}_n - \mathcal{P}_A$ be the
projection onto the subspace that is orthogonal to  ${\rm span}[\ddot X[A]]$.  For a vector $Z \in \mathbb{R}^{nT}$, let
$
\tilde \beta_Z(A) := \arg\min_{\{b \in \mathbb{R}^p : \ b_j = 0, \ \forall j \not \in A\}} \|Z- \ddot Xb\|^2
$
be the coefficient of the linear projection of $Z$ onto ${\rm span}[\ddot  X[A]]$.   If $A = \varnothing$,
interpret $\mathcal{P}_A = 0_n$, and $\tilde \beta_Z = 0_p$.  Finally, denote $\semin{m} = \semin{m}\(\frac{1}{nT} \sum_{i=1}^n \sum_{t=1}^T \ddot x_{it}\ddot x_{it}' \)$ and  $\semax{m} = \semax{m}\(\frac{1}{nT} \sum_{i=1}^n \sum_{t=1}^T \ddot x_{it}\ddot x_{it}'\)$.  

The proof of the theorem proceeds in seven steps.  The first step proves asymptotic normality.  The second through sixth steps provide supporting results for the first step.  The seventh step establishes consistency of the variance estimator.

\textbf{Step 1.}
Write
$
\hat\alpha = \( \ddot D' \MX  \ddot D/nT\)^{-1}\(\ddot D'\MX \ddot Y/nT\)
$
so that
$$
\sqrt{n\iota_T}(\hat\alpha - \alpha_0) = \( \ddot D' \MX  \ddot D/nT\)^{-1}\( \frac{\sqrt{n \iota_T}}{nT} V^{-1/2} \ddot D'\MX (\ddot g + \ddot \zeta) \) =: A_2^{-1} A_1.
$$
By Steps 2 and 3, we conclude that $$A_2 =  \ddot U' \ddot U/nT + o_P(1) $$ $$ A_1 =\frac{\sqrt{n \iota_T}}{nT}V^{-1/2} \ddot U' \ddot \zeta + o_P(1).$$ Next note that because $ \ddot U' \ddot U/nT = \Ep[ \ddot U' \ddot U/nT] + o_P(1)$, and because $\Ep[ \ddot U' \ddot U/nT]$ is bounded away from zero and from above uniformly in $n$ by Condition SMPLM(ii)(a), we have $A_2^{-1} = \Ep[ \ddot U' \ddot U/nT]^{-1} + o_P(1)$.

By reasoning similar to that in the proof of Theorem 2, the desired central limit theorems for $n,T \rightarrow \infty$ jointly and $n\rightarrow \infty, T$ fixed follow.

\textbf{Step 2.}

Here we show that the term $A_1$ as defined in Step 1 satisfies $A_1= \frac{\sqrt{n \iota_T^{u \zeta}}}{nT} V^{-1/2} \ddot U' \ddot \zeta/  + o_P(1)$.

Decompose, using $ \ddot D= \ddot m+ \ddot  U$,
\begin{eqnarray*}
A_1 =  \frac{\sqrt{n \iota_T^{u \zeta}}}{nT} \ddot U' \ddot \zeta+ \underset{=:A_{1.1}}{ \frac{\sqrt{n \iota_T^{u \zeta}}}{nT} \ddot m' \MX \ddot g} + \underset{=:A_{1.2}}{ \frac{\sqrt{n \iota_T^{u \zeta}}}{nT} \ddot m'\MX \ddot \zeta} + \underset{=:A_{1.3}}{ \frac{\sqrt{n \iota_T^{u \zeta}}}{nT} 
\ddot U'\MX \ddot g} - \underset{=:A_{1.4}}{  \frac{\sqrt{n \iota_T^{u \zeta}}}{nT} \ddot U'\PX \ddot \zeta}.
\end{eqnarray*}
 
We proceed by showing that $A_{1.1}$, $A_{1.2}$, $A_{1.3}$, and $A_{1.4}$ are all $o_P(1)$.
First, bound $A_{1.1}$ by 

\begin{align*}
 |A_{1.1}| &=  \frac{\sqrt{n \iota_T^{u \zeta}}}{nT}  |\ddot m'\MX \ddot g| \leq  \sqrt{n \iota_T^{u \zeta}} \|\MX \ddot g/\sqrt{nT}\|\|\MX \ddot m/\sqrt{nT}\|  \\
& = \sqrt{n \iota_T^{u \zeta}} O_\Pr\(\sqrt{ \frac{s \log (p \vee nT)}{n \min \{ \iota_T^{FS},\iota_T^{RF} \} }} \) O_\Pr\(\sqrt{ \frac{s \log (p \vee nT)}{n \iota_T^{FS} } } \) \ \ \ \text{(by Steps 5 and 6 below)} \\
&=o_\Pr(1) \ \ \ \text{(by SMPLM (ii)(d))}
\end{align*}

Second, bound $|A_{1.2}|$.  Using that $\ddot m = \ddot X\beta_{m} ,$ then  $\ddot m'\MX\ddot \zeta = -(\tilde \beta_m(\hat I) - \beta_{m})'\ddot X'\ddot \zeta$ , we have the following bounds, beginning with H\"older's inequality:

\begin{align*}
|A_{1.2}| &=\frac{\sqrt{n \iota_T^{u \zeta}}}{nT}|(\tilde \beta_m(\hat I) - \beta_{m})'\ddot X'\ddot \zeta| \\
 &\leq \frac{\sqrt{n \iota_T^{u \zeta}}}{nT}\|\tilde \beta_m (\hat I) - \beta_{m}\|_1 \|\ddot X'\ddot \zeta\|_{\infty}  \\
& \leq  \frac{\sqrt{n \iota_T^{u \zeta}}}{nT} \|\tilde \beta_m (\hat I) - \beta_{m0}\|_1 O_\Pr \( \sqrt{ \log p } \max_{j \leq p} \sqrt{\frac{T}{\iota_T^{x_j \zeta}}} \) \\
& =\frac{\sqrt{n \iota_T^{u \zeta}}}{nT} O_\Pr\( \sqrt{ \frac{s \log(p \vee nT)}{n\iota_T^{FS}}}\)
O_\Pr \( \sqrt{ \log p } \max_{j \leq p}\sqrt{\frac{T}{\iota_T^{x_j \zeta}}} \) \\
&=o_\Pr(1)  \ \ \ \text{(by SMPLM (ii)(d))}
\end{align*}

Third, bound $|A_{1.3}|$. Using similar reasoning as for $A_{2.1}$, conclude
$$|A_{1.3}| = O_\Pr \( \sqrt{\frac{\iota_T^{u\zeta}}{T}}\sqrt{ \frac{s \log (p\vee nT)}{n \min \{  \iota_T^{FS}, \iota_T^{RF} \} }} \sqrt{\log p} \max_{j \leq p} \sqrt{ \frac{T}{\iota_T^{x_j u}}} \) = o_P(1). $$

Fourth, we have
\begin{align*}
|A_{1.4}| &\leq \frac{\sqrt{n \iota_T^{u \zeta}}}{nT} |\tilde \beta_{\ddot U}(\hat I)' \ddot X' \ddot  \zeta| \\ 
&\leq  \frac{\sqrt{n \iota_T^{u \zeta}}}{nT}  \| \tilde \beta_{\ddot U}(\hat I)\|_1 \|\ddot  X' \ddot \zeta\|_{\infty}  \\  
&=  \frac{\sqrt{n \iota_T^{u \zeta}}}{nT}  \|\tilde \beta_{\ddot U}(\hat I) \|_1 O_\Pr \(\sqrt{nT}   \sqrt{ \log p } \max_{j \leq p}\sqrt{\frac{T}{\iota_T^{x_j \zeta}}} \) \ \ \ \text{(by Step 4 below)} \\
& \leq \frac{\sqrt{n \iota_T^{u \zeta}}}{nT}  \sqrt {\hat s} \|\tilde \beta_{\ddot U}(\hat I) \|_2 O_\Pr \( \sqrt{nT}  \sqrt{ \log p } \max_{j \leq p}\sqrt{\frac{T}{\iota_T^{x_j \zeta}}} \) \\
& \leq \frac{\sqrt{n \iota_T^{u \zeta}}}{nT}  \sqrt{\hat s} \| (\ddot  X[\hat I]'\ddot  X[\hat I]/nT)^{-1} \ddot  X[\hat I]'\ddot  U/nT\| O_\Pr \( \sqrt{nT}  \sqrt{ \log p } \max_{j \leq p}\sqrt{\frac{T}{\iota_T^{x_j \zeta}}} \) \\
\end{align*}
\begin{align*}
& \leq   \frac{\sqrt{n \iota_T^{u \zeta}}}{nT}   \sqrt{\hat s} \phi_{\min}^{-1}(\hat s) \sqrt{\hat s}  \| \ddot  X'\ddot U/\sqrt{nT}\|_{\infty}/\sqrt{nT} O_\Pr \( \sqrt{nT}  \sqrt{ \log p } \max_{j \leq p}\sqrt{\frac{T}{\iota_T^{x_j \zeta}}} \)\\
& = \frac{\sqrt{n \iota_T^{u \zeta}}}{nT}  O_\Pr(s) O_\Pr\(\sqrt{ \log p} \max_{j \leq p} \sqrt{\frac{T}{\iota_T^{x_j u}}} \)/\sqrt{nT} O_\Pr \( \sqrt{nT} \sqrt{ \log p } \max_{j \leq p}\sqrt{\frac{T}{\iota_T^{x_j \zeta}}} \)\\
& = o_\Pr(1) \ \ \ \text{(by SMPLM)}
\end{align*}

\textbf{Step 3} The term $A_2$ satisfies $A_2 =\ddot  U' \ddot U/nT + o_P(1)$.

Decompose
\begin{align*}
A_2 &= \frac{\sqrt{n \iota_T^{u \zeta}}}{nT}  (\ddot m+\ddot U)'\MX (\ddot m+\ddot U)/nT \\
&= \frac{\sqrt{n \iota_T^{u \zeta}}}{nT}  \ddot U'\ddot U/nT + \underset{=:A_{2.1}}{\frac{\sqrt{n \iota_T^{u \zeta}}}{nT}  \ddot m' \MX \ddot m/nT} +
\underset{=:A_{2.2}}{ \frac{\sqrt{n \iota_T^{u \zeta}}}{nT}  2\ddot m'\MX \ddot U/nT } - \underset{=:A_{2.3}}{\frac{\sqrt{n \iota_T^{u \zeta}}}{nT}  \ddot U'\PX \ddot U/nT}.
\end{align*}
and use similar reasoning as in Step 2 to conclude that $A_{2.1}$, $A_{2.2}$, $A_{2.3} = o_\Pr(1)$.

\textbf{Step 4.}
 The following bounds on $\|\ddot X' \ddot \zeta\|_{\infty}$ and $\|\ddot X' \ddot U\|_{\infty}$ hold:  

$$\|\ddot X' \ddot \zeta/\sqrt{nT}\|_{\infty} = O_\Pr \(\sqrt {\log p}\max_{j \leq p}\sqrt{\frac{T}{\iota_T^{x_j\zeta}}} \), \ \ \ \|\ddot X' \ddot U/\sqrt{nT}\|_{\infty} = O_\Pr \(\sqrt {\log p}\max_{j \leq p}\sqrt{\frac{T}{\iota_T^{x_j u}}} \).$$

The argument follows the same logic as was required to bound $\max_{j\leq p} \left | \sum_{i=1}^n \sum_{t=1}^T \ddot x_{itj} \ddot \zeta_{it} \right |$ in Theorem 2.  For details, see Step 3 in the proof of Theorem 2.

\textbf{Step 5.} The following bounds hold:
$$
\text{(a)}   \  \| \MX \ddot m/\sqrt{nT}\|  =O_\Pr \left(\sqrt{ s \log (p\vee nT)/n\iota_T^{FS}} \right)$$
$$  \text{(b)} \  \| \tilde \beta_{\ddot m}(\hat I) - \beta_{m}\|  =O_\Pr \left(\sqrt{ s \log (p\vee nT)/n\iota_T^{FS}} \right).
$$

To show this note that $\| \MX \ddot  m /\sqrt{nT} \| \leq \| \mathcal M_{\hat I_{FS}} \ddot m / \sqrt{nT} \| \ \ \text{by $\hat I_{FS} \subset \hat I$.  Next, we have }$
$\| \mathcal M_{\hat I_{FS}} \ddot m / \sqrt{nT} \| = O_\Pr\(\sqrt{ \frac{s \log(p \vee nT)}{n \iota_T^{FS}}}\) \ \ \ \text{by Theorem 1}.$

Turning to claim (b), note that 
$$\sqrt{\phi_{\min} (\hat s + s)} \| \tilde \beta_{\ddot  m}(\hat I) - \beta_m \| \leq  \|\ddot X(\tilde \beta_{\ddot m}(\hat I) - \beta_m)/\sqrt{nT} \| = O_\Pr\(\sqrt{ \frac{s \log(p \vee nT)}{n \iota_T^{FS}}}\).$$
Then by $\hat s = O_\Pr(s)$ by Theorem 1 and $1/\phi_{\min}(\hat  s + s) = O_\Pr(1)$ by Condition SE, conclude claim (b) above.

\textbf{Step 6.} The following bound on $\|\MX \ddot  g\|$ and related quantities hold:
$$
\text{(a)}   \  \|\MX \ddot  g/\sqrt{nT} \| = O_\Pr \(\sqrt{\frac{s \log (p \vee nT)}{n \min\{\iota^{RF}_T , \iota^{FS}_T \}}}\)$$
$$ \text{(b)} \  \| \tilde \beta_{\ddot g}(\hat I) - \beta_{g}\| = O_\Pr \(\sqrt{\frac{s \log (p \vee nT)}{n \min\{\iota^{RF}_T , \iota^{FS}_T \}}}\).
$$

To show claim (a), observe that because $\hat I_{RF} \subset \hat I$
\begin{align*}
\|\MX (\alpha \ddot  m +\ddot   g)/\sqrt{nT} \| & \leq \|\mathcal M_{\hat I_{RF}} (\alpha \ddot  m + \ddot  g)/\sqrt{nT} \|\\
&=O_\Pr \(\sqrt{\frac{s \log (p \vee nT)}{n \iota^{RF}_T}}\)
\end{align*}
using Theorem 1.
Since $|\alpha|$ is uniformly bounded, it follows from the line above, Step 5, and the triangle inequality that  
$$\|\MX \ddot  g/\sqrt{nT} \| = O_\Pr \(\sqrt{\frac{s \log (p \vee nT)}{n \min\{\iota^{RF}_T , \iota^{FS}_T \}}}\)$$

Claim (b) follows similarly to Claim (b) in Step 5.

\textbf{Step 7.} This step establishes
consistency of the variance estimator
of Theorem 3.

 Recall that
$$\hat \Omega :=\EnTT \hat \zeta_{it} \hat \zeta_{it'} \hat u_{it} \hat u_{it'}\ \ \ \text{ and} \ \ \ \Omega :=\EnTT \Ep[\ddot \zeta_{it} \ddot \zeta_{it'} \ddot u_{it} \ddot u_{it'}].$$   For this step, it is sufficient to show $\frac{\imath_T^{u\zeta}}{T}\hat \Omega - \frac{\imath_T^{u\zeta}}{T} \Omega \to_{\mathrm{P}} 0.$

Through appropriate use of the Cauchy-Schwarz and triangle inequalities, we note that
 it is sufficient to show the three following statements:
\begin{align*}
& \left | \frac{\imath_T^{u\zeta}}{T}\EnTT \ddot \zeta_{it} \ddot \zeta_{it'} \hat u_{it} \hat u_{it'}  - \frac{\imath_T^{u\zeta}}{T}\EnTT \hat \zeta_{it} \hat  \zeta_{it'} \hat  u_{it} \hat u_{it'} \right | =o_\Pr(1), \\
&\left | \frac{\imath_T^{u\zeta}}{T}\EnTT \ddot \zeta_{it} \ddot \zeta_{it'} \ddot u_{it} \ddot u_{it'} - \frac{\imath_T^{u\zeta}}{T}\EnTT \ddot \zeta_{it} \ddot \zeta_{it'} \hat u_{it} \hat u_{it'} \right | = o_\Pr(1), \\
&\left | \frac{\imath_T^{u\zeta}}{T}\EnTT \ddot \zeta_{it} \ddot \zeta_{it'} \ddot u_{it} \ddot u_{it'} - \frac{\imath_T^{u\zeta}}{T}\EnTT \Ep \[ \ddot \zeta_{it} \ddot \zeta_{it'} \ddot u_{it} \ddot u_{it'} \] \right | = o_\Pr(1). \\
\end{align*}

Given the above short outline, we bound each term separately starting with the first.  In this exercise, it will be convenient to let $\delta_{it}^u = \hat u_{it}- \ddot u_{it} = \ddot x_{it}'\beta_m - \ddot x_{it}'\hat \beta_m$ and $\delta_{it}^g = \ddot x_{it}'\beta_g - \ddot x_{it}' \hat \beta_g$.

We begin by noting that by making the substitution $\hat \zeta_{it} = \ddot \zeta_{it} + \ddot d_{it}(\alpha - \hat \alpha) + \ddot x_{it}'(\beta_g - \hat\beta_g) = \ddot \zeta_{it} + (\alpha - \hat\alpha) \ddot d_{it} + \delta_{it}^g$ and expanding, we have
\begin{eqnarray*}
&  & \hspace{-4mm} \left | \frac{\imath_T^{u\zeta}}{T} \EnTT  (\ddot \zeta_{it} \ddot \zeta_{it'} - \hat \zeta_{it} \hat \zeta_{it'}) \hat u_{it} \hat u_{it'} \right |   \\
& & \leq  \frac{\imath_T^{u\zeta}}{T} \frac{1}{nT} \left  | \sum_{i=1}^n \sum_{t=1}^T \sum_{t'=1}^T \ddot d_{it} \ddot d_{it'}(\hat \alpha - \alpha)^2 \hat u_{it} \hat u_{it'} \right |\\
& & \ \ \ \ +  \frac{\imath_T^{u\zeta}}{T}  \frac{1}{nT} \left  | \sum_{i=1}^n \sum_{t=1}^T \sum_{t'=1}^T\ddot \delta_{it'}^g \delta_{it}^g \hat u_{it} \hat u_{it'} \right |\\
& & \ \ \ \ +2  \frac{\imath_T^{u\zeta}}{T}  \frac{1}{nT} \left  | \sum_{i=1}^n \sum_{t=1}^T \sum_{t'=1}^T\delta_{it'}^g \ddot d_{it}(\hat \alpha - \alpha) \hat u_{it} \hat u_{it'} \right |\\
& & \ \ \ \ +2  \frac{\imath_T^{u\zeta}}{T}  \frac{1}{nT} \left  | \sum_{i=1}^n \sum_{t=1}^T \sum_{t'=1}^T\ddot \zeta_{it'} \delta_{it}^g \hat u_{it} \hat u_{it'} \right |\\
& & \ \ \ \ +2  \frac{\imath_T^{u\zeta}}{T}  \frac{1}{nT} \left  | \sum_{i=1}^n \sum_{t=1}^T \sum_{t'=1}^T\ddot \zeta_{it'} \ddot d_{it}(\hat \alpha - \alpha) \hat u_{it} \hat u_{it'} \right |\\
& & :=B_1 + B_2 + 2B_3 + 2B_4 + 2B_5\\
\end{eqnarray*}
We bound the terms $B_1,...,B_5$ separately.  Beginning with $B_1$, note that:
\begin{eqnarray*}
&B_1&= \frac{\imath_T^{u\zeta} }{T} \frac{1}{nT} \left  | \sum_{i=1}^n \sum_{t=1}^T \sum_{t'=1}^T \ddot d_{it} \ddot d_{it'}(\hat \alpha - \alpha)^2 \hat u_{it} \hat u_{it'} \right | \\
& &=  (\hat \alpha - \alpha)^2 \frac{\imath_T^{u\zeta}}{T}\frac{1}{nT} \left  | \sum_{i=1}^n \sum_{t=1}^T \sum_{t'=1}^T \ddot d_{it} \ddot d_{it'} \hat u_{it} \hat u_{it'} \right | \\
& &=:(\hat \alpha - \alpha)^2 B_{1.1} = O_\Pr(1/(n\imath_T^{u\zeta}))O_\Pr(\imath_T^{u\zeta}) = o_\Pr(1)
\end{eqnarray*}
where we use $B_{1.1}:= \frac{\imath_T^{u\zeta}}{T} \frac{1}{nT} \sum_{i} \sum_{t} \sum_{t'} \ddot d_{it}\ddot d_{it'} \hat u_{it} \hat u_{it'} = O_P(\imath_T^{u\zeta})$ which is shown in \textbf{Step 7. Bounds}.

Next consider $B_2$.  
\begin{eqnarray*}
&B_2&=  \frac{\imath_T^{u\zeta}}{T}  \frac{1}{nT}  \sum_{i=1}^n \sum_{t=1}^T \sum_{t'=1}^T\delta_{it}^g \delta_{it'}^g \hat u_{it} \hat u_{it'} \\
& &\leq 2 \frac{\imath_T^{u\zeta}}{T} \EnTT
\delta_{it}^g \delta_{it'}^g\delta_{it}^u\delta_{it'}^u +  2\frac{\imath_T^{u\zeta}}{T}  \frac{1}{nT} \sum_{i=1}^n \sum_{t=1}^T \sum_{t'=1}^T\delta_{it}^g \delta_{it'}^g\ddot u_{it}  \ddot u_{it'} \\
&  &\leq 2\frac{\imath_T^{u\zeta}}{T} \frac{1}{n} \sum_{i=1}^n \frac{1}{T} \( \sum_{t=1}^T (\delta_{it}^{u})^2 \)\( \sum_{t=1}^T (\delta_{it}^{g})^2 \)  +2 \frac{\imath_T^{u\zeta}}{T} \frac{1}{n} \sum_{i=1}^n \frac{1}{T}  \( \sum_{t=1}^T \ddot u_{it}^2 \)\( \sum_{t=1}^T (\delta_{it}^{g})^2 \) \\
& &\leq 2\imath_T^{u\zeta} \(  \max_{i} \frac{1}{T}\sum_{t=1}^T(\delta_{it}^{u})^2 + \max_{i} \frac{1}{T}\sum_{t=1}^T \ddot u_{it}^2\) \frac{1}{nT} \sum_{i=1}^n \sum_{t=1}^T  (\delta_{it}^{g})^2 \\
& &=:2 \imath_T^{u\zeta} (B_{2.1} + B_{2.3}) B_{2.2} \\
& &= o_\Pr(1) \ \ \ \text{(by SMPLM (ii)(d), where $B_{2.1}, B_{2.2}, B_{2.3}$ are handled in } \textbf{Step 7. Bounds}).
\end{eqnarray*}

Next, we consider $B_3$.
\begin{align*}
B_3 &= \frac{\imath_T^{u\zeta}}{T}  \frac{1}{nT} \left | \sum_{i=1}^n \left(\sum_{t=1}^T (\alpha - \hat \alpha)\ddot d_{it}\hat u_{it}\right) \left( \sum_{t=1}^T \delta_{it}^g \hat u_{it} \right) \right | \\
& \leq |\hat\alpha - \alpha| \left | \left( \frac{\imath_T^{u\zeta}}{T}  \frac{1}{nT} \sum_{i=1}^n \left(\sum_{t=1}^T \ddot d_{it}\hat u_{it}\right)^2   \right)^{1/2} \right | \left | \left( \frac{\imath_T^{u\zeta}}{T}  \frac{1}{nT} \sum_{i=1}^n \left( \sum_{t=1}^T \delta_{it}^g \hat u_{it} \right)^2  \right)^{1/2} \right | \\
&= |\hat\alpha - \alpha| (B_{1.1})^{1/2} (B_2)^{1/2} = O_\Pr((n\imath_T^{u\zeta})^{-1/2})O_\Pr((\imath_T^{u\zeta})^{1/2})o_\Pr(1) = o_\Pr(1).
\end{align*}

We then bound $B_4$.
\begin{align*}
B_4 &= \frac{\imath_T^{u\zeta}}{T} \frac{1}{nT} \left | \sum_{i=1}^n \left( \sum_{t=1}^T \ddot \zeta_{it} \hat u_{it} \right) \left( \sum_{t=1}^T \delta_{it}^g \hat u_{it} \right) \right | \\
& \leq \left | \left( \frac{\imath_T^{u\zeta}}{T}  \frac{1}{nT} \sum_{i=1}^n \left(\sum_{t=1}^T \ddot \zeta_{it}\hat u_{it}\right)^2   \right)^{1/2} \right | \left | \left( \frac{\imath_T^{u\zeta}}{T}  \frac{1}{nT} \sum_{i=1}^n \left( \sum_{t=1}^T \delta_{it}^g \hat u_{it} \right)^2  \right)^{1/2} \right | \\
&= (B_{4.1})^{1/2} (B_2)^{1/2} = O_\Pr(1) o_\Pr(1)
\end{align*}
where we show $B_{4.1} = O_\Pr(1)$ in \textbf{Step 7. Bounds}.


Finally, we bound $B_5$.

\begin{eqnarray*}
&B_5&= \frac{\imath_T^{u\zeta}}{T}  \frac{1}{nT} \left  | \sum_{i=1}^n \sum_{t=1}^T \sum_{t'=1}^T\ddot \zeta_{it'} \ddot d_{it}(\hat \alpha - \alpha) \hat u_{it} \hat u_{it'} \right | \\
& & \leq  |\hat \alpha - \alpha| \frac{\imath_T^{u\zeta}}{T}  \frac{1}{nT} \left  | \sum_{i=1}^n \( \sum_{t=1}^T \ddot \zeta_{it} \hat u_{it} \) \( \sum_{t=1}^T \ddot d_{it} \hat u_{it} \) \right |\\
& &\leq |\hat \alpha - \alpha| \sqrt{\frac{\imath_T^{u\zeta}}{T}  \frac{1}{nT}   \sum_{i=1}^n \( \sum_{t=1}^T \ddot \zeta_{it} \hat u_{it} \)^2} \sqrt{ \frac{\imath_T^{u\zeta}}{T}  \frac{1}{nT}  \sum_{i=1}^n \( \sum_{t=1}^T \ddot d_{it} \hat u_{it} \)^2} \\
& &:= |\hat \alpha - \alpha|\sqrt{B_{4.1}}\sqrt{B_{1.1}} \\
& &=O_\Pr((n\imath_T^{u\zeta})^{-1/2})O_\Pr(1)O_\Pr((\imath_T^{u\zeta})^{1/2}) = o_\Pr(1).
\end{eqnarray*}


Now note that
\begin{eqnarray*}
&  &\left |\frac{\imath_T^{u\zeta}}{T} \EnTT  \ddot \zeta_{it} \ddot \zeta_{it'}\widehat u_{it}\widehat u_{it'}- \frac{\imath_T^{u\zeta}}{T}  \EnTT   \ddot \zeta_{it} \ddot \zeta_{it'} \ddot u_{it} \ddot u_{it'} \right | \\
& &\leq 2\left | \frac{\imath_T^{u\zeta}}{T} \frac{1}{nT} \sum_{i=1}^n \( \sum_{t=1}^T \ddot \zeta_{it} \ddot u_{it} \) \( \sum_{t=1}^T \ddot \zeta_{it} \delta_{it}^u \) \right | + \left | \frac{\imath_T^{u\zeta}}{T} \frac{1}{nT}  \sum_{i=1}^n \( \sum_{t=1}^T \ddot \zeta_{it} \delta_{it}^u \)^2 \right |\\
& & =: 2C_1 + C_2
\end{eqnarray*}

Looking at term $C_2$, we have that 
\begin{eqnarray*}
& C_2&= \frac{\imath_T^{u\zeta}}{T} \frac{1}{nT}  \sum_{i=1}^n \( \sum_{t=1}^T \ddot \zeta_{it} \delta_{it}^u \)^2 \\
&  &\leq  \imath_T^{u\zeta} \frac{1}{n} \sum_{i=1}^n \( \frac{1}{T}\sum_{t=1}^T \ddot \zeta_{it}^2 \) \( \frac{1}{T} \sum_{t=1}^T  (\delta_{it}^u)^2 \) \\
& &\leq \imath_T^{u\zeta} \(  \max_{i} \frac{1}{T}\sum_{t=1}^T \ddot \zeta_{it}^2 \) \frac{1}{nT} \sum_{i=1}^n \sum_{t=1}^T  (\delta_{it}^u)^2 \\
& &=\imath_T^{u\zeta}O_\Pr(n^{2/q}) C_{2.1}
\end{eqnarray*}
follows using SMPLM (ii)(a).  We show in \textbf{Step 7. Bounds} that $C_{2.1} = O_\Pr\( \frac{s \log(p \vee nT)}{n \imath_T^{FS}}\)$, and $C_2 = o_p(1)$ then follows under SMPLM (ii)(d).

Turning to $C_1$, we have
\begin{eqnarray*}
& C_1  &\leq \sqrt{ \frac{\imath_T^{u\zeta}}{T} \frac{1}{nT} \sum_{i=1}^n \( \sum_{t=1}^T  
\ddot \zeta_{it} \ddot  u_{it} \)^2}\sqrt{ \frac{\imath_T^{u\zeta}}{T}  \frac{1}{nT} \sum_{i=1}^n \( \sum_{t=1}^T  
\ddot \zeta_{it}  \delta_{it}^u \)^2 } \\
& &=: \sqrt{C_{1.1}}\sqrt{C_{2}} = O_\Pr(1) o_\Pr(1) \\
\end{eqnarray*}
\noindent using the bound for $C_2$ given immediately above and that 
$$ C_{1.1} = \frac{\imath_T^{u\zeta} }{T} \frac{1}{nT} \sum_{i=1}^n \( \sum_{t=1}^T \ddot \zeta_{it}  \ddot u_{it} \)^2 =  \frac{\imath_T^u}{T} \phi_{u\zeta} ^2 = O_\Pr(1)
$$
from SMPLM(ii)(c).



Finally, $\frac{\imath_T^{u\zeta}}{T} \EnTT \ddot \zeta_{it} \ddot \zeta_{it'} \ddot  u_{it} \ddot u_{it'} - \frac{\imath_T^{u\zeta}}{T} \EnTT \Ep  \[\ddot \zeta_{it} \ddot \zeta_{it'} \ddot  u_{it} \ddot u_{it'}\] \to_{\mathrm{P}} 0$ is a consequence of SMPLM(ii)(c).

We conclude that $\frac{\imath_T^{u\zeta}}{T} \hat \Omega - \frac{\imath_T^{u\zeta}}{T} \Omega \rightarrow_{\Pr} \Omega$. \qed

\noindent \textbf{Step 7. Bounds} 

Here, we provide bounds for the terms $B_{1.1}$, $B_{2.1}$, $B_{2.2}$, $B_{2.3}$, $B_{4.1}$ and $C_{2.1}$ from the above argument.  

We start with 
\begin{align*}
C_{2.1} = \frac{1}{nT} \sum_{i=1}^n \sum_{t=1}^T (\delta_{it}^u)^2 &= (\ddot D - \ddot X\hat\beta_m - \ddot U)'(\ddot D - \ddot X\hat\beta_m - \ddot U) \\
&= \frac{1}{nT}(\ddot X\beta_m - \ddot X\hat\beta_m)'(\ddot X\beta_m - \ddot X\hat\beta_m) \\
&= \frac{1}{nT}(\ddot X\beta_m - \ddot X_{\hat I}(\ddot X_{\hat I}'\ddot X_{\hat I})^{-1}\ddot X_{\hat I}'(\ddot X\beta_m) 
- \ddot X_{\hat I}(\ddot X_{\hat I}'\ddot X_{\hat I})^{-1}\ddot X_{\hat I}'\ddot U)' \times \\
& \qquad (\ddot X\beta_m - \ddot X_{\hat I}(\ddot X_{\hat I}'\ddot X_{\hat I})^{-1}\ddot X_{\hat I}'(\ddot X\beta_m) 
- \ddot X_{\hat I}(\ddot X_{\hat I}'\ddot X_{\hat I})^{-1}\ddot X_{\hat I}'\ddot U) \\
&= \frac{1}{nT}(\mathcal{M}_{\hat I} \ddot m - \mathcal{P}_{\hat I} \ddot U)'(\mathcal{M}_{\hat I} \ddot m - \mathcal{P}_{\hat I} \ddot U) \\
&= \frac{1}{nT} \ddot m'\mathcal{M}_{\hat I} \ddot m + \frac{1}{nT} \ddot U' \mathcal{P}_{\hat I} \ddot U.
\end{align*}
From Step 5, we have that $\frac{1}{nT} \ddot m'\mathcal{M}_{\hat I} \ddot m = O_\Pr\left(\frac{s \log(p \vee nT)}{n \imath_T^{FS}}\right)$. Using the same argument as used in Step 2 to bound term $A_{1.4}$, we can also show that $\frac{1}{nT} \ddot U' \mathcal{P}_{\hat I} \ddot U = O_\Pr\left(\frac{s \log(p)}{n \imath_T^{FS}}\right)$.  It follows that $C_{2.1} = O_\Pr\left(\frac{s \log(p \vee nT)}{n \imath_T^{FS}}\right)$.

Next, we provide bounds for $B_{1.1}$ and $B_{4.1}$.
\begin{align*}
B_{4.1} &= \frac{\imath_T^{u\zeta}}{T}\frac{1}{nT} \sum_{i=1}^{n}\left(\sum_{t=1}^{T} \ddot \zeta_{it} \hat u_{it}\right)^2 \\
& \leq 2 \frac{\imath_T^{u\zeta}}{T}\frac{1}{nT} \sum_{i=1}^{n} \left(\sum_{t=1}^{T} \ddot \zeta_{it} \delta_{it}^u\right)^2 + 2 \frac{\imath_T^{u\zeta}}{T}\frac{1}{nT} \sum_{i=1}^{n} \left(\sum_{t=1}^{T} \ddot \zeta_{it} \ddot u_{it}\right)^2 \\
&= 2C_2 + 2\frac{\imath_T^{u\zeta}}{T} \phi_{u\zeta} ^2 = o_\Pr(1) + O_\Pr(1) = O_\Pr(1).
\end{align*} 
$B_{1.1} = \frac{\imath_T^{u\zeta}}{T}\frac{1}{nT} \sum_{i=1}^{n}(\sum_{t=1}^{T} \ddot d_{it} \hat u_{it})^2$ has a similar structure to $B_{4.1}$, and following the argument for $B_{4.1}$ and $C_2$ yields
\begin{align*}
B_{1.1} &\leq 2 \imath_T^{u\zeta} \(  \max_{i} \frac{1}{T}\sum_{t=1}^T \ddot d_{it}^2 \) \frac{1}{nT} \sum_{i=1}^n \sum_{t=1}^T  (\delta_{it}^u)^2      + 2\frac{\imath_T^{u\zeta}}{T} \phi_{ud} ^2  \\
&= \imath_T^{u\zeta}O_\Pr(n^{2/q}) O_\Pr\( \frac{s \log(p \vee nT)}{n \imath_T^{FS}}\) + \imath_T^{u\zeta} O_\Pr(1) = O_\Pr(\imath_T^{u\zeta}).
\end{align*}

Finally, we turn to the three terms in the bound for term $B_2$.  We begin by noting that $B_{2.3} = O_\Pr\(n^{2/q}\)$ by $\Ep \[ \(\frac{1}{T} \sum_{t=1}^T \ddot u_{it}^2 \)^q \] = O(1)$.  Turning next to $B_{2.1}$, we have
\begin{align*}
B_{2.1} & = \max_i \frac{1}{T}\sum_{t=1}^T (\delta_{it}^{u})^2 \\
& = \max_i \frac{1}{T}\sum_{t=1}^T \sum_{j \in \hat I \cup \textrm{supp}(\beta_m)} (\ddot x_{itj} (\hat \beta_{m,j} - \beta_{m,j}))^2 \\
 &\leq \|\hat \beta_m - \beta_m\|_2^2 (\hat s + s) \max_{i,t,j} \ddot x_{itj}^2 \\
 &= \max_{i,t,j} \ddot x_{itj}^2 O_\Pr(s) \|\hat \beta_m - \beta_m\|_2^2.
\end{align*}
That $B_{2.1} = \max_{i,t,j} \ddot x_{itj}^2 O_\Pr\left(\frac{s^2 \log(p \vee nT)}{n \imath_T^{FS}}\right)$ follows from
\begin{align*}
\|\hat \beta_m - \beta_m\|_2^2 &= \|(\ddot X_{\hat I}'\ddot X_{\hat I})^{-1}\ddot X_{\hat I}'(\ddot X\beta_m + \ddot U) - \beta_m\|^2  \\
&\leq 2\|(\ddot X_{\hat I}'\ddot X_{\hat I})^{-1}\ddot X_{\hat I}'(\ddot X\beta_m) - \beta_m\|^2 + 2\|(\ddot X_{\hat I}'\ddot X_{\hat I})^{-1}\ddot X_{\hat I}'\ddot U\|^2 \\
&= O_\Pr\left(\frac{s \log(p \vee nT)}{n \imath_T^{FS}}\right) + 2\|(\ddot X_{\hat I}'\ddot X_{\hat I})^{-1}\ddot X_{\hat I}'\ddot U\|^2 \qquad (\textrm{follows from Step 5}) \\
&\leq O_\Pr\left(\frac{s \log(p \vee nT)}{n \imath_T^{FS}}\right) + 2(\sqrt{\hat s} \phi_{\min}^{-1}(\hat s) \|\ddot X' \ddot U/\sqrt{nT}\|_{\infty}/\sqrt{nT})^2 \\
& \qquad \qquad (\textrm{using the same argument as in bounding} \ A_{1.4}) \\
&= O_\Pr\left(\frac{s \log(p \vee nT)}{n \imath_T^{FS}}\right) + O_\Pr\left(\frac{s \log(p)}{n \imath_T^{FS}}\right).
\end{align*}

For the final term, $B_{2.2}$, we have that 
\begin{eqnarray*}
&B_{2.2} &= \frac{1}{nT}\sum_{i=1}^n \sum_{t=1}^T \(\ddot x_{it}'\beta_g - \ddot x_{it}'\hat \beta_g \)^2 \\
& &= \frac{1}{nT}\|\ddot X\beta_g - \ddot X\hat \beta_g \|^2 = \frac{1}{nT}\|\ddot X\beta_g - \mathcal P_{\hat I} (\ddot y - \hat \alpha \ddot D) \|^2 \\
& &=\frac{1}{nT}\| \ddot g - \mathcal P_{\hat I}(\ddot y - \alpha \ddot D - \ddot \zeta) + \mathcal (\alpha - \hat \alpha)P_{\hat I}\ddot D + \mathcal P_{\hat I} \ddot \zeta\|^2\\
& & \leq 3\frac{1}{nT}\| \mathcal M_{\hat I} \ddot g\|^2 + 3\frac{1}{nT}(\alpha - \hat \alpha)^2\|\mathcal P_{\hat I} \ddot D\|^2 + 3 \frac{1}{nT} \| \mathcal P_{\hat I} \ddot \zeta\|^2.
\end{eqnarray*}
From Step 6, we have that 
$$
\frac{1}{nT}\| \mathcal M_{\hat I} \ddot g\|^2 = O_\Pr\left( \frac{s \log (p \vee nT)}{n \min \{ \imath_T^{RF}, \imath_T^{FS}\}}\right),
$$
and we can show that 
$$
\frac{1}{nT} \| \mathcal P_{\hat I} \ddot \zeta\|^2 = O_\Pr\left(\frac{s \log p}{n \min_j \{\imath_T(x_j \zeta)\}}\right)
$$
using the same argument as for bounding $A_{1.4}$ in Step 2.


Finally, we have that 
\begin{align*}
\frac{1}{nT}\|\mathcal P_{\hat I} \ddot D\|^2 \leq \frac{1}{nT} \| \ddot D \|^2 & = \frac{1}{nT} \sum_{i=1}^{n}\sum_{t=1}^{T}\ddot d_{it}^2 = O_\Pr(1)
\end{align*}
where the inequality follows from $\mathcal P_{\hat I} \ddot D$ being a projection and the result then follows under SMPLM(ii)(a).
Putting these terms together yields $B_{2.2} = O_\Pr\left( \frac{s \log (p \vee nT)}{n \min \{ \imath_T^{RF}, \imath_T^{FS}\}}\right) +  O_\Pr\left(\frac{s \log p}{n \min_j \{\imath_T^{x_j \zeta}\}}\right) +  O_\Pr((n\imath_T^{u\zeta})^{-1})$.

\clearpage
\bibliographystyle{econometrica}
\bibliography{mybibVOLUME}

\clearpage
\setlength{\tabcolsep}{10pt}
\begin{table}[h!]
\footnotesize{\caption{\footnotesize{Estimates of the Effect of Gun Prevalence on homicide Rates}}\label{CLTableResults}
\begin{center}
\begin{tabular}{lccc}
  & Overall  & Gun & non-Gun  \\
\cline{2-4}
Cook and Ludwig (2006) Baseline & 0.086 (0.038)  & 0.173 (0.049)  & -0.033 (0.040) \\
FSS + Census Baseline           & 0.070 (0.035)  & 0.178 (0.046)  & -0.071 (0.038) \\
Full Set of Controls            & -0.010 (0.033) & 0.000 (0.044)  & -0.033 (0.042) \\
Cluster Post-Double Selection   & 0.079 (0.043)  & 0.171 (0.047)  & -0.019 (0.040) \\
\hline
\end{tabular}
\end{center}}
\begin{flushleft}
\footnotesize{This table presents estimates of the effect of gun ownership on homicide rates for a panel of 195 US Counties over the years 1980-1999.  The columns ``Overall'', ``Gun'', and ``non-Gun'' respectively report the estimated effect of gun prevalence on the log of the overall homicide rate, the log of the gun homicide rate, and the log of the non-gun homicide rate.  Each row corresponds to a different specficiation as described in the text.  In each specification, the outcome corresponding to the column label is regressed on lagged log(FSS) (a proxy for gun ownership) and additional covariates as described in the text.  Each specification includes a full set of year and county fixed effects.  Standard errors clustered by county are provided in parentheses.}
\end{flushleft}
\end{table}

\clearpage

\setlength{\tabcolsep}{10pt}
\begin{table}[h!]
\footnotesize{\caption{\footnotesize{Variables Selected}}\label{CLSelFS}
\begin{center}
\begin{tabular}{c}
\hline
A. $\log$(FSS) \\
\hline
Owner occupied housing units  \\
Renter occupied housing units \\
Males 15 yrs widowed   \\
Institutionalized population  \\
$t \times $ (Total bank deposits)$_0$  \\
$t \times $ (\% Change in households)$_0$  \\
\hline
B. Overall homicide \\
\hline
Persons 5 yrs and over by residence - Same house for last 5 yrs \\
Vote cast for president, third party candidate \\
$t^3 \times$ (Valuation of new housing by building permits)$_0$ \\
\hline
C. Gun homicide \\
\hline
Resident population age 50 - 54 years \\
Vote cast for president, third party candidate \\
Owner occupied housing units\\
Families with income 15,000 - 19,999\\
\hline
D. non-Gun homicide \\
\hline
Resident population median age \\
Persons per household\\
$t \times$  (Hispanic persons 25 years and over)$_0$\\
\hline
\end{tabular}
\end{center}
}
\begin{flushleft}
\footnotesize{The table presents selected variables by Cluster-Lasso in the gun example using our extended list of controls variables.  Variables selected with lagged $\log(FSS)$, the log of the overall homicide rate, the log of the gun homicide rate, and the log of the non-gun homicide rate are given in Panels A, B, C, and D respectively.}
\end{flushleft}
\end{table}

\clearpage

\setlength{\tabcolsep}{10pt}
\begin{table}[H]
\footnotesize{
\begin{center}
\caption{\footnotesize{Panel IV Design 1, $p=n \times (T-2)$} }\label{IVDesign11}
\begin{tabular}{l c c c c  }
\hline \hline
 & $n=50$ &$n=100$     & $n=150$   & $n=200$  \\
			\cline{2-5} & \multicolumn{4}{c} {Replications with No Instruments Selected} \\ \cline{2-5}							
Heteroscedastic Loadings		&	0	&	0	&	0	&	0	\\
Clustered Loading		&	0	&	0	&	0	&	0	\\
		\cline{2-5} & \multicolumn{4}{c} {A. Bias } \\ \cline{2-5}								
Oracle		&	-0.001	&	-0.007	&	0.000	&	-0.005	\\
FE Oracle		&	-0.001	&	-0.010	&	-0.004	&	-0.004	\\
All		&	0.382	&	0.517	&	0.508	&	0.522	\\
Heteroscedastic Loadings		&	0.087	&	0.121	&	0.099	&	0.093	\\
Clustered Loading		&	0.004	&	0.000	&	-0.001	&	-0.001	\\
			\cline{2-5} & \multicolumn{4}{c} {B. RMSE } \\ \cline{2-5}							
Oracle		&	0.043	&	0.075	&	0.065	&	0.057	\\
FE Oracle		&	0.077	&	0.078	&	0.060	&	0.053	\\
All		&	0.384	&	0.518	&	0.508	&	0.522	\\
Heteroscedastic Loadings		&	0.120	&	0.151	&	0.120	&	0.111	\\
Clustered Loading		&	0.081	&	0.078	&	0.062	&	0.053	\\
		\cline{2-5} & \multicolumn{4}{c} {C. Size (Cluster s.e.) } \\ \cline{2-5}								
Oracle		&	0.067	&	0.048	&	0.057	&	0.052	\\
FE Oracle		&	0.062	&	0.065	&	0.053	&	0.056	\\
All		&	1.000	&	1.000	&	1.000	&	1.000	\\
Heteroscedastic Loadings		&	0.328	&	0.526	&	0.504	&	0.519	\\
Clustered Loading		&	0.079	&	0.065	&	0.059	&	0.057	\\
		\cline{2-5} & \multicolumn{4}{c} {D. Size (Heteroscedastic s.e.) } \\ \cline{2-5}								
Oracle		&	0.421	&	0.289	&	0.329	&	0.320	\\
FE Oracle		&	0.249	&	0.240	&	0.234	&	0.214	\\
All		&	1.000	&	1.000	&	1.000	&	1.000	\\
Heteroscedastic Loadings		&	0.586	&	0.710	&	0.709	&	0.715	\\
Clustered Loading		&	0.275	&	0.247	&	0.236	&	0.221	\\

\hline
\end{tabular}
\end{center}
}
\begin{flushleft}
\footnotesize{ This table presents simulation results for the high dimensional instrumental variables model with fixed effects.  Estimators include our proposed Cluster-Lasso estimator (Clustered Loadings) and alternative estimators: heteroscedastic-Lasso (Heteroscedastic Loadings), 2SLS with all instruments (All), an oracle estimator that knows the values of the first-stage coefficients (Oracle), and an oracle estimator that knows the values of the first-stage coefficients and the fixed effects (FE Oracle).  Bias, RMSE, and statistical size for 5\% level tests using clustered standard errors and heteroscedastic standard errors are reported based on 1000 simulation replications.}
\end{flushleft}
\end{table}

\setlength{\tabcolsep}{10pt}
\begin{table}[H]
\footnotesize{
\begin{center}
\caption{\footnotesize{Panel IV Design 1, $p=n \times (T+2)$ } }\label{IVDesign12}
\begin{tabular}{l c c c c  }
\hline \hline
 & $n=50$ &$n=100$     & $n=150$   & $n=200$  \\
			\cline{2-5} & \multicolumn{4}{c} {Replications with No Instruments Selected} \\ \cline{2-5}				Heteroscedastic Loadings		&	0	&	0	&	0	&	0	\\
Clustered Loading		&	0	&	1	&	0	&	0	\\
	\cline{2-5} & \multicolumn{4}{c} {A. Bias } \\ \cline{2-5}						
Oracle		&	0.000	&	-0.008	&	0.000	&	-0.003	\\
FE Oracle		&	0.000	&	-0.009	&	-0.004	&	-0.003	\\
All		&		&		&		&		\\
Heteroscedastic Loadings		&	0.114	&	0.164	&	0.129	&	0.123	\\
Clustered Loading		&	0.005	&	0.002	&	0.000	&	0.001	\\
					\cline{2-5} & \multicolumn{4}{c} {B. RMSE } \\ \cline{2-5}					
Oracle		&	0.043	&	0.075	&	0.064	&	0.055	\\
FE Oracle		&	0.076	&	0.076	&	0.060	&	0.053	\\
All		&		&		&		&		\\
Heteroscedastic Loadings		&	0.143	&	0.186	&	0.146	&	0.140	\\
Clustered Loading		&	0.081	&	0.075	&	0.061	&	0.054	\\
		\cline{2-5}		\cline{2-5} & \multicolumn{4}{c} {C. Size (Cluster s.e.) } \\ \cline{2-5}						
Oracle		&	0.064	&	0.048	&	0.053	&	0.048	\\
FE Oracle		&	0.060	&	0.059	&	0.051	&	0.062	\\
All		&		&		&		&		\\
Heteroscedastic Loadings		&	0.473	&	0.706	&	0.662	&	0.690	\\
Clustered Loading		&	0.079	&	0.067	&	0.056	&	0.060	\\
	 \cline{2-5}		\cline{2-5} & \multicolumn{4}{c} {D. Size (Heteroscedastic s.e.) } \\ \cline{2-5}						
Oracle		&	0.414	&	0.292	&	0.324	&	0.307	\\
FE Oracle		&	0.246	&	0.222	&	0.238	&	0.214	\\
All		&		&		&		&		\\
Heteroscedastic Loadings		&	0.680	&	0.848	&	0.824	&	0.836	\\
Clustered Loading		&	0.274	&	0.227	&	0.234	&	0.221	\\

\hline
\end{tabular}
\end{center}
}
\begin{flushleft}
\footnotesize{ This table presents simulation results for the high dimensional instrumental variables model with fixed effects.  Estimators include our proposed Cluster-Lasso estimator (Clustered Loadings) and alternative estimators: heteroscedastic-Lasso (Heteroscedastic Loadings), 2SLS with all instruments (All), an oracle estimator that knows the values of the first-stage coefficients (Oracle), and an oracle estimator that knows the values of the first-stage coefficients and the fixed effects (FE Oracle).  Bias, RMSE, and statistical size for 5\% level tests using clustered standard errors and heteroscedastic standard errors are reported based on 1000 simulation replications.}
\end{flushleft}
\end{table}

\setlength{\tabcolsep}{10pt}
\begin{table}[H]
\footnotesize{
\begin{center}
\caption{\footnotesize{Panel IV Design 2, $p=n \times (T-2)$ } }\label{IVDesign21}
\begin{tabular}{l c c c c  }
\hline \hline
 & $n=50$ &$n=100$     & $n=150$   & $n=200$  \\
			\cline{2-5} & \multicolumn{4}{c} {Replications with No Instruments Selected} \\ \cline{2-5}				Heteroscedastic Loadings	&	0	&	0	&	0	&	0	\\
Clustered Loading	&	0	&	2	&	0	&	0	\\
		\cline{2-5} & \multicolumn{4}{c} {A. Bias } \\ \cline{2-5}							
Oracle	&	0.000	&	-0.008	&	-0.002	&	-0.004	\\
FE Oracle	&	0.000	&	-0.005	&	-0.004	&	-0.003	\\
All	&	0.300	&	0.417	&	0.416	&	0.424	\\
Heteroscedastic Loadings	&	0.081	&	0.126	&	0.110	&	0.095	\\
Clustered Loading	&	0.003	&	0.004	&	0.000	&	-0.001	\\
			\cline{2-5} & \multicolumn{4}{c} {B. RMSE } \\ \cline{2-5}						
Oracle	&	0.036	&	0.062	&	0.057	&	0.049	\\
FE Oracle	&	0.063	&	0.055	&	0.048	&	0.041	\\
All	&	0.302	&	0.418	&	0.417	&	0.425	\\
Heteroscedastic Loadings	&	0.110	&	0.148	&	0.127	&	0.111	\\
Clustered Loading	&	0.074	&	0.076	&	0.061	&	0.052	\\
		\cline{2-5} & \multicolumn{4}{c} {C. Size (Cluster s.e.) } \\ \cline{2-5}							
Oracle	&	0.060	&	0.043	&	0.058	&	0.060	\\
FE Oracle	&	0.072	&	0.052	&	0.049	&	0.063	\\
All	&	1.000	&	1.000	&	1.000	&	1.000	\\
Heteroscedastic Loadings	&	0.346	&	0.560	&	0.583	&	0.558	\\
Clustered Loading	&	0.069	&	0.062	&	0.048	&	0.057	\\
		\cline{2-5} & \multicolumn{4}{c} {D. Size (Heteroscedastic s.e.) } \\ \cline{2-5}							
Oracle	&	0.411	&	0.273	&	0.328	&	0.284	\\
FE Oracle	&	0.256	&	0.202	&	0.232	&	0.215	\\
All	&	1.000	&	1.000	&	1.000	&	1.000	\\
Heteroscedastic Loadings	&	0.584	&	0.755	&	0.791	&	0.749	\\
Clustered Loading	&	0.273	&	0.232	&	0.236	&	0.214	\\

\hline
\end{tabular}
\end{center}
}
\begin{flushleft}
\footnotesize{ This table presents simulation results for the high dimensional instrumental variables model with fixed effects.  Estimators include our proposed Cluster-Lasso estimator (Clustered Loadings) and alternative estimators: heteroscedastic-Lasso (Heteroscedastic Loadings), 2SLS with all instruments (All), an oracle estimator that knows the values of the first-stage coefficients (Oracle), and an oracle estimator that knows the values of the first-stage coefficients and the fixed effects (FE Oracle).  Bias, RMSE, and statistical size for 5\% level tests using clustered standard errors and heteroscedastic standard errors are reported based on 1000 simulation replications.}
\end{flushleft}
\end{table}

\setlength{\tabcolsep}{10pt}
\begin{table}[H]
\footnotesize{
\begin{center}
\caption{\footnotesize{Panel IV Design 2, $p=n \times (T+2)$ } }\label{IVDesign22}
\begin{tabular}{l c c c c  }
\hline \hline
 & $n=50$ &$n=100$     & $n=150$   & $n=200$  \\
			\cline{2-5} & \multicolumn{4}{c} {Replications with No Instruments Selected} \\ \cline{2-5}				Heteroscedastic Loadings	&	0	&	0	&	0	&	0	\\
Clustered Loading	&	0	&	2	&	0	&	0	\\
				\cline{2-5} & \multicolumn{4}{c} {A. Bias } \\ \cline{2-5}					
Oracle	&	-0.001	&	-0.008	&	-0.003	&	-0.005	\\
FE Oracle	&	0.001	&	-0.005	&	-0.004	&	-0.004	\\
All	&		&		&		&		\\
Heteroscedastic Loadings	&	0.098	&	0.149	&	0.148	&	0.120	\\
Clustered Loading	&	0.004	&	0.005	&	0.000	&	-0.001	\\
					\cline{2-5} & \multicolumn{4}{c} {B. RMSE } \\ \cline{2-5}				
Oracle	&	0.037	&	0.064	&	0.058	&	0.047	\\
FE Oracle	&	0.063	&	0.056	&	0.047	&	0.040	\\
All	&		&		&		&		\\
Heteroscedastic Loadings	&	0.123	&	0.169	&	0.162	&	0.134	\\
Clustered Loading	&	0.073	&	0.075	&	0.062	&	0.052	\\
				\cline{2-5} & \multicolumn{4}{c} {C. Size (Cluster s.e.) } \\ \cline{2-5}					
Oracle	&	0.067	&	0.054	&	0.056	&	0.051	\\
FE Oracle	&	0.062	&	0.055	&	0.050	&	0.064	\\
All	&		&		&		&		\\
Heteroscedastic Loadings	&	0.445	&	0.688	&	0.769	&	0.717	\\
Clustered Loading	&	0.073	&	0.062	&	0.055	&	0.057	\\
				\cline{2-5} & \multicolumn{4}{c} {D. Size (Heteroscedastic s.e.) } \\ \cline{2-5}					
Oracle	&	0.406	&	0.276	&	0.320	&	0.268	\\
FE Oracle	&	0.262	&	0.228	&	0.214	&	0.227	\\
All	&		&		&		&		\\
Heteroscedastic Loadings	&	0.652	&	0.837	&	0.895	&	0.857	\\
Clustered Loading	&	0.275	&	0.226	&	0.222	&	0.216	\\

\hline
\end{tabular}
\end{center}
}
\begin{flushleft}
\footnotesize{ This table presents simulation results for the high dimensional instrumental variables model with fixed effects.  Estimators include our proposed Cluster-Lasso estimator (Clustered Loadings) and alternative estimators: heteroscedastic-Lasso (Heteroscedastic Loadings), 2SLS with all instruments (All), an oracle estimator that knows the values of the first-stage coefficients (Oracle), and an oracle estimator that knows the values of the first-stage coefficients and the fixed effects (FE Oracle).  Bias, RMSE, and statistical size for 5\% level tests using clustered standard errors and heteroscedastic standard errors are reported based on 1000 simulation replications.}
\end{flushleft}
\end{table}

\setlength{\tabcolsep}{10pt}
\begin{table}[H]
\footnotesize{
\begin{center}
\caption{\footnotesize{Panel IV Design 3, $p=n \times (T-2)$ } }\label{IVDesign31}
\begin{tabular}{l c c c c  }
\hline \hline
 & $n=50$ &$n=100$     & $n=150$   & $n=200$  \\
			\cline{2-5} & \multicolumn{4}{c} {Replications with No Instruments Selected} \\ \cline{2-5}				Heteroscedastic Loadings	&	0	&	0	&	0	&	0	\\
Clustered Loading	&	412	&	57	&	0	&	0	\\
		\cline{2-5} & \multicolumn{4}{c} {A. Bias } \\ \cline{2-5}							
Oracle	&	-0.008	&	-0.009	&	-0.003	&	-0.006	\\
FE Oracle	&	-0.010	&	-0.008	&	-0.005	&	-0.005	\\
All	&	0.523	&	0.548	&	0.537	&	0.545	\\
Heteroscedastic Loadings	&	0.184	&	0.193	&	0.135	&	0.122	\\
Clustered Loading	&	0.153	&	0.051	&	0.015	&	0.013	\\
			\cline{2-5} & \multicolumn{4}{c} {B. RMSE } \\ \cline{2-5}						
Oracle	&	0.123	&	0.090	&	0.074	&	0.066	\\
FE Oracle	&	0.100	&	0.082	&	0.063	&	0.057	\\
All	&	0.525	&	0.549	&	0.538	&	0.546	\\
Heteroscedastic Loadings	&	0.213	&	0.213	&	0.152	&	0.137	\\
Clustered Loading	&	0.217	&	0.134	&	0.086	&	0.076	\\
		\cline{2-5} & \multicolumn{4}{c} {C. Size (Cluster s.e.) } \\ \cline{2-5}							
Oracle	&	0.063	&	0.057	&	0.050	&	0.044	\\
FE Oracle	&	0.053	&	0.050	&	0.045	&	0.052	\\
All	&	1.000	&	1.000	&	1.000	&	1.000	\\
Heteroscedastic Loadings	&	0.616	&	0.745	&	0.674	&	0.687	\\
Clustered Loading	&	0.173	&	0.125	&	0.088	&	0.085	\\
		\cline{2-5} & \multicolumn{4}{c} {D. Size (Heteroscedastic s.e.) } \\ \cline{2-5}							
Oracle	&	0.287	&	0.330	&	0.286	&	0.303	\\
FE Oracle	&	0.184	&	0.234	&	0.211	&	0.214	\\
All	&	1.000	&	1.000	&	1.000	&	1.000	\\
Heteroscedastic Loadings	&	0.768	&	0.857	&	0.843	&	0.833	\\
Clustered Loading	&	0.284	&	0.265	&	0.235	&	0.268	\\

\hline
\end{tabular}
\end{center}
}
\begin{flushleft}
\footnotesize{ This table presents simulation results for the high dimensional instrumental variables model with fixed effects.  Estimators include our proposed Cluster-Lasso estimator (Clustered Loadings) and alternative estimators: heteroscedastic-Lasso (Heteroscedastic Loadings), 2SLS with all instruments (All), an oracle estimator that knows the values of the first-stage coefficients (Oracle), and an oracle estimator that knows the values of the first-stage coefficients and the fixed effects (FE Oracle).  Bias, RMSE, and statistical size for 5\% level tests using clustered standard errors and heteroscedastic standard errors are reported based on 1000 simulation replications.}
\end{flushleft}
\end{table}

\setlength{\tabcolsep}{10pt}
\begin{table}[H]
\footnotesize{
\begin{center}
\caption{\footnotesize{Panel IV Design 3, $p=n \times (T+2)$ } }\label{IVDesign32}
\begin{tabular}{l c c c c  }
\hline \hline
 & $n=50$ &$n=100$     & $n=150$   & $n=200$  \\
			\cline{2-5} & \multicolumn{4}{c} {Replications with No Instruments Selected} \\ \cline{2-5}				Heteroscedastic Loadings	&	0	&	0	&	0	&	0	\\
Clustered Loading	&	461	&	80	&	0	&	0	\\
				\cline{2-5} & \multicolumn{4}{c} {A. Bias } \\ \cline{2-5}					
Oracle	&	-0.009	&	-0.009	&	-0.003	&	-0.006	\\
FE Oracle	&	-0.010	&	-0.008	&	-0.004	&	-0.004	\\
All	&		&		&		&		\\
Heteroscedastic Loadings	&	0.232	&	0.239	&	0.174	&	0.160	\\
Clustered Loading	&	0.167	&	0.054	&	0.015	&	0.012	\\
					\cline{2-5} & \multicolumn{4}{c} {B. RMSE } \\ \cline{2-5}				
Oracle	&	0.121	&	0.089	&	0.074	&	0.067	\\
FE Oracle	&	0.101	&	0.082	&	0.063	&	0.057	\\
All	&		&		&		&		\\
Heteroscedastic Loadings	&	0.254	&	0.256	&	0.188	&	0.173	\\
Clustered Loading	&	0.233	&	0.132	&	0.086	&	0.078	\\
				\cline{2-5} & \multicolumn{4}{c} {C. Size (Cluster s.e.) } \\ \cline{2-5}					
Oracle	&	0.059	&	0.055	&	0.049	&	0.047	\\
FE Oracle	&	0.050	&	0.052	&	0.049	&	0.055	\\
All	&		&		&		&		\\
Heteroscedastic Loadings	&	0.770	&	0.871	&	0.829	&	0.846	\\
Clustered Loading	&	0.181	&	0.123	&	0.085	&	0.088	\\
				\cline{2-5} & \multicolumn{4}{c} {D. Size (Heteroscedastic s.e.) } \\ \cline{2-5}					
Oracle	&	0.275	&	0.325	&	0.279	&	0.307	\\
FE Oracle	&	0.188	&	0.225	&	0.210	&	0.219	\\
All	&		&		&		&		\\
Heteroscedastic Loadings	&	0.880	&	0.933	&	0.914	&	0.935	\\
Clustered Loading	&	0.280	&	0.255	&	0.230	&	0.274	\\

\hline
\end{tabular}
\end{center}
}
\begin{flushleft}
\footnotesize{ This table presents simulation results for the high dimensional instrumental variables model with fixed effects.  Estimators include our proposed Cluster-Lasso estimator (Clustered Loadings) and alternative estimators: heteroscedastic-Lasso (Heteroscedastic Loadings), 2SLS with all instruments (All), an oracle estimator that knows the values of the first-stage coefficients (Oracle), and an oracle estimator that knows the values of the first-stage coefficients and the fixed effects (FE Oracle).  Bias, RMSE, and statistical size for 5\% level tests using clustered standard errors and heteroscedastic standard errors are reported based on 1000 simulation replications.}
\end{flushleft}
\end{table}

\setlength{\tabcolsep}{10pt}
\begin{table}[H]
\footnotesize{
\begin{center}
\caption{\footnotesize{Panel PLM Design 1, $p=n \times (T-2)$ } }\label{LMDesign11}
\begin{tabular}{l c c c c }
\hline \hline
 & $n=50$ &$n=100$     & $n=150$   & $n=200$  \\
\cline{2-5} & \multicolumn{4}{c} {A. Bias } \\ \cline{2-5}						
Oracle		&	0.003	&	0.002	&	0.001	&	0.000	\\
FE Oracle		&	0.004	&	-0.002	&	0.001	&	0.001	\\
All		&	0.005	&	-0.001	&	-0.005	&	0.000	\\
Select over FE		&	0.070	&	0.014	&	-0.021	&	-0.020	\\
Heteroscedastic Loadings		&	0.007	&	-0.023	&	-0.015	&	-0.012	\\
Clustered Loading		&	0.040	&	0.006	&	0.010	&	0.009	\\
			\cline{2-5} & \multicolumn{4}{c} {B. RMSE } \\ \cline{2-5}							
Oracle		&	0.089	&	0.058	&	0.051	&	0.042	\\
FE Oracle		&	0.074	&	0.051	&	0.042	&	0.037	\\
All		&	0.150	&	0.099	&	0.085	&	0.073	\\
Select over FE		&	0.108	&	0.087	&	0.052	&	0.046	\\
Heteroscedastic Loadings		&	0.074	&	0.057	&	0.045	&	0.038	\\
Clustered Loading		&	0.084	&	0.051	&	0.043	&	0.038	\\
			\cline{2-5} & \multicolumn{4}{c} {C. Size (Cluster s.e.) } \\ \cline{2-5}							
Oracle		&	0.075	&	0.056	&	0.067	&	0.050	\\
FE Oracle		&	0.060	&	0.052	&	0.047	&	0.057	\\
All		&	0.514	&	0.467	&	0.494	&	0.458	\\
Select over FE		&	0.194	&	0.174	&	0.085	&	0.092	\\
Heteroscedastic Loadings		&	0.085	&	0.101	&	0.076	&	0.081	\\
Clustered Loading		&	0.093	&	0.062	&	0.059	&	0.057	\\
			\cline{2-5} & \multicolumn{4}{c} {D. Size (Heteroscedastic s.e.) } \\ \cline{2-5}							
Oracle		&	0.330	&	0.289	&	0.329	&	0.297	\\
FE Oracle		&	0.236	&	0.213	&	0.230	&	0.216	\\
All		&	0.562	&	0.532	&	0.554	&	0.519	\\
Select over FE		&	0.414	&	0.328	&	0.310	&	0.313	\\
Heteroscedastic Loadings		&	0.227	&	0.262	&	0.259	&	0.235	\\
Clustered Loading		&	0.294	&	0.212	&	0.226	&	0.236	\\
\hline
\end{tabular}
\end{center}
}
\begin{flushleft}
\footnotesize{This table presents simulation results from a linear fixed effects model.  Estimators include our proposed Cluster-Lasso estimator (Clustered Loadings), heteroscedastic-Lasso (Heteroscedastic Loadings), a double-selection estimator that includes the fixed effects in the set of variables to be selected over (Select over FE), fixed effects using all controls (All), an oracle estimator that knows the values of the coefficients on the control variables (Oracle), and an oracle estimator that knows the values of the coefficients on the controls variables and the fixed effects (FE Oracle).  Bias, RMSE, and statistical size for 5\% level tests using clustered standard errors and heteroscedastic standard errors are reported based on 1000 simulation replications.}
\end{flushleft}
\end{table}

\setlength{\tabcolsep}{10pt}
\begin{table}[H]
\footnotesize{
\begin{center}
\caption{\footnotesize{Panel PLM Design 1, $p=n \times (T+2)$ } }\label{LMDesign12}
\begin{tabular}{l c c c c }
\hline \hline
 & $n=50$ &$n=100$     & $n=150$   & $n=200$  \\
\cline{2-5} & \multicolumn{4}{c} {A. Bias } \\ \cline{2-5}							

Oracle	&	-0.002	&	-0.002	&	0.000	&	0.001	\\
FE Oracle	&	0.000	&	-0.001	&	0.003	&	0.000	\\
All	&		&		&		&		\\
Select over FE	&	0.075	&	0.011	&	-0.020	&	-0.019	\\
Heteroscedastic Loadings	&	-0.006	&	-0.034	&	-0.022	&	-0.020	\\
Clustered Loading	&	0.035	&	0.007	&	0.011	&	0.008	\\
	\cline{2-5} & \multicolumn{4}{c} {B. RMSE } \\ \cline{2-5}								
Oracle	&	0.084	&	0.060	&	0.050	&	0.042	\\
FE Oracle	&	0.074	&	0.053	&	0.042	&	0.037	\\
All	&		&		&		&		\\
Select over FE	&	0.109	&	0.089	&	0.054	&	0.045	\\
Heteroscedastic Loadings	&	0.074	&	0.061	&	0.047	&	0.042	\\
Clustered Loading	&	0.081	&	0.053	&	0.043	&	0.038	\\
	\cline{2-5} & \multicolumn{4}{c} {C. Size (Cluster s.e.) } \\ \cline{2-5}								
Oracle	&	0.061	&	0.060	&	0.071	&	0.049	\\
FE Oracle	&	0.060	&	0.062	&	0.046	&	0.058	\\
All	&		&		&		&		\\
Select over FE	&	0.210	&	0.180	&	0.096	&	0.091	\\
Heteroscedastic Loadings	&	0.088	&	0.151	&	0.119	&	0.127	\\
Clustered Loading	&	0.093	&	0.071	&	0.066	&	0.062	\\
	\cline{2-5} & \multicolumn{4}{c} {D. Size (Heteroscedastic s.e.) } \\ \cline{2-5}								
Oracle	&	0.311	&	0.316	&	0.322	&	0.301	\\
FE Oracle	&	0.218	&	0.223	&	0.212	&	0.215	\\
All	&		&		&		&		\\
Select over FE	&	0.448	&	0.355	&	0.303	&	0.290	\\
Heteroscedastic Loadings	&	0.216	&	0.301	&	0.257	&	0.281	\\
Clustered Loading	&	0.277	&	0.228	&	0.230	&	0.229	\\

\hline
\end{tabular}
\end{center}
}
\begin{flushleft}
\footnotesize{This table presents simulation results from a linear fixed effects model.  Estimators include our proposed Cluster-Lasso estimator (Clustered Loadings), heteroscedastic-Lasso (Heteroscedastic Loadings), a double-selection estimator that includes the fixed effects in the set of variables to be selected over (Select over FE), fixed effects using all controls (All), an oracle estimator that knows the values of the coefficients on the control variables (Oracle), and an oracle estimator that knows the values of the coefficients on the controls variables and the fixed effects (FE Oracle).  Bias, RMSE, and statistical size for 5\% level tests using clustered standard errors and heteroscedastic standard errors are reported based on 1000 simulation replications.}
\end{flushleft}
\end{table}

\setlength{\tabcolsep}{10pt}
\begin{table}[H]
\footnotesize{
\begin{center}
\caption{\footnotesize{Panel PLM Design 2, $p=n \times (T-2)$ } }\label{LMDesign21}
\begin{tabular}{l c c c c }
\hline \hline
 & $n=50$ &$n=100$     & $n=150$   & $n=200$  \\
\cline{2-5} & \multicolumn{4}{c} {A. Bias } \\ \cline{2-5}						

Oracle		&	0.001	&	0.002	&	0.000	&	0.001	\\
FE Oracle		&	0.001	&	0.004	&	0.000	&	0.001	\\
All		&	0.004	&	0.000	&	0.003	&	0.000	\\
Select over FE		&	0.028	&	0.003	&	-0.027	&	-0.029	\\
Heteroscedastic Loadings		&	-0.027	&	-0.021	&	-0.020	&	-0.014	\\
Clustered Loading		&	0.007	&	0.007	&	0.002	&	0.005	\\
		\cline{2-5} & \multicolumn{4}{c} {B. RMSE } \\ \cline{2-5}								
Oracle		&	0.084	&	0.060	&	0.047	&	0.042	\\
FE Oracle		&	0.073	&	0.053	&	0.043	&	0.036	\\
All		&	0.150	&	0.105	&	0.081	&	0.073	\\
Select over FE		&	0.074	&	0.069	&	0.051	&	0.046	\\
Heteroscedastic Loadings		&	0.070	&	0.050	&	0.044	&	0.035	\\
Clustered Loading		&	0.065	&	0.046	&	0.038	&	0.032	\\
		\cline{2-5} & \multicolumn{4}{c} {C. Size (Cluster s.e.) } \\ \cline{2-5}								
Oracle		&	0.063	&	0.058	&	0.040	&	0.051	\\
FE Oracle		&	0.066	&	0.064	&	0.060	&	0.055	\\
All		&	0.472	&	0.494	&	0.441	&	0.452	\\
Select over FE		&	0.104	&	0.147	&	0.122	&	0.139	\\
Heteroscedastic Loadings		&	0.114	&	0.118	&	0.112	&	0.077	\\
Clustered Loading		&	0.072	&	0.066	&	0.061	&	0.051	\\
			\cline{2-5} & \multicolumn{4}{c} {D. Size (Heteroscedastic s.e.) } \\ \cline{2-5}							
Oracle		&	0.300	&	0.299	&	0.298	&	0.293	\\
FE Oracle		&	0.225	&	0.226	&	0.224	&	0.207	\\
All		&	0.543	&	0.552	&	0.510	&	0.513	\\
Select over FE		&	0.278	&	0.321	&	0.342	&	0.384	\\
Heteroscedastic Loadings		&	0.259	&	0.262	&	0.294	&	0.257	\\
Clustered Loading		&	0.231	&	0.226	&	0.234	&	0.218	\\

\hline
\end{tabular}
\end{center}
}
\begin{flushleft}
\footnotesize{This table presents simulation results from a linear fixed effects model.  Estimators include our proposed Cluster-Lasso estimator (Clustered Loadings), heteroscedastic-Lasso (Heteroscedastic Loadings), a double-selection estimator that includes the fixed effects in the set of variables to be selected over (Select over FE), fixed effects using all controls (All), an oracle estimator that knows the values of the coefficients on the control variables (Oracle), and an oracle estimator that knows the values of the coefficients on the controls variables and the fixed effects (FE Oracle).  Bias, RMSE, and statistical size for 5\% level tests using clustered standard errors and heteroscedastic standard errors are reported based on 1000 simulation replications.}
\end{flushleft}
\end{table}

\setlength{\tabcolsep}{10pt}
\begin{table}[H]
\footnotesize{
\begin{center}
\caption{\footnotesize{Panel PLM Design 2, $p=n \times (T+2)$ } }\label{LMDesign22}
\begin{tabular}{l c c c c }
\hline \hline
 & $n=50$ &$n=100$     & $n=150$   & $n=200$  \\
\cline{2-5} & \multicolumn{4}{c} {A. Bias } \\ \cline{2-5}						

Oracle	&	-0.002	&	-0.003	&	-0.001	&	0.001	\\
FE Oracle	&	0.000	&	-0.001	&	0.001	&	0.001	\\
All	&		&		&		&		\\
Select over FE	&	0.035	&	-0.005	&	-0.023	&	-0.028	\\
Heteroscedastic Loadings	&	-0.038	&	-0.035	&	-0.028	&	-0.023	\\
Clustered Loading	&	0.002	&	0.002	&	0.004	&	0.005	\\
\cline{2-5} & \multicolumn{4}{c} {B. RMSE } \\ \cline{2-5}									
Oracle	&	0.084	&	0.060	&	0.050	&	0.042	\\
FE Oracle	&	0.074	&	0.053	&	0.041	&	0.037	\\
All	&		&		&		&		\\
Select over FE	&	0.079	&	0.071	&	0.048	&	0.046	\\
Heteroscedastic Loadings	&	0.077	&	0.058	&	0.047	&	0.039	\\
Clustered Loading	&	0.064	&	0.046	&	0.036	&	0.032	\\
\cline{2-5} & \multicolumn{4}{c} {C. Size (Cluster s.e.) } \\ \cline{2-5}									
Oracle	&	0.063	&	0.061	&	0.065	&	0.047	\\
FE Oracle	&	0.060	&	0.059	&	0.046	&	0.059	\\
All	&		&		&		&		\\
Select over FE	&	0.129	&	0.161	&	0.099	&	0.147	\\
Heteroscedastic Loadings	&	0.157	&	0.180	&	0.151	&	0.143	\\
Clustered Loading	&	0.058	&	0.067	&	0.058	&	0.055	\\
	\cline{2-5} & \multicolumn{4}{c} {D. Size (Heteroscedastic s.e.) } \\ \cline{2-5}								
Oracle	&	0.318	&	0.306	&	0.314	&	0.298	\\
FE Oracle	&	0.218	&	0.227	&	0.212	&	0.209	\\
All	&		&		&		&		\\
Select over FE	&	0.308	&	0.342	&	0.312	&	0.396	\\
Heteroscedastic Loadings	&	0.291	&	0.360	&	0.337	&	0.320	\\
Clustered Loading	&	0.207	&	0.240	&	0.223	&	0.209	\\

\hline
\end{tabular}
\end{center}
}
\begin{flushleft}
\footnotesize{This table presents simulation results from a linear fixed effects model.  Estimators include our proposed Cluster-Lasso estimator (Clustered Loadings), heteroscedastic-Lasso (Heteroscedastic Loadings), a double-selection estimator that includes the fixed effects in the set of variables to be selected over (Select over FE), fixed effects using all controls (All), an oracle estimator that knows the values of the coefficients on the control variables (Oracle), and an oracle estimator that knows the values of the coefficients on the controls variables and the fixed effects (FE Oracle).  Bias, RMSE, and statistical size for 5\% level tests using clustered standard errors and heteroscedastic standard errors are reported based on 1000 simulation replications.}
\end{flushleft}
\end{table}

\setlength{\tabcolsep}{10pt}
\begin{table}[H]
\footnotesize{
\begin{center}
\caption{\footnotesize{Panel PLM Design 3, $p=n \times (T-2)$ } }\label{LMDesign31}
\begin{tabular}{l c c c c}
\hline \hline
 & $n=50$ &$n=100$     & $n=150$   & $n=200$  \\
\cline{2-5} & \multicolumn{4}{c} {A. Bias } \\ \cline{2-5}						

Oracle	&	-0.001	&	-0.001	&	-0.002	&	0.002	\\
FE Oracle	&	0.001	&	0.000	&	0.000	&	0.001	\\
All	&	-0.003	&	0.004	&	-0.002	&	0.000	\\
Select over FE	&	-0.011	&	0.048	&	0.000	&	-0.028	\\
Heteroscedastic Loadings	&	-0.029	&	-0.027	&	-0.022	&	-0.019	\\
Clustered Loading	&	-0.001	&	0.000	&	-0.001	&	0.000	\\
		\cline{2-5} & \multicolumn{4}{c} {B. RMSE } \\ \cline{2-5}							
Oracle	&	0.085	&	0.059	&	0.049	&	0.042	\\
FE Oracle	&	0.075	&	0.052	&	0.042	&	0.037	\\
All	&	0.145	&	0.101	&	0.086	&	0.073	\\
Select over FE	&	0.069	&	0.076	&	0.068	&	0.054	\\
Heteroscedastic Loadings	&	0.079	&	0.057	&	0.048	&	0.041	\\
Clustered Loading	&	0.068	&	0.050	&	0.039	&	0.035	\\
		\cline{2-5} & \multicolumn{4}{c} {C. Size (Cluster s.e.) } \\ \cline{2-5}							
Oracle	&	0.067	&	0.053	&	0.064	&	0.049	\\
FE Oracle	&	0.063	&	0.054	&	0.048	&	0.058	\\
All	&	0.473	&	0.470	&	0.460	&	0.455	\\
Select over FE	&	0.087	&	0.189	&	0.240	&	0.197	\\
Heteroscedastic Loadings	&	0.117	&	0.111	&	0.101	&	0.092	\\
Clustered Loading	&	0.075	&	0.076	&	0.053	&	0.047	\\
		\cline{2-5} & \multicolumn{4}{c} {D. Size (Heteroscedastic s.e.) } \\ \cline{2-5}							
Oracle	&	0.323	&	0.302	&	0.313	&	0.300	\\
FE Oracle	&	0.226	&	0.221	&	0.211	&	0.214	\\
All	&	0.527	&	0.548	&	0.541	&	0.516	\\
Select over FE	&	0.218	&	0.441	&	0.510	&	0.458	\\
Heteroscedastic Loadings	&	0.238	&	0.259	&	0.257	&	0.273	\\
Clustered Loading	&	0.212	&	0.232	&	0.216	&	0.217	\\

\hline
\end{tabular}
\end{center}
}
\begin{flushleft}
\footnotesize{This table presents simulation results from a linear fixed effects model.  Estimators include our proposed Cluster-Lasso estimator (Clustered Loadings), heteroscedastic-Lasso (Heteroscedastic Loadings), a double-selection estimator that includes the fixed effects in the set of variables to be selected over (Select over FE), fixed effects using all controls (All), an oracle estimator that knows the values of the coefficients on the control variables (Oracle), and an oracle estimator that knows the values of the coefficients on the controls variables and the fixed effects (FE Oracle).  Bias, RMSE, and statistical size for 5\% level tests using clustered standard errors and heteroscedastic standard errors are reported based on 1000 simulation replications.}
\end{flushleft}
\end{table}

\setlength{\tabcolsep}{10pt}
\begin{table}[H]
\footnotesize{
\begin{center}
\caption{\footnotesize{Panel PLM Design 3, $p=n \times (T+2)$ } }\label{LMDesign32}
\begin{tabular}{l c c c c  }
\hline \hline
 & $n=50$ &$n=100$     & $n=150$   & $n=200$  \\
\cline{2-5} & \multicolumn{4}{c} {A. Bias } \\ \cline{2-5}						

Oracle	&	-0.001	&	-0.002	&	0.000	&	0.001	\\
FE Oracle	&	0.001	&	0.000	&	0.003	&	0.000	\\
All	&		&		&		&		\\
Select over FE	&	-0.007	&	0.052	&	0.010	&	-0.024	\\
Heteroscedastic Loadings	&	-0.037	&	-0.035	&	-0.027	&	-0.026	\\
Clustered Loading	&	-0.002	&	0.000	&	0.002	&	0.000	\\
	\cline{2-5} & \multicolumn{4}{c} {B. RMSE } \\ \cline{2-5}								
Oracle	&	0.083	&	0.059	&	0.050	&	0.042	\\
FE Oracle	&	0.074	&	0.052	&	0.042	&	0.037	\\
All	&		&		&		&		\\
Select over FE	&	0.067	&	0.077	&	0.070	&	0.055	\\
Heteroscedastic Loadings	&	0.082	&	0.061	&	0.050	&	0.045	\\
Clustered Loading	&	0.067	&	0.052	&	0.039	&	0.035	\\
	\cline{2-5} & \multicolumn{4}{c} {C. Size (Cluster s.e.) } \\ \cline{2-5}								
Oracle	&	0.061	&	0.056	&	0.069	&	0.048	\\
FE Oracle	&	0.059	&	0.058	&	0.048	&	0.062	\\
All	&		&		&		&		\\
Select over FE	&	0.083	&	0.199	&	0.259	&	0.211	\\
Heteroscedastic Loadings	&	0.137	&	0.158	&	0.127	&	0.149	\\
Clustered Loading	&	0.072	&	0.085	&	0.055	&	0.049	\\
	\cline{2-5} & \multicolumn{4}{c} {D. Size (Heteroscedastic s.e.) } \\ \cline{2-5}								
Oracle	&	0.309	&	0.297	&	0.318	&	0.299	\\
FE Oracle	&	0.218	&	0.222	&	0.213	&	0.221	\\
All	&		&		&		&		\\
Select over FE	&	0.217	&	0.446	&	0.530	&	0.488	\\
Heteroscedastic Loadings	&	0.251	&	0.309	&	0.285	&	0.333	\\
Clustered Loading	&	0.202	&	0.231	&	0.223	&	0.225	\\

\hline
\end{tabular}
\end{center}
}
\begin{flushleft}
\footnotesize{This table presents simulation results from a linear fixed effects model.  Estimators include our proposed Cluster-Lasso estimator (Clustered Loadings), heteroscedastic-Lasso (Heteroscedastic Loadings), a double-selection estimator that includes the fixed effects in the set of variables to be selected over (Select over FE), fixed effects using all controls (All), an oracle estimator that knows the values of the coefficients on the control variables (Oracle), and an oracle estimator that knows the values of the coefficients on the controls variables and the fixed effects (FE Oracle).  Bias, RMSE, and statistical size for 5\% level tests using clustered standard errors and heteroscedastic standard errors are reported based on 1000 simulation replications.}
\end{flushleft}
\end{table}

\end{document}